\documentclass[10pt]{IIMAS-conm-p-l}
\makeatletter
\def\@stpelt#1{\stepcounter{#1}\global\csname c@#1\endcsname \z@}
\makeatother

\usepackage[english]{babel}
\usepackage{enumerate}
\usepackage{amssymb}
\usepackage{amsthm}
\usepackage{exscale}

\newcommand{\be}{\begin{equation}}
\newcommand{\ee}{\end{equation}}
\newcommand{\bea}{\begin{eqnarray*}}
\newcommand{\eea}{\end{eqnarray*}}
\newcommand{\beq}{\begin{eqnarray}}
\newcommand{\eeq}{\end{eqnarray}}
\newcommand{\nn}{\nonumber}
\newcommand{\RR}{\mathbb{R}}
\newcommand{\CC}{\mathbb{C}}

\newcommand{\NN}{\mathbb{N}}
\newcommand{\ZZ}{\mathbb{Z}}
\newcommand{\TT}{\mathbb{T}}
\newcommand{\EE}{\, \mathbb{E} \,}
\newcommand{\PP}{\mathbb{P}}
\newcommand{\QQ}{\mathbb{Q}}

\newcommand{\cB}{\mathcal{B}}

\newcommand{\cD}{\mathcal{D}}
\newcommand{\cE}{\mathcal{E}}
\newcommand{\cF}{\mathcal{F}}
\newcommand{\cH}{\mathcal{H}}
\newcommand{\cJ}{\mathcal{J}}
\newcommand{\cL}{\mathcal{L}}
\newcommand{\cN}{\mathcal{N}}
\newcommand{\cO}{\mathcal{O}}
\newcommand{\cP}{\mathcal{P}}

\newcommand{\cV}{\mathcal{V}}

\renewcommand{\oe}{\o}

\newcommand{\eff}{\mathop{\mathrm{eff}}}

\newcommand{\la}{\langle}
\newcommand{\ra}{\rangle}
\newcommand{\rank}[1]{\mathop{\mathrm{rank}}(#1) }

\newcommand{\mindex}[1]{\hspace{0em}}

\newcommand{\modcap}{\, \cap \,}
\newcommand{\modsubset}{\, \subset \,}
\renewcommand{\Im}{\mathop{\mathrm{Im}}}

\newcommand{\diam}{{\mathop{\mathrm{diam}}}}

\newcommand{\supp}{{\mathop{\mathrm{supp}}}}
\newcommand{\Id}{{\mathop{\mathrm{Id}}}}
\newcommand{\Tr}{{\mathop{\mathrm{Tr}}}}
\newcommand{\vol}{{\mathop{\mathrm{vol}}}}
\newcommand{\Span}{{\mathop{\mathrm{span}}}}

\font\medtimesfont=cmsy10 scaled\magstep3
    \def\medtimes{\mathop{\mathstrut\smash{\vcenter     {\hbox{\medtimesfont\char"02}}}\!}\limits}

\newcommand{\loc}{{\mathop{\mathrm{loc}}}}
\newcommand{\unif}{{\mathop{\mathrm{unif}}}}
\newcommand{\per}{{\mathop{\mathrm{per}}}}

\newcommand{\dist}{{\mathop{\mathrm{dist}}}}
\DeclareMathOperator{\Ric}{\mathrm{Ric}}
\DeclareMathOperator{\Var}{\mathrm{Var}}

\DeclareMathOperator{\inter}{\mathrm{int}}
\newcommand{\half}{\textstyle\frac{1}{2}}
\newtheorem{thm}{Theorem}[subsection]
\newtheorem{lem}[thm]{Lemma}
\newtheorem{prp}[thm]{Proposition}

\theoremstyle{definition}
\newtheorem{dfn}[thm]{Definition}

\theoremstyle{remark}
\newtheorem{rem}[thm]{Remark}
\newtheorem{exm}[thm]{Example}
\theoremstyle{theorem}
\newtheorem{athm}{Theorem}[section]
\newtheorem{alem}[athm]{Lemma}

\theoremstyle{definition}

\theoremstyle{remark}
\newtheorem{arem}[athm]{Remark}

\numberwithin{equation}{section}

\renewcommand{\L}{\Lambda}
\newcommand{\tL}{\tilde\Lambda}
\newcommand{\Lp}{\mathop{\Lambda^+}}

\newcommand{\dirH}{\oplus H}

\newcommand{\Erw}{ \mbox{\bf E}_x }

\newcommand{\CcsmoothD}{C_c^{\infty} (\Lambda)}

\usepackage{backref}
\makeindex
\begin{document}

\title[Integrated density of states and Wegner estimates]{Integrated density of states and Wegner estimates for random Schr\"odinger Operators}

\author[I.~Veseli\'c]{Ivan Veseli\'c}
\address{Forschungsstipendiat der Deutschen Forschungsgemeinschaft}
\email{ivan.veselic@ruhr-uni-bochum.de}
\urladdr{http://homepage.ruhr-uni-bochum.de/Ivan.Veselic/}
\curraddr{Dept.~of Mathematics, California Institute of
Technology, CA 91125, USA}

\date{\today}

\thanks{\hspace*{-2em}See also e-print math-ph/0307062 on arxiv.org and no.~03-473 on www.ma.utexas.edu/mp\_arc/}


\begin{abstract}
We survey recent results on spectral properties of random
Schr\"o\-dinger operators. The focus is set on the integrated
density of states (IDS). First we present a proof of the existence
of a self-averaging IDS which is general enough to be applicable to
random  Schr\"o\-dinger and Laplace-Beltrami operators on manifolds.
Subsequently we study more specific models in Euclidean space,
namely of alloy type, and concentrate on the regularity properties
of the IDS. We discuss  the role of the integrated density of
states and its regularity properties for the  spectral analysis of
random Schr\"o\-dinger operators, particularly in relation to
localisation.
Proofs of the central results are given in detail. Whenever there are alternative proofs,
the different approaches are compared.
\\[1em]
\textsc{Resumen.}
Revisamos  resultados recientes en propiedades espectrales de operadores de Schr\"o\-dinger aleatorios.
Nos enfocamos principialmente en la densidad integrada de estados (IDS).
Primero presentamos una prueba de la existencia de la IDS y su propriedad auto-promediadora (self-averaging).
El m\'etodo es suficientemente general para ser aplicable a operadores de Schr\" odinger
y  de Laplace-Beltrami aleatorios en variedades de Riemann.
Posteriormente estudiamos los modelos m\'as espec\'ificos en el espacio Euclidiano, a saber de tipo aleaci\'on,
y nos concentramos en las propiedades de la regularidad de la IDS.
Discutimos el papel de la densidad integrada de estados y sus propiedades de regularidad para el an\'alisis espectral de operadores de Schr\" odinger aleatorios, particularmente con relaci\'on a localizaci\'on espectral.
Las pruebas de los resultados centrales son descritas en detalle. Cuando hay pruebas alternativas, los enfoques diferentes se comparan.
\end{abstract}
\maketitle

\tableofcontents

\section{Random operators}

\subsection{Physical background}
\index{Schr\"odinger operator!random ---}Random Schr\"o\-dinger operators are used as models of disordered solids within the framework of quantum mechanics.

A macroscopic solid consists of an order of magnitude of $10^{23}$ of nuclei and electrons.
The resulting Hamiltonian taking into account all interactions is highly complicated.
To arrive at a Schr\"o\-dinger operator which can be studied in some detail one neglects the electron-electron interaction and treats the nuclei in the infinite mass approximation.
Thus one arrives at an \index{one-electron Hamiltonian} one-electron Schr\"o\-dinger operator with an external potential due to the electric forces between the electron and the  nuclei, which are assumed to be fixed in space.

In the case that the nuclei are arranged \mindex{periodic}periodically on a lattice, the potential energy of the electron is a periodic function of the space variable.

On the other hand, the electron could be moving in an amorphous medium, in which case there is no large group of symmetries of the Hamiltonian. However, from the physical point of view it is reasonable to assume
that the local structure of the medium will be translation invariant \emph{on average}.
This means that we consider the potential which the electron experiences as a particular realisation of a random process and assume stationarity with respect to some group of translations.
Moreover, physical intuition suggests to assume that the local properties of the medium in two regions far apart (on the microscopic scale) are approximately independent from each other. Therefore the stochastic process describing the potential should have a correlation function which decays to zero, or --- more precisely --- should be \index{ergodic} ergodic.

There are interesting models which lie between the two extreme cases of lattice-periodic and amorphous media.
They still have an underlying lattice structure which is, however, modified by disorder. Probably the best studied Hamiltonian with this properties is the \index{alloy type!--- model}alloy type model. We leave its precise definition for the next paragraph and
introduce here a special case on the intuitive level. Consider first the potential
\[
V_\omega (x):= \sum_{k\in \ZZ^d}  u_k(\omega,x)
\]
Each $k$ corresponds to a nucleus sitting on a lattice point. The function $u_k(\omega,\cdot)$ describes the atomic or nuclear potential at the site $k$ and depends on the random parameter $\omega$ which models the different realisations  of the configuration of the nuclei.
If there is only one type of atom present, which has a spherically symmetric potential,
all the $u_k(\omega,\cdot)$ are the same, and $V_\omega$ is periodic. Now assume that there are two kinds $a$ and $b$ of atoms present, which have spherically symmetric atomic potentials of the same shape, but which differ in their \mindex{nuclear charge numbers}nuclear charge numbers.

In this case the potential looks like
\[
V_\omega (x)
:= \sum_{\text{$k$ occupied by $a$}}  q_a \, u(x-k)  + \sum_{\text{$k$ occupied by $b$}}  q_b \, u(x-k)
\]
If the two sorts of atoms are arranged on the lattice in a regular pattern, this again gives  rise to a periodic potential.

However, there are physically interesting situations (e.g.~\index{binary alloy} binary alloys) where the type of atom sitting on site $k$ is random, for example obeying the law
\[
\PP\{k \text{ is occupied by atom } a  \}= P, \qquad \PP\{k \text{ is occupied by atom } b  \}= 1-P \quad P>0
\]
Here $\PP\{\dots\}$ denotes the probability of an event. If we furthermore assume that the above probabilities are independent at each site and the parameter $P$ is the same for all $k$, we arrive at the continuum
\index{Bernoulli!--- -Anderson potential}\emph{Bernoulli-Anderson potential}
\[
V_\omega(x) = \sum_{k} q_k(\omega) \, u(x-k)
\]
Here $q_k(\omega) \in \{q_a,q_b\}, k\in \ZZ^d$ denotes a collection of \mindex{independent, identically distributed}independent, identically distributed \index{Bernoulli!--- random variables}Bernoulli random variables and $u$ is some atomic potential.

This model is a prototype which has motivated much research in the physics and mathematics literature, a part of which we will review in the present work.
\smallskip

Properties of disordered systems are discussed in the books \cite{Bonch-BruevichEEKMZ-1984,EfrosS-84,LifshitzGP-88} from the point of view of theoretical physics. The mathematical literature on random Schr\"o\-dinger operators includes the books \cite{CarmonaL-1990,PasturF-92}, the introductory article \cite{Kirsch-89a}, a section on random Jacobi matrices in \cite{CyconFKS-87}, the Lifshitz memorial issue \cite{Lifschitz-1985}, and  a monograph specialised on localisation phenomena \cite{Stollmann-01}.

In the present work we focus on the spectral features of random
Hamiltonians which are encoded in, or in close relation to the
properties of the integrated density of states (IDS). We present the
proofs of most of the results in detail. Indeed, for the two
central theorems --- the existence of an self-averaging integrated
density of states and for the Wegner estimate --- two independent
proofs are given and compared.

\index{IDS|see{integrated density of states}}

\subsection{Model and Notation}
\label{ss-MN}
Let us start with some mathematical notation. The symbols $\RR,\ZZ, \NN, \NN_0$ denote the set of reals, the set of integers, the set of natural numbers, and the set on non-negative integers, respectively. For  a set $A$ we denote by $A^c$ its complement. An open subset of $\RR^d$
will be denoted by $\Lambda$, and if there is a sequence of such sets its members will be denoted $\Lambda_1, \dots, \Lambda_l,\dots$. The symbol $|\Lambda|$ is used for the Lebesgue measure of $\Lambda$. We write $|x|$ for the norm of $x\in \RR^d$, while the norm of a vector $f$ in a function space is denoted by $\|f\|$.

The Hilbert space of (equivalence classes of) measurable functions on $\Lambda$ which are square integrable with respect to Lebesgue measure is denoted by  $L^2(\Lambda)$. Similarly, \mindex{$L^p$}$L^p(\Lambda)$ with $p\ge 0$ stands for the space of measurable functions $f$ such that $|f|^p$ is integrable, while $L^\infty(\Lambda)$ is the set of measurable functions which are essentially bounded with respect to Lebesgue measure. The space of sequences $\{a_n\}_{ n \in \NN}$ such that $|a_n| ^p$ is summable is denoted by $l^p(\NN)$. Note that the case $p \in ]0,1[$  is included in our notation.
In our context we will often choose the exponent $p$ dependent on the dimension of the configuration space. In the following we denote by \mindex{$p(d)$}$p(d)$ any number in $[1,\infty[$ which satisfies
    \begin {equation}
    \label{e-defp(d)}
    p(d)
    \begin {cases}
    \ge   2 & \text{if }  d \le 3,\\
    >    d/2  & \text{if }  d \ge  4  \\
    \end {cases}
    \end {equation}
The symbols $C(\Lambda),C^\infty(\Lambda)$ stand for the continuous, respectively smooth, functions on $\Lambda$. The subscript $_c$ in  \mindex{$C_c$}$C_c(\Lambda),C_c^\infty(\Lambda),L_c^p(\Lambda)$ means that we consider only those functions which have compact support in $\Lambda$. In the sequel we will often consider potentials from the class of functions which are \mindex{uniformly locally $L^p$}\emph{uniformly locally} in $L^p$. More precisely, $ f$ is in the set of uniformly locally $L^p$-functions, denoted by \mindex{$L_{\unif, \loc}^p(\RR^d)$}$L_{\unif, \loc}^p(\RR^d)$, if and only if there is a constant $C$ such that for each $y\in \RR^d$
\[
\int_{|x-y|<1} |f(x)|^p  \,dx \le C
\]
The infimum over all such constants $C$ is \mindex{$\|f\|_{p, \, \unif, \loc}^p$}$\|f\|_{p, \, \unif, \loc}^p$.

Let $\Delta$ denote the Laplacian on $\RR^d$. If we choose its \index{operator domain}\emph{operator domain} $\cD(\Delta)$ to be the Sobolev space $ W_2^2(\RR^d)$ of functions in $L^2(\RR^d)$ whose second derivatives (in the sense of distributions) are square integrable, it becomes a selfadjoint operator. The restriction of $\Delta$ to a true open subset $\Lambda\subset \RR^d$ becomes selfadjoint only if we specify appropriate 
\index{boundary!conditions@--- conditions}\index{b.c.|see{boundary conditions}}boundary conditions (b.c.). \mindex{Dirichlet b.c.}Dirichlet b.c.~are defined in Remark \ref{r-QFsa}. For the definition of \mindex{Neumann b.c.}Neumann and \mindex{periodic b.c.}periodic b.c.~see for instance \cite{ReedS-78}.

Let $A,B$ be two symmetric operators on a Hilbert space $\cH$, whose norm we denote by $\|\cdot\|$. We say that $B$ is \index{relatively bounded}\emph{(relatively) $A$-bounded} if the domains obey the inclusion $\cD(A)\subset \cD(B)$ and there are finite constants  $a$ and $c_a$ such that for all $f \in \cD(A)$
\be
\label{e-relb}
\|Bf\| \le  a \|Af\| +c_a \|f\|
\ee
The infimum over all $a$ such that the estimate holds with some $c_a$ is called \mindex{relative bound}\emph{relative bound} (of $B$ with respect to $A$).
If $B$ is $A$-bounded with relative bound zero, we call it \index{infinitesimally bounded}\emph{infinitesimally $A$-bounded}.
Let $A$ be selfadjoint, and $B$ symmetric and relatively $A$-bounded with relative bound smaller than one. Then the operator sum $A+B$ on the domain $\cD(A)$ is selfadjoint by the \index{Kato-Rellich Theorem}Kato-Rellich Theorem.
We will apply this result to the sum of the negative Laplacian and a potential. A multiplication operator by a function
$V\in L_{\unif, \loc}^p(\RR^d)$ is infinitesimally $\Delta$-bounded if $p=p(d)$, cf.~\cite[Thm.~XIII.96]{ReedS-78}. Moreover, the constant $c_a$ in \eqref{e-relb} depends only on
$ \|V\|_{p, \, \unif, \loc}$.
 Thus the sum $H:=-\Delta +V$ is selfadjoint on $ W_2^2(\RR^d)$.
If $B$ is relatively $A$-bounded in operator sense with relative bound $a$ it implies
\[
\la f, Bf\ra \le a \la f, Af\ra + C_a \la f, f\ra  \text{ for some } C_a\in \RR
\]
which is called \index{relative form-boundedness}\emph{relative form-boundedness} of the form $Q_B(f,g):=\la f, Bg\ra$
with respect to $Q_A(f,g):=\la f, Ag\ra$. See \S~VI.1.7 in \cite{Kato-80} for more details.
\medskip

\smallskip

The triple $(\Omega, \cB_\Omega, \PP)$ stands for a probability space with associated $\sigma$-algebra and probability measure,
while $\EE\{\dots\}$ denotes the expectation value with respect to $\PP$. A collection $T_j\colon \Omega \to \Omega, j \in \cJ$ of measure preserving transformations is called \index{ergodic!--- action}\emph{ergodic} if
all measurable sets in $\Omega$ which are invariant under the action of all $T_j, j \in \cJ$ have measure zero or one.

\begin{dfn}
\label{d-AM}
Let  $p=p(d)$ be as in \eqref{e-defp(d)}, $u \in L_c^p(\RR^d)$ and $q_k\colon \Omega \to \RR, k \in \ZZ^d$
be a sequence of bounded, independent, identically distributed random variables, called \index{coupling constants}\emph{coupling constants}.
Then the family of multiplication operators given by the stochastic process
\be
\label{e-APdfn}
V_\omega(x)  :=  \sum_{k\in \ZZ^d} q_k(\omega) \, u(x-k)
\ee
is called \index{alloy type!--- potential}\emph{alloy type potential}. 
The function $u$ is called \index{single site!--- potential}\emph{single site potential}.
Let $H_0 :=   -\Delta+ V_\per $ be a periodic Schr\"o\-dinger operator with $V_\per \in L_{\unif,\loc}^p(\RR^d)$. The family of operators
\be
\label{e-AMdfn}
H_\omega  :=   H_0+ V_\omega, \quad \omega \in \Omega
\ee
is called \index{alloy type!--- model}\emph{alloy type model}.

The distribution measure of the random variable $q_0$ will be called \index{single site!--- distribution}\emph{single site distribution} and denoted by $\mu$.  \index{distribution!of coupling constant@--- of coupling constant}
If not stated otherwise, in the sequel we assume that $\mu$ is \index{absolutely continuous!--- coupling constants}absolutely continuous with respect to the Lebesgue measure and has a bounded density. The density function is denoted by $f$.
\end{dfn}
\smallskip

Due to our assumptions on the boundedness of the coupling constants, for each $a>0$ there is a constant $c_a$ such that for all $\omega$ and all $\psi \in \cD(\Delta)$
\[
\|V_\omega \psi\| \le  a  \|\Delta \psi\| +c_a \|\psi\|,
\quad
\|V_\per \psi\| \le  a  \|\Delta \psi\| +c_a \|\psi\|
\]
In particular $H_0$ and all $H_\omega$ are selfadjoint on the operator domain of $\Delta$.
It will be of importance to us that the constant $c_a$ may be chosen independently of the random parameter $\omega$.

\begin{rem}
\label{r-AM}
(a) \
In several paragraphs we study Hamiltonians as in Definition  \ref{d-AM}, but where some of the hypotheses on the single site potential  or the coupling constants are relaxed. More precisely, we will consider single site potentials with non-compact support and coupling constants which are unbounded, correlated, or do not have a bounded density.

(b) \
If the \index{coupling constants!unbounded ---}coupling constants are not bounded, one has to impose some moment condition to make sure that the alloy type model still makes sense.
The main difference (to the bounded case) is that for $\omega$ in a set $\Omega'\subset\Omega$ of full measure the operator $H_\omega$ will be (essentially) selfadjoint, however this will fail to hold for $\omega$ in  the complement $\Omega \setminus \Omega'$. See for example \cite{KirschM-82a,KirschM-82b,KirschM-83b} for more details.

(c) \
\label{r-AMerg}
There is a group of \index{measure preserving transformations}measure preserving transformations $T_k, k\in \ZZ^d$ on $(\Omega, \cB_\Omega, \PP)$
such that \eqref{e-APdfn} obeys
\[
V_\omega(x-k)= V_{T_k\omega}(x)
\]
In other words the stochastic process $V\colon \Omega \times \RR \to \RR$ is \index{stationary}\emph{stationary} with respect to translations by vectors in $\ZZ^d$. Moreover, the group $T_k, k\in \ZZ^d$ acts ergodically on $\Omega$, therefore we call $V$ an \mindex{ergodic potential}\emph{$\ZZ^d$-ergodic potential}.

To see that the above statements are true we pass over to the canonical probability space $\Omega = \times_{k \in \ZZ^d} \, \RR$, equipped with the product measure $\PP := \otimes_{k \in \ZZ^d} \, \mu$. Now the stochastic process
$\{\pi_k\}_{k\in \ZZ^d}$,  defined by $ \pi_k(\omega) =\omega_k$ for all $ k\in \ZZ^d$, has the same finite dimensional \index{distribution!finite dimensional ---s}distributions as $\{q_k\}_k$.
It is easily seen that the transformations $ \left (T_k (\omega) \right )_j := \omega_{j-k} $ are  measure preserving and that the group $\ZZ^d$ acts ergodically on $\Omega$.
\smallskip

Using the stochastic process $\{\pi_k\}_k$ the alloy type potential can be written as
\be
\label{e-AMrprst}
V_\omega(x)  :=  \sum_{k\in \ZZ^d} \omega_k \, u(x-k)
\ee
which we will use without distinction in the sequel.
\end{rem}

Abstracting the properties of stationarity and ergodicity we define general random potentials and operators with $\ZZ^d$-ergodic structure.

\begin{dfn}
\label{d-ergO}
Let $V\colon \Omega \times \RR^d \to \RR$ be a stochastic process such that for almost all  $\omega\in \Omega$ the realisation of the potential obeys $V_\omega \in L_{\unif, \loc}^p(\RR^d)$, $p=p(d)$ and additionally $\EE \{\|V_\omega \chi_\Lambda \|_p^p\}< \infty$, where $\Lambda$ is an unit cube. Let $T_k, k\in \ZZ^d$ be a group of measure preserving transformations acting ergodically on $(\Omega, \cB_\Omega, \PP)$ such that
\[
V_\omega(x-k)= V_{T_k\omega}(x)
\]
Then we call $\{V_\omega\}_\omega$  a  (\emph{$\ZZ^d$-ergodic}) \index{ergodic!--- potential}\emph{random potential}
and $\{H_\omega\}_\omega$ with $H_\omega = -\Delta + V_\omega$ a (\emph{$\ZZ^d$-ergodic}) 
\index{ergodic!--- operator}\emph{random operator}.
\end{dfn}

The restriction of $H_\omega$ to an open subset $\Lambda$ will be denoted by $H_\omega^\Lambda$ if we impose Dirichlet boundary conditions and by $H_\omega^{\Lambda,N}$ in the case of Neumann b.c.
While we will be mainly concerned with $\ZZ^d$-ergodic operators
we will give some comments as asides on their counterparts which are ergodic with respect to the group $\RR^d$. The recent overview \cite{LeschkeMW-02} is devoted to such models that model amorphous media. Insight in the research on almost periodic operators can be obtained for instance in \cite{Shubin-1979,Shubin-1982,AvronS-1983,BellissardLT-1985,CyconFKS-87,PasturF-92}  and the references therein.

\begin{rem}
All $\ZZ^d$-ergodic potentials can be represented in a form which resembles \index{alloy type!--- potential}alloy type potentials.
Namely, for such $V\colon \Omega \times \RR^d \to \RR$   there exists a sequence $f_k, k\in \ZZ^d$ of random variables on $\Omega$ taking values in the separable Banach space $L^p(\RR^d)$ such that $V$ can be written as
\begin{equation}
\label{e-kirschmodell}
V_\omega(x) = \sum_{k \in \ZZ^d} f_k (\omega, x-k).
\end{equation}
This representation is of interest because it ensures that after passing to an equivalent probability space and stochastic process one may assume that the sigma algebra on $\Omega $ is countably generated.
See \cite{Kirsch-81} and Remark 2.8 in \cite{LenzPV-2002?} for more information.
\end{rem}

\subsection{Transport properties and spectral types}
\label{ss-transport}

The main interest in the study of random operators is to understand the transport properties of the
media they model. In the particular case of the quantum mechanical Hamiltonian of an electron in a disordered solid
the electric conductance properties are the main object of interest.

The Hamiltonian governs the equation of motion, i.e.~the time dependent 
\index{Schr\"odinger equation}Schr\"o\-dinger equation
\[
\frac{\partial \psi(t)}{\partial t} = -i H_\omega \psi(t)
\]
The time evolution of the vector $\psi(t)$ in Hilbert space describes the movement of the electron. Since we
chose the space representation in the Schr\"o\-dinger picture, we can can think of $\psi(t)$ as a wave packet which evolves in time.
The square of its absolute value $|\psi(t,\cdot)|^2\in L^1(\RR^d)$ is a probability density. More precisely, $\int_A|\psi(t,x)|^2dx$ is the probability to find the electron in the set $A \subset \RR^d$ at time $t$.

For a given initial state $\psi_0:=\psi(0)$ supported in a compact set
$A$  one would like to know whether for large times the
function $\psi(t)$ stays (essentially) supported near $A$, or
moves away to infinity. In the first case one speaks of a
\index{bound state}\emph{bound state}, since it remains localised near its original support for all times.
The other extreme case would be that $\psi(t)$ leaves any compact region in $\RR^d$
(and never comes back) as time goes to infinity. Such a state is called a
\index{scattering state}\emph{scattering} or \index{extended state|see{scattering state}}\emph{extended state}. By the \mindex{RAGE theorem}RAGE theorem,
cf.~e.g.~\cite{ReedS-79,CyconFKS-87,Stollmann-01}, it is possible
to relate the dynamical properties of states just described to the
spectral properties of the Hamiltonian. Roughly speaking, bound states correspond to
\index{pure point spectrum}pure point spectrum and scattering states to 
\index{absolutely continuous!--- spectrum}(absolutely) continuous spectrum. For a more precise statement consult for instance \cite{ReedS-79,CyconFKS-87,Stollmann-01}.


This motivates the systematic study of spectral properties of the introduced Schr\"o\-dinger operators. 
If a random Schr\"odinger operator exhibits almost surely only \index{pure point spectrum}pure point spectrum 
in an energy region one speaks of 
\index{localisation!Anderson ---}\emph{Anderson} or 
\index{spectral!localisation@ --- localisation!}\index{localisation!spectral ---}\emph{spectral localisation}. 
The name goes back to Anderson's seminal work
\cite{Anderson-58}. This property has been established for a
variety of random models. In most of those cases one can
additionally prove that the corresponding eigenfunctions \index{decay!--- of eigenfunctions|(}decay
exponentially in configuration space, a phenomenon called
\index{localisation!exponential ---}\emph{exponential (spectral) localisation}. The situation is
different for random potentials with \index{long range!--- correlations}long range correlations,
where sometimes only power-law decay of the eigenfunctions has
been established \cite{KirschSS-97,FischerLM-00,Zenk-02}.

If an energy interval contains almost surely only pure point spectrum, we call it 
\index{localisation!interval@--- interval}\emph{localisation interval}.
An  eigenfunction of $H_\omega$ which decays exponentially is called an \emph{exponentially localised eigenstate}.
The region or point in space where the localised state has its highest amplitude will be called \emph{localisation centre} (we will not need a mathematically precise definition of this notion).

However, it turns out that the spectrum captures only a rough view on the dynamical properties of the quantum mechanical system. A more detailed understanding can be obtained by studying the time evolution of the moments of the position operators.
This led to a formulation of several criteria of \index{localisation!dynamical ---}\emph{dynamical localisation}.
The strongest characterisation of this phenomenon, namely \emph{strong dynamical
localisation in Hilbert-Schmidt} \index{Hilbert-Schmidt class}topology  means that
for all $q >0$
\begin{equation}
\label{e-sdL}
\EE \left\{\sup_{\|f \|_\infty \leq 1}  \left\| |X|^{q/2} f( H_\omega)
P_\omega^l(I) \chi_K \right\|_{HS}^2 \right\} < \infty
\end{equation}
Here  $P_\omega^l(I)$ denotes the \index{spectral!--- projection}spectral projection onto the energy interval $I$ associated to the operator $H_\omega^l$, $\| \cdot\|_{HS}$ denotes the \index{Hilbert-Schmidt class}Hilbert-Schmidt norm, $K\subset\RR^d$ is any compact set,
and $|X|$ denotes the operator of multiplication with the function $|x|$.
For the interpretation of \eqref{e-sdL} as non-spreading of
wave-packets one chooses $f(y) = e^{-it y}$. Dynamical localisation \eqref{e-sdL}
implies in particular that the random Hamiltonian $H_\omega$ exhibits spectral localisation in $I$.
The reader can consult e.g.~\cite{DelRioJLS-95,GerminetDB-98b,Stollmann-01,GerminetK-2001a,GerminetK-2001b,AizenmanENSS} to get an insight how the notion of dynamical localisation evolved and for recent developments.
In \cite{DelRioJLS-95,DeBievreG-00,JitomirskayaSBS-03} examples are discussed where spectral localisation occurs, but certain dynamical criteria for localisation are not satisfied.

For the purpose of the present paper these distinctions are not crucial. In the case of the 
\index{alloy type!--- model}alloy type model, to which we devote most attention,  spectral and dynamical localisation coincide, cf.~\cite{DamanikS-2001,GerminetK-2001b}. In the sequel we mean by localisation that the considered operator exhibits in a certain energy interval only pure point spectrum, and that the corresponding eigenfunctions \index{decay!--- of eigenfunctions|)}decay sufficiently fast.

Since we are dealing not just with a single Hamiltonian, but with a whole family of them, we have to say something on how the spectral properties depend on the parameter $\omega$ describing the randomness:
most properties of the spectrum of an operator pertaining to the family $\{H_\omega\}_\omega$ hold almost surely, i.e.~for $\omega$ in a set such that its complement has measure zero in $\Omega$. This is at least true for the properties we discuss in the present paper.

%

We shortly describe what kinds of spectral types one expects from the physical point of view for random Schr\"o\-dinger operators, say of alloy type. In case there are rigorous  results which have confirmed this intuition we quote the reference.

In one space dimension the spectrum is pure point for all energies almost surely. Rigorous proofs of this statement can for instance be found in  \cite{GoldsheidMP-77,KotaniS-87}.

In three or more dimensions it is expected that the spectrum is pure point near the boundaries of the spectrum while in the interior it is purely \index{absolutely continuous!--- spectrum}absolutely continuous.
In the latter case one speaks also of an energy region with \index{delocalised states}\emph{delocalised} states. However, for alloy type Hamiltonians the proof of delocalisation is open. Some results on existence of \index{absolutely continuous!--- spectrum}absolutely continuous spectrum
for random models of a different type can be found in \cite{Klein-1996,FedotovK-?a,KirschKO-2000}.
 The two regions with localised, respectively delocalised states are separated by  a threshold, the so called \index{mobility edge}\emph{mobility edge}, for partial results see \cite{JaksicL-2000a,KirschKO-2000,GerminetK-2001b,FedotovK-02a}.

The literature on the existence of pure point spectrum is extensive. We discuss it in more detail in \S~\ref{ss-MSA}.
\smallskip

How large the intervals with point or continuous spectrum are, depends on the disorder present in the model. For instance, in \eqref{e-AMdfn} one could introduce a global coupling constant $\lambda$ in front of the potential
\[
H_\omega = H_0 + \lambda V_\omega
\]
Now large $\lambda$ means large disorder,  small $\lambda$ small disorder.
The larger the disorder, the larger is the portion of the spectrum which contains localised states.
For other types of random Schr\"o\-dinger operators there are similar ways to introduce a disorder parameter.


The phenomenon that localised states emerge at the edges of the spectrum can be understood in terms of the so-called \index{fluctuation boundaries of the spectrum}\emph{fluctuation boundaries}. These are the regions of the spectrum which correspond to extremely rare configurations of the potential. Consequently, the \emph{density of states} (or the spectral density function, see the next section for a precise definition) is very thin in this region. This has been first understood on physical grounds by Lifshitz. Today the tails of the density of states at the fluctuation boundaries bear the name of \emph{Lifshitz-asymptotics} or \index{Lifshitz tails}\emph{Lifshitz-tails}. We give a (non-exhaustive) list of works devoted the study of this asymptotics: \cite{Pastur-74,Pastur-77,Nakao-77,Fukushima-81,KirschM-83a,Simon-85b,KirschS-86,Mezincescu-86,Mezincescu-87,Simon-87,Sznitman-90a,Mezincescu-93,Klopp-1999,Stollmann-99a,KloppP-1999,KloppW-2001,Klopp-02c,Klopp-02d} and the references at the end of \S~\ref{ss-MagF}.

The existence of localised states for random Schr\"o\-dinger operators is in sharp contrast to the features of periodic operators. Indeed, for operators with \index{periodic!--- potential}periodic potential, satisfying some mild regularity assumptions, it is known that the spectrum is purely \index{absolutely continuous!--- spectrum}absolutely continuous, \cite{BirmanS-99,Sobolev-99,Shen-02,Shterenberg-01,KuchmentL-02}. This difference might seem somewhat surprising, given the similarity of the structure of an alloy type and a periodic operator.

\subsection{Outline of the paper}
In the present section we discussed the physical motivation to study random operators and introduced the models we will analyse in the sequel. The next section is devoted to the proof of the existence of the IDS for these models. In the third section we discuss why regularity properties of the IDS are of interest. The following
two sections give two independent proofs of the continuity of the IDS. Both approaches are suitable to extensions in various directions.
We review some recently obtained results.
For more details see the table of contents.

\subsection*{Acknowledgements}
I would like to thank R.~del Rio and C.~Villegas for organising the Workshop on Schr\"o\-dinger operators
and the stimulating atmosphere at the IIMAS.
This is an opportunity to thank D.~Hundertmak, R.~Killip, V.~Kostrykin, D.~Lenz, N.~Peyerimhoff, and
O.~Post for the pleasure of collaborating with them on some of the results presented here, and in particular W.~Kirsch who introduced me to the topic of random Schr\"odinger operators.
I am grateful to E.~Giere, P.~M\"uller, R.~Muno, K.~Schnee, K.~Veseli\' c, S.~Warzel and the referees for valuable comments.
Financial support by a fellowship of the DFG and hospitality at CalTech by B.~Simon is gratefully acknowledged.

\section{Existence of the integrated density of states}

Intuitively, the integrated density of states (IDS) measures how many electron energy
levels can be found below a given energy per unit volume of a solid.
It can be used to calculate the free energy and hence all basic thermodynamic quantities of the
corresponding non-interacting many-particle system.

To define the IDS mathematically one uses an exhaustion procedure.
More precisely, one takes an increasing sequence $\Lambda_l$ of open subsets of $\RR^d$ such that each $\Lambda_l$ has finite volume and $\bigcup_l \Lambda_l=\RR^d$.
Then the operator $H_\omega^l$, which is the restriction of $H_\omega$ to $\Lambda_l$
with Dirichlet boundary conditions, is selfadjoint, bounded below and its spectrum consists
of discrete eigenvalues
$\lambda_1(H_\omega^l) \le \lambda_2(H_\omega^l) \le \dots \le \lambda_n(H_\omega^l) \to \infty$.
Here $\lambda_n=\lambda_{n+1}$ means that the eigenvalue is degenerate and we take this into account in the enumeration.

The \index{normalised eigenvalue counting function}\emph{normalised eigenvalue counting function} or \index{finite volume integrated density of states|see{normalised eigenvalue counting function}}\emph{finite volume integrated density of states} $N_\omega^l$
is defined as
    \be
    \label{d-fvIDS}
    N_\omega^l(E):= \frac{\#\{n| \, \lambda_n(H_\omega^l) < E\}}{|\Lambda_l|}
    \ee
The numerator can equally well be expressed using the trace of a spectral projection
    \[
    \#\{n| \, \lambda_n(H_\omega^l) < E\} = \Tr  \Big [ P_\omega^l\big(]-\infty,E[\big) \Big ]
    \]
Note that $N_\omega^l\colon \RR \to [0,\infty[$ is a \index{distribution!function@--- function|(}\emph{distribution function} of a point measure  for all $l \in \NN$, i.e. $N_\omega^l(E)=\nu_\omega^l(]-\infty,E[)$. Here $\nu_\omega^l$ is the \index{density of states!--- measure}\emph{finite volume density of states measure}
defined by
    \[
    \nu_\omega^l(I):= |\Lambda_l|^{-1}   \, \#\{n| \, \lambda_n(H_\omega^l) \in I\}
    \]
By definition a distribution function is non-negative, left-continuous and monotone increasing. In particular, it
has at most countably many points of discontinuity.
\smallskip

Under some additional conditions on the random operator and the exhaustion sequence $\Lambda_l, l\in\NN$
one can prove that
\begin{enumerate}[\rm (i)]
\item
\label{i-IDSconv}
For almost all $\omega \in \Omega$ the sequence $N_\omega^l$ converges to a distribution function $N_\omega$ as $l$ goes to infinity.
This means that we have $N_\omega^l(E)\to N_\omega(E)$ for all continuity points $E$ of the limit distribution
$N_\omega$.
\item
For almost all $\omega\in \Omega$ the distribution functions $N_\omega$ coincide, i.e.~there is an $\omega$-independent
distribution function $N$ such that $N=N_\omega$ for almost all $\omega$.
This function $N$ is called the \index{integrated density of states}\emph{integrated density of states}.
Note that its independence of $\omega$ is not due to an explicit integration over the probability space $\Omega$, but only to the exhaustion procedure. This is the reason why the IDS is called \index{self-averaging}\emph{self-averaging}.
\item
In most cases there is a formula for the IDS as an expectation value of a trace of a localised projection.
For $\ZZ^d$-ergodic operators it reads
    \be
    \label{e-IDSformula}
    N(E):= \EE \Big \{\Tr \big [\chi_\Lambda P_\omega(]-\infty,E[) \big] \Big\}
    \ee
Here $\Lambda$ denotes the unit box $]0,1[^d$, which is the periodicity cell of the lattice $\ZZ^d$. Actually, one could choose certain other functions instead of  $\chi_\Lambda$, yielding all the same result, cf.~Formula \eqref{e-tau-u}.
The equality \eqref{e-IDSformula} holds for $\RR^d$-ergodic operators, too. It is sometimes called 
\index{Pastur-\v Subin trace formula}\emph{Pastur-\v Subin trace formula}.
\end{enumerate}
\smallskip

In the following we prove the properties of the IDS just mentioned by two methods. In \S\S~\ref{ss-RSOd} -- \ref{ss-EtLt} a complete proof is given using the \index{Laplace transform}Laplace transforms of the distribution functions $N_\omega^l$, while \S~\ref{ss-DNb} is devoted to a short sketch of an alternative method. It uses \index{Dirichlet-Neumann bracketing}Dirichlet-Neumann bracketing estimates for Schr\"o\-dinger operators, which carry over to the corresponding eigenvalue counting functions. 
These are thus \index{superadditive process}super- or \index{subadditive process}subadditive processes to which an \index{ergodic!--- theorem}ergodic theorem \cite{AkcogluK-1981} can be applied.
\smallskip

Actually the proof using Laplace transforms will apply to more general situations than discussed so far,
namely to  more general geometries than Euclidean space.
To be precise, we will consider random Schr\"o\-dinger operators on Riemannian covering manifolds, where both the potential and the metric may depend on the randomness. This includes random Laplace-Beltrami operators.

We follow the presentation and proofs in \cite{PeyerimhoffV-2002,LenzPV-2002?}\nocite{LenzPV-03?}. The general strategy we use was developed by Pastur and \v Subin in \cite{Pastur-1971} and \cite{Shubin-1979} for random and almost-periodic operators in Euclidean space. A particular idea of this approach is to prove the convergence of the Laplace transforms $\cL_\omega^l$ of the normalised finite volume eigenvalue counting functions $N_\omega^l$ instead of proving the convergence of $N_\omega^l$ directly. This is actually the main difference to the second approach we outline in \S~\ref{ss-DNb}, which is taken from \cite{KirschM-82c}. The Pastur-\v Subin strategy seems to be better suited for geometries with underlying group structure which is non-abelian.

Indeed, one of the differences between random operators on manifolds and those on $\RR^d$ is that the operator
is equivariant with respect to a group which does not need to be commutative. This means that one has to use a
non-abelian ergodic theorem to derive the convergence of the \index{distribution!function@--- function|)}distribution functions $N_\omega^l$ or, alternatively, of their Laplace transforms $\cL_\omega^l$.

This imposes some restriction on the strategy of the proof since the ergodic theorems which apply to non-abelian
groups need more restrictive assumptions than their counterparts for commutative groups, cf.~also Remark \ref{r-ergthr}

For processes which are not additive, but only super- or subadditive, there is  a non-abelian maximal ergodic theorem at disposal (cf.~6.4.1 Theorem in \cite{Krengel-1985}) but so far no pointwise theorem. This is also the reason why the Dirichlet-Neumann bracketing approach of \S~\ref{ss-DNb} does not seem applicable to random operators living on a covering manifold with non-abelian \mindex{deck-transformation group}deck-transformation group (covering transformation group).

\subsection{Schr\"o\-dinger operators on manifolds: motivation}
\label{ss-RSOm}
In this section we study the IDS of random Schr\"odinger \index{Schr\"odinger operator!--- on manifold}operators on manifolds. Let us first explain the physical motivation for this task.

Consider a particle or a system of particles which are constrained to a sub-manifold of the ambient (configuration) space. The classical and quantum Hamiltonians for such systems have been studied e.g.~in \cite{Mitchell-01,FroeseH-01} (see also the references therein).
To arrive at an \index{effective Hamiltonian on a manifold}effective Hamiltonian describing the constrained motion on the sub-manifold, a limiting procedure is used:
a (sequence of) confining high-barrier potential(s) is added to the Hamiltonian defined on the ambient space to restrict the particle (system) to the sub-manifold.
In \cite{Mitchell-01,FroeseH-01} one can find a discussion of the similarities and differences between the obtained effective quantum Hamiltonian
and its classical analogue.

A important feature of the effective quantum Hamiltonian is the appearance of a so-called \index{extra-potential}\emph{extra-potential} depending on the extrinsic curvature of the sub-manifold and the curvature of the ambient space.
This means that even if we disregard external electric forces the relevant quantum mechanical Hamiltonian of the constrained system is not the pure Laplacian but contains (in general) a potential energy term.
This fact explains the existence of \mindex{curvature-induced bound states}curvature-induced bound states in 
\index{quantum!--- waveguides}quantum waveguides and layers, see \cite{ExnerS-89,DuclosE-95,LonderganCM-99,DuclosEK-01} and the references therein.

As is mentioned in \cite{Mitchell-01}, the study of effective Hamiltonians of constrained systems is motivated by specific physical applications. They include stiff molecular bonds in (clusters of) rigid molecules and molecular systems evolving along reaction paths.
From the point of view of the present paper quantum wires, wave guides and layers are particularly interesting physical examples.
Indeed, for these models (in contrast to quantum dots) at least one dimension of  the constraint sub-manifold is of macroscopic size. Moreover, it is natural to assume that the resulting Hamiltonian exhibits some form of translation invariance in the macroscopic direction. E.g.~it may be periodic, quasiperiodic or --- in the case of a random model --- ergodic.

For random quantum waveguides and layers the existence of dense point spectrum is expected, cf. the discussion of localisation in Paragraph \ref{ss-transport}. For a waveguide embedded in Euclidean space this has been rigorously proven in  \cite{Kleespies-1999a,KleespiesS-2000}.
The question of spectral localisation due to random geometries has been raised already in \cite{Davies-1990a}.
There the behaviour of \index{Laplace-Beltrami operator}Laplace-Beltrami operators under non-smooth perturbations of the metric is studied.

A second motivation for the analysis of operators on manifolds studied in this section comes from differential geometry. 
The spectral properties of \index{periodic!--- operator on manifold}periodic Schr\"odinger operators on manifolds have attracted the interest of various authors. A non-exhaustive list is  \cite{Atiyah-76,Sunada-1988,KobayashiOS-1989,Sunada-1990,Sunada-1992b,BrueningS-1992b,BrueningS-1992a,Shubin-96,KarpP-98,Post-00,Post-03}.
By 'periodic' we mean that the operator acts on a covering manifold and is invariant under the unitary operators induced by the deck-transformations.

Particular attention was devoted to the analysis of the gap structure of the spectrum of periodic operators of Schr\"odinger type.
More precisely, one is interested whether the spectrum in interrupted by \index{spectral!--- gaps}\emph{spectral gaps}, i.e.~intervals on the real line which belong to the resolvent set.
In case there are gaps: can one establish upper and lower bounds for the width and number of gaps and the spectral bands separating them?
Although the gap structure of the spectrum is a mathematically intriguing question for its own sake, it is also important from the physical point of view. The features of gaps in the energy spectrum are relevant for the conductance properties of the physical system. 
Even for periodic Schr\"odinger operators in Euclidean space the gap structure is highly non-trivial. This is maybe best illustrated by works
devoted to the \index{Bethe-Sommerfeld conjecture}Bethe-Sommerfeld conjecture, e.g.~\cite{SommerfeldB-33,Skriganov-84,Skriganov-85,Skriganov-87,HelfferM-98}.
Another interesting feature of some periodic Laplace-Beltrami operators is the existence of \mindex{eigenfunctions of periodic operators}$L^2$-eigenfunctions, cf.~the discussion in Remark \ref{r-cIDSgeo}.

These periodic operators on manifolds are generalised by their random analogues studied in this section.

\subsection{Random Schr\"o\-dinger operators on manifolds: definitions}
\label{ss-RSOd}
Let us explain the geometric setting in which we are working precisely: let $X$ be a complete $d$-dimensional Riemannian \mindex{complete manifold}manifold with metric $g_0$. We denote the volume form of $g_0$ by $\vol_0$. Let $\Gamma$ be a \mindex{discrete group}discrete, \mindex{finitely generated group}finitely generated subgroup of the isometries of $(X,g_0)$ which acts \mindex{free action}freely and \mindex{properly discontinuous action}properly discontinuously on $X$ such that the quotient $M:=X/\Gamma$ is a compact ($d$-dimensional) Riemannian manifold.
Let $(\Omega,\cB_\Omega, \PP)$ be a probability space on which $\Gamma$ acts by measure preserving transformations.
Assume moreover that the action of $\Gamma$ on $\Omega$ is ergodic. Now we are in the position to define what we mean by a random metric and consequently a random Laplace-Beltrami operator.

\begin{dfn}
\label{d-randommetric}
Let $\{g_\omega\}_{\omega\in\Omega}$ be a family of Riemannian metrics on $X$.
Denote the corresponding volume forms by $\vol_\omega$.
We call the family $\{g_\omega\}_{\omega\in\Omega}$  a \index{random metric}\emph{random metric on $(X,g_0)$}
if the following five properties are satisfied:
\stepcounter{equation}
\begin{enumerate}
\item[{\rm (\theequation)} \label{M1}]
The map $\Omega \times TX \to \RR$, $(\omega,v) \mapsto g_\omega(v,v)$
is jointly measurable.
\item[\refstepcounter{equation} {\rm (\theequation)} \label{M2}]
There is a $C_g \in \, ]0,\infty[$ such that
\begin{equation*}
\label{quasiisom}
C_g^{-1} \, g_0(v,v) \le g_\omega(v,v) \le C_g  \, g_0(v,v) \  \text{ for all } \,
\, v \in TX.
\end{equation*}
\item[\refstepcounter{equation} {\rm (\theequation)} \label{M3}]
There is a $C_\rho  \in \, ]0,\infty[$ such that
\[
\vert \nabla_0 \, \rho_\omega(x) \vert_0 \le C_\rho \  \mbox{for all} \,
\, x \in X,
\]
where $\nabla_0$ denotes the gradient with respect to $g_0$, $\rho_\omega$ is the unique smooth density of $\vol_0$ with
respect to $\vol_\omega$, and $\vert v \vert_0^2 = g_0(v,v)$.
\item[\refstepcounter{equation} {\rm (\theequation)} \label{M4}]
There is a uniform lower bound $(d-1)K \in \RR$ for the \index{Ricci curvature}Ricci curvatures of all
Riemannian manifolds $(X,g_\omega)$. Explicitly, $\Ric (g_\omega )\ge (d-1) K g_\omega$
for all $\omega \in \Omega$ and on the whole of $X$.
\item[\refstepcounter{equation} {\rm (\theequation)} \label{M5}]
The metrics are compatible in the sense that the deck transformations
\[
\gamma\colon (X,g_\omega) \to (X,g_{\gamma \omega}) ,\quad \gamma
\colon x \mapsto \gamma x
\]
are isometries.
\end{enumerate}
\end{dfn}
\hspace*{-1.6em}%
Property \eqref{M5} implies in particular that the induced maps
\[
U_{(\omega,\gamma)}\colon L^2(X,\vol_{\gamma^{-1}\omega}) \to
L^2(X,\vol_\omega), \quad (U_{(\omega,\gamma)} f)(x) = f(\gamma^{-1}x)
\]
are unitary operators.

The \index{volume!--- density}density $\rho_\omega$ appearing in \eqref{M3} satisfies by definition
\[
\int_X f(x) \, d\vol_0(x) = \int_X f(x) \rho_\omega(x) \, d\vol_\omega(x).
\]
It is a smooth function and can be written as
\[
\rho_\omega(x) = \left (\det g_0(e_\omega^i, e_\omega^j) \right )^{1/2} =
\left (\det g_\omega(e_0^i, e_0^j)\right )^{-1/2}
\]
Here  $e_0^1,\dots,e_0^d $ denotes any basis of $T_xX$ which is orthonormal
with respect to the scalar product $g_0(x)$, and $e_\omega^1,\dots,e_\omega^d \in T_xX$ is any basis
orthonormal with respect to $g_\omega(x)$.
It follows from \eqref{M2} that
\begin {equation}
\label{e-rhoest}
C_g^{-d/2} \le \rho_\omega(x) \le C_g^{d/2} \quad \text{ for all }  x \in X, \
 \omega\in \Omega
\end{equation}
which in turn, together with property \eqref{M3} and the chain rule, implies
\be
\label{e-sqrtrho}
\vert \nabla_0 \, \rho_\omega^{\, \pm \, 1/2}(x) \vert_0
\le
C_g^{3d/4} \, \vert \nabla_0 \, \rho_\omega(x) \vert_0 \quad \text{ for all }  x \in X, \
 \omega\in \Omega
\ee
Moreover, for any measurable $\Lambda \subset X$ by \eqref{e-rhoest} we have the volume estimate
\be
\label{e-volest}
C_g^{-d/2} \, \vol_0(\Lambda) \le \vol_\omega(\Lambda) \le C_g^{d/2} \, \vol_0(\Lambda)
\ee
We denote the \index{Laplace-Beltrami operator}Laplace-Beltrami operator with respect to the metric $g_\omega$ by
$\Delta_\omega$.
\medskip

Associated to the random metric just described we define a random family of operators.
\begin{dfn}
\label{d-randomoperator}
Let $\{ g_\omega \}$ be a random metric on $(X,g_0)$.
Let $V\colon \Omega \times X\to \RR$ be a jointly measurable mapping such that
for all $\omega\in\Omega$ the \emph{potential} $V_\omega:= V(\omega,\cdot) \ge 0$ is in $L_{\loc}^1(X)$.
For each $\omega \in \Omega$ let $H_\omega = -\Delta_\omega + V_\omega$ be a
Schr\"o\-dinger operator defined on a dense subspace $\cD_\omega$ of the Hilbert space
$L^2(X,\vol_\omega)$. The family $\{H_\omega\}_{\omega\in\Omega}$ is called a
\index{Schr\"odinger operator!random ---}\emph{random Schr\"o\-dinger operator} if it satisfies for all $\gamma \in \Gamma$ and $\omega \in \Omega$
 the following \index{equivariance}\emph{equivariance} condition
\begin{equation}
\label{compcomp}
H_\omega = U_{(\omega,\gamma)} H_{\gamma^{-1} \omega} U_{(\omega,\gamma)}^*
\end{equation}
\end{dfn}


\begin{rem}[Restrictions, quadratic forms and selfadjointness]
\label{r-QFsa}
Some remarks  are in order why the sum of the Laplace-Beltrami operator and the
potential is selfadjoint. We consider the two cases of an operator on the whole manifold $X$ and on a proper open subset of $X$ simultaneously.
The set of all smooth    functions with compact support in an open set $\Lambda \subset X$
is denoted by $C_c^\infty(\Lambda)$.
For each $\omega\in \Omega$ we define the \index{quadratic form}quadratic form
    \begin{gather}
    \widetilde{Q} (H_\omega^\Lambda) \colon C_c^\infty(\Lambda) \times \CcsmoothD  \rightarrow \RR,
    \\
    \nn
    (f,h)
    \mapsto  \int_{\Lambda}  g_\omega(x)\Big (\nabla f (x), \nabla h(x)\Big) \, d\vol_\omega (x)
    + \int_{\Lambda}  f(x) V_\omega(x) h(x) \, d\vol_\omega(x)
    \end{gather}
We infer from Theorem 1.8.1 in \cite{Davies-1989} that this quadratic from is closable and
its closure $Q(H_\omega^\Lambda)$ gives rise to a densely defined, non-negative selfadjoint operator $H_\omega^\Lambda$. Actually, $Q(H_\omega^\Lambda)$ is the form sum of the quadratic forms of the negative Laplacian and the potential.
The  result in \cite{Davies-1989} is stated for the Euclidean case $X=\RR^d$ but the proof works
equally well for general Riemannian manifolds.

The unique selfadjoint operator associated to the above quadratic form is called \emph{Schr\"o\-dinger operator with Dirichlet boundary conditions}\index{boundary!conditions@--- conditions}. It is the \index{extension!Friedrichs ---}Friedrichs extension of the restriction $H_\omega^\Lambda|_{\displaystyle C_c^\infty(\Lambda)}$. If the potential term is absent we call it negative \emph{Dirichlet Laplacian}.
\end{rem}

There are special subsets of the manifold which will play a prominent role later:

\begin{dfn}
A subset $\cF \subset X$ is called \index{fundamental domain}\emph{$\Gamma$-fundamental domain} if it contains
exactly one element of each \mindex{orbit}orbit $O(x):= \{y\in X| \, \exists \gamma \in \Gamma : y =\gamma x \} $, $x\in X$.
\end{dfn}

In \cite[Section 3]{AdachiS-1993} it is explained how to obtain a connected, \index{polyhedral domain}polyhedral
$\Gamma$-fundamental domain $\cF \subset X$ by lifting simplices of a
\mindex{triangularisation}triangularisation of $M$ in a suitable manner. $\cF$ consists of finitely many smooth images of
simplices which can overlap only at their boundaries. In particular, it has piecewise smooth boundary.

\bigskip

To illustrate the above definitions we will look at some examples.
Firstly, we consider covering manifolds with abelian deck-transformation group.

\begin{exm}[Abelian covering manifolds] \index{covering manifold!abelian ---}
\label{x-AbelianCover}
Consider a covering manifold $(X,g_0)$ with a finitely generated, abelian subgroup $\Gamma$ of the isometries of $X$. If the number of generators of the group  $\Gamma$ equals $r$, it is isomorphic to $\ZZ^{r_0} \times  \ZZ_{p_1}^{r_0} \times  \dots  \ZZ_{p_n}^{r_n}$. Here $\sum r_i=r$ and $\ZZ_p$ is the \mindex{cyclic group}cyclic group of order $p$.  Assume as above that the quotient $X/\Gamma$ is compact. Periodic Laplace-Beltrami and Schr\"odinger operators on such spaces have been analised e.g.~in \cite{Sunada-1990,Post-00,Post-03}.

In the following we will discus some examples studied by Post in \cite{Post-00,Post-03}. The aim of this papers was to construct covering manifolds, such that the corresponding Laplace operator has open \index{spectral!--- gaps}spectral gaps. More precisely, for any given natural number $N$, manifolds are constructed with at least $N$ spectral gaps. For technical reasons the study is restricted to abelian coverings. In this case the Floquet decomposition of the periodic operator can be used effectively.
Post studies two classes of examples with spectral gaps. In the first case a \mindex{conformal perturbation}conformal perturbation of a given covering manifold is used to open up gaps in the energy spectrum of the Laplacian.
The second type of examples in \cite{Post-03} is of more interest to us. There, one starts with infinitely many translated copies of a compact manifold and joins them by cylinders to form a periodic network of 'pipes'. By shrinking the radius of the connecting cylinders, more and more gaps emerge in the spectrum. Such manifolds have in particular a non-trivial \mindex{fundamental group}fundamental group and are thus topologically not equivalent to $\RR^d$.
On the other hand their deck-transformation group is rather easy to understand, since it is abelian. In particular,
it is amenable (cf.~Definition \ref{d-amen}), which is a crucial condition in the study conducted later in this section. Some of the examples in \cite{Post-00,Post-03} are manifolds which can be embedded in $\RR^3$ as surfaces. They can be thought of as periodic quantum waveguides and networks. See \cite{Post-00} for some very illustrative figures.

Furthermore, in \cite{Post-03} perturbations techniques for Laplace operators on covering manifolds have been developed, respectively carried over from earlier versions suited for compact manifolds, cf.~\cite{ChavelF-81,Anne-87,Fukaya-87}.
They include conformal perturbations and local geometric deformations. Floquet decomposition\mindex{Floquet decomposition} is used to reduce the problem to an operator on a fundamental domain with quasi-periodic boundary conditions and discrete spectrum.  Thereafter the min-max principle is applied to geometric perturbations of the Laplacian.

Related random perturbations of Laplacians are studied in \cite{LenzPV-03?,LenzPPV}, cf.~also Example \ref{x-alloy-on-mnf}. In particular a Wegner estimate for such operators is derived.

\end{exm}

Now we give an instance of a covering manifold $X$ with non-abelian deck-transformation group $\Gamma$.
\begin{exm}[Heisenberg group]\index{Heisenberg group}
\label{x-Hbg}
The Heisenberg group $H_3$ is the manifold of $3\times 3$-matrices given by
\begin{equation}
\label{Heisenberg}
H_3 = \left \{  \left (
\begin{array}{ccc}
1 & x & y
\\
0 & 1 & z
\\
0 & 0 & 1
\end{array}
\right )  \left |  \right .  x,y,z \in  \RR \right \}
\end{equation}
equipped with a left-invariant metric. The Lie-group $H_3$ is
diffeomorphic to $\RR^3$. Its group structure is not abelian, but
nilpotent.

The subset $\Gamma = H_3 \cap M(3,\ZZ)$ forms a discrete subgroup.
It acts from the left on $H_3$ by isometries and the quotient manifold $H_3/\Gamma $ is compact.
\end{exm}

Next we give examples of a random potential and a random metric which give rise to a random Schr\"o\-dinger operator
as in Definition \ref{d-randomoperator}.
Both have an underlying structure which resembles alloy-type models (in Euclidean space).

\begin{exm}
\label{x-alloy-on-mnf}
(a) Consider the case where the metric is fixed, i.e.~$g_\omega=g_0$ for all $\omega \in \Omega$, and only the potential depends on the randomness in the following way:
\begin{equation}
V_\omega (x) := \sum_{\gamma \in \Gamma} q_\gamma(\omega) \, u(\gamma^{-1} x),
\end{equation}
Here $u: X \to \RR$ is a bounded, compactly supported measurable function and
$q_\gamma\colon \Omega \to \RR$ is a sequence of independent, identically distributed random variables.
By considerations as in Remark \ref{r-AMerg} the random operator $H_\omega:= -\Delta +V_\omega, \omega\in \Omega$ is seen to satisfy the equivariance condition.

(b) Now we consider the situation where the metric has an alloy like structure. Let $(g_0,X)$ be a Riemannian covering manifold and let a family of metrics $\{g_\omega\}_\omega$ be given by
\[
g_\omega(x)= \bigg(\sum_{\gamma\in \Gamma}\, r_\gamma(\omega) \,
u(\gamma^{-1} x) \bigg) \, g_0(x)
\]
where $u\in C_c^\infty(X)$ and the $r_\gamma\colon \Omega\to  \, ]0,\infty[, \gamma \in \Gamma$ are a
collection of independent, identically distributed random variables. Similarly as in the previous example one sees that the operators $\Delta_\omega$ are equivariant.
\end{exm}

\subsection{Non-randomness of spectra and existence of the IDS}
\label{ss-manifthms}
Here we state the main theorems on the \index{non-randomness of spectra}non-randomness of the spectral components and
the existence and the non-randomness of the IDS. They refer to random Schr\"o\-dinger operators as defined in
\ref{d-randomoperator}.

\begin{thm}
\label{t-nrSpec}
There exists a subset $\Omega'$ of full measure in $(\Omega,\cB_\Omega, \PP)$ and subsets of the real line
$\Sigma$ and $ \Sigma_\bullet$, where $\bullet \in\{ disc,ess,ac, sc,pp\}$ such that  for all
$\omega\in \Omega' $
 \[
  \sigma(H_\omega)=\Sigma \quad \text{ and } \quad
  \sigma_\bullet (H_\omega)= \Sigma_\bullet
 \]
for any $\bullet = disc,ess,ac, sc,pp$. If $\Gamma$ is infinite,  $\Sigma_{disc}=\emptyset$.
\end{thm}

The theorem is proven in \cite{LenzPV-2002}, see Theorem 5.1. The arguments go to a large part along the lines of \cite{Pastur-80,KunzS-80,KirschM-82a}. Compare also the literature on  almost periodic Schr\"o\-dinger operators, for instance   \cite{Shubin-1979,AvronS-83}.

For the proof of the theorem one has to find random variables which encode the spectrum of $\{H_\omega\}_\omega$
and which are invariant under the action of $\Gamma$. By ergodicity they will be constant almost surely.
The natural random variables to use are spectral projections, more precisely, their \mindex{trace}traces.
However, since $\RR$ is uncountable and one has to deal also with the different spectral components,
some care is needed.
\medskip

Random operators introduced in Definition \ref{d-randomoperator} are naturally affiliated to a
\index{von Neumann algebra}von Neumann algebra of operators which we specify in

\begin{dfn}
A family $\{B_\omega\}_{\omega\in \Omega}$ of bounded operators
$B_\omega\colon L^2(X,\vol_\omega)\to L^2(X,\vol_\omega)$ is called a
\index{bounded random operator}\emph{bounded random operator} if it satisfies:
\begin{enumerate}[\rm (i)]
\item
$\omega\mapsto \langle g_\omega, B_\omega f_\omega\rangle$ is measurable for arbitrary
$f,g\in L^2(\Omega\times X, \PP\circ \vol)$.
\item
There exists a $\omega$-uniform bound on the norms $\|B_\omega\|$ for almost all $\omega \in \Omega$.
\item
For all $\omega\in\Omega, \gamma \in \Gamma$ the equivariance condition
\[
B_\omega = U_{(\omega,\gamma)} B_{\gamma^{-1} \omega} U_{(\omega,\gamma)}^*
\]
holds.
\end{enumerate}
\end{dfn}
By the results of the next paragraph \S~\ref{ss-meas}, $\{F(H_\omega)\}_\omega$ is a bounded random operator
for any measurable, bounded function $F$.

It turns out that (equivalence classes of) bounded random operators form a von Neumann algebra. More precisely, consider two bounded random operators $\{A_\omega\}_\omega$ and $ \{B_\omega\}_\omega$ as \emph{equivalent} if they differ only on a subset of $\Omega$ of measure zero.
Each equivalence class gives rise to a bounded operator on $L^2(\Omega\times X, \PP\circ \vol)$
by $(B f) (\omega,x) := B_{\omega} f_{\omega}(x)$, see Appendix A in \cite{LenzPV-2002}. This set of operators is a
von Neumann algebra $\cN$ by Theorem 3.1 in \cite{LenzPV-2002}.
On $\cN$ a \mindex{trace}trace $\tau$ of type II$_\infty$ is given by
\[
\tau(B) :=   \EE \left[ \Tr ( \chi_\cF \, B_\bullet) \right]
\]
Here $\Tr:=\Tr_\omega$ denotes the trace on the Hilbert space $L^2(X, \vol_\omega)$.
Actually for any choice of $ u\colon\Omega\times X \to \RR^+$ with
$\sum_{\gamma \in\Gamma} u_{\gamma^{-1}\omega}(\gamma^{-1}x)\equiv 1$ for all
$(\omega,x) \in\Omega\times X$
we have
\be
\label{e-tau-u}
\tau(B) =   \EE \left[ \Tr ( u_\bullet \, B_\bullet) \right]
\ee
In analogy with the case of operators which are $\Gamma$-invariant \cite{Atiyah-76} we call $\tau$ the \index{Gamma-trace@$\Gamma$-trace}\emph{$\Gamma$-trace}.
The spectral projections $\{P_\omega\big (]-\infty, \lambda[\big )\}_\omega$ of $\{H_\omega\}_\omega$ onto the interval $]-\infty, \lambda[$ form a bounded random operator. Thus it
corresponds to an element of $\cN$ which we denote by $P(]-\infty, \lambda[)$.
Consider the normalised $\Gamma$-trace of  $P$
\begin{equation}
\label{e-absIDS}
N_H(\lambda) := \frac{\tau(P\big (]-\infty, \lambda[\big )}{\EE \left[ \vol_\bullet(\cF)\right] }
\end{equation}

The following is Theorem 3 in \cite{LenzPV-2002?}, see also \cite{LenzPV-2002}.
\begin{thm}
\label{t-bigH}
$P(]-\infty, \lambda[) $ is the spectral projection of the direct integral operator
\[
H:= {\int_\Omega}^\oplus H_\omega \, d \PP(\omega)
\]
and $N_H$ is the \index{distribution!function@--- function}distribution function of its \index{spectral!--- measure}spectral measure. In particular, the almost sure spectrum $\Sigma$
of $\{H_\omega\}_\omega$ coincides with the \index{points of increase}points of increase
\[
\{ \lambda\in \RR | \, N_H(\lambda+\epsilon) > N_H(\lambda -\epsilon) \text{ for all $\epsilon >0$}\}
\]
of $N_H$.
\end{thm}
That the IDS can be expressed in terms of a trace on a von Neumann Algebra was known long ago.
In \cite{Shubin-77} and \cite{Shubin-1979} \v Subin establishes this relation for almost periodic elliptic diffential operators in Euclidean space. He attributes the idea of such an interpretation to Berezin, see the last sentence in Section 3 of \cite{Shubin-1979}.
\medskip

We want to describe the self-averaging IDS by an exhaustion of the whole manifold $X$
along a sequence $\Lambda_l \to X, \, l \in \NN$ of subsets of $X$. To ensure the existence of a sequence of subsets
which is appropriate for the exhaustion procedure, we have to impose additional conditions on the group $\Gamma$.

\begin{dfn}
\label{d-amen}
A group $\Gamma$ is called \index{amenable group}\emph{amenable} if it has an left invariant mean $m_L$.
\end{dfn}

Amenability enters as a key notion in Definition \ref{d-admex} and Theorem \ref{t-selfaverIDS}. For readers acquainted only with Euclidean geometry, its role is motivated in Remark \ref{r-amen}.

Under some conditions on the group amenability can be expressed in other ways. A \emph{locally compact} group $\Gamma$ is amenable if for any $\epsilon>0$ and compact $K\subset \Gamma$ there is a compact $G\subset \Gamma$ such that
\[
m_L (G \Delta KG) < \epsilon \, m_L (G)
\]
where $m_L$ denotes the left invariant Haar measure, cf.~Theorem 4.13 in \cite{Paterson-88}. This is a geometric description of amenability of $\Gamma$.
If $\Gamma$ is a discrete, finitely generated group we chose $m_L$ to be the counting measure and write instead $|\cdot|$. In this case $\Gamma$ is amenable if and only if a \index{F\oe lner sequence}\emph{{F\oe lner} sequence} exists.

\begin{dfn}
Let $\Gamma$ be a discrete, finitely generated group.
\begin{enumerate}[\rm (i)]
\item
A sequence $\{I_l\}_l$ of finite, non-empty subsets of $\Gamma$ is called a \emph{{F\oe lner} sequence}
if for any finite $K\subset \Gamma$ and $\epsilon >0$
\[
 \vert I_l \Delta K I_l \vert  \le \epsilon \,{\vert I_l\vert}
\]
for all $ l$ large enough.
\item
We say that a sequence $I_l\subset \Gamma, l\in \NN$ of finite sets has the \index{Tempelman property}\emph{Tempelman} or \index{doubling property|see{Tempelman property}}\emph{doubling property} if it obeys
\[
\sup_{l \in \NN} \frac{\vert I_{l} I_l^{-1} \vert}{\vert I_{l} \vert} < \infty
\]
\item
We say that a sequence $I_l\subset \Gamma, l\in \NN$ of finite sets has the \index{Shulman property}\emph{Shulman property}
if it obeys
\[
\sup_{l \in \NN} \frac{\vert I_{l} I_{l-1}^{-1} \vert}{\vert I_{l} \vert} < \infty
\]
\item
A {F\oe lner} sequence $\{I_l\}_l$ is called a \index{tempered sequence}\emph{tempered {F\oe lner} sequence} if it
has the Shulman property.
\end{enumerate}
\end{dfn}
In our setting $\Gamma$ is discrete and finitely generated.
(Actually, $K:= \{\gamma \in \Gamma \mid \gamma\cF \cap \cF \neq \emptyset \}$ is a finite \mindex{generator set}generator set for $\Gamma$. This follows from the fact that the quotient manifold $X / \Gamma $ is compact, cf.~\S~3 in \cite{AdachiS-1993}.)
Under this circumstances a {F\oe lner} sequence exists if and only if there is  a sequence $J_l \subset \Gamma, l \in \NN$ such that $ \lim_{l \to \infty} \frac{\vert J_l \Delta \gamma J_l \vert} {\vert J_l\vert} = 0$ for all $\gamma \in   \Gamma$.
Moreover, for discrete, finitely generated, amenable groups there exists  a {F\o lner} sequence
which is increasing and \index{exhaustion}exhausts $\Gamma$, cf.~Theorem 4 in \cite{Adachi-1993}.

Both properties (ii) and (iii) control the growth of the group $\Gamma$.
Lindenstrauss observed in \cite{Lindenstrauss-2001} that each {F\oe lner} sequence has a tempered subsequence.
Note that this implies that every amenable group contains a tempered {F\oe lner} sequence.
One of the deep results of Lindenstrauss' paper is, that this condition is actually sufficient for a 
\index{ergodic!--- theorem!pointwise ---}pointwise ergodic theorem, cf.~Theorem \ref{ergthm}.
Earlier it was known that such theorems can be established under the more restrictive Tempelman property \cite{Tempelman-1972,Krengel-1985,Tempelman-1992}. Shulman \cite{Shulman-1988} first realised the usefulness  of the relaxed condition (iii).
\smallskip

In the class of countably generated, discrete groups there are several properties which ensure amenability.
Abelian groups are amenable. More generally, all solvable groups and groups of subexponential growth, in particular nilpotent groups, are amenable. This includes the (discrete) Heisenberg group considered in Example \ref{x-Hbg}. Subgroups and quotient groups of amenable groups are amenable. On the other hand, the free group with two generators is not amenable.

For the discussion of combinatorial properties of {F\oe lner} sequences in discrete amenable groups see \cite{Adachi-1993}.
\smallskip

Any finite subset $I \subset \Gamma$ defines a corresponding
set
\[
\phi(I) := {\rm int}\bigg ( \bigcup_{\gamma \in I} \gamma \overline{\cF}
\bigg ) \subset X
\]
where ${\rm int} (\cdot)$ stands for the open interior of a set.

In the following we will need some notation for the thickened boundary. Denote by $d_0$ the distance function on $X$ associated to the Riemannian metric $g_0$. For $ h >0$, let $\partial_{h} \Lambda:= \{x \in X| \, d_0(x,\partial \Lambda) \le h \}$ be the \index{boundary!--- tube}\emph{boundary tube of width $h$} and $\Lambda_h$ be the interior of the set $\Lambda \setminus \partial_{h} \Lambda$.

\begin{dfn}
\label{d-admex}
(a)
A sequence $\{\Lambda_l\}_l$ of subsets of $X$ is called \index{exhaustion!admissible ---|(}\emph{admissible exhaustion} if there exists an increasing, tempered {F\oe lner} sequence $\{I_l\}_l$ with $\bigcup_l I_l =\Gamma$ such that $\Lambda_l = \phi (I_l^{-1})$, $l\in \NN$.

(b) A sequence $\Lambda_l, l\in \NN$ of subsets of $(X,g_0)$ is
said to satisfy the \index{van Hove property}\emph{Van Hove property} \cite{VanHove-49} if
   \be
   \label{e-isop}
   \lim_{l \to \infty} \frac{\vol_0( \partial_h \Lambda_l )}{\vol_0(\Lambda_l)} = 0 \text{ for all $h > 0$}
   \ee
\end{dfn}

\begin{rem}

In our setting it is always possible to chose the sequences $\{I_l\}_l$ and $\{\Lambda_l\}_l$ in such a way that they \index{exhaustion}exhausts the group, respectively the manifold. However, this is not really necessary for our results.

A simple instance where  $\cup_l \Lambda_l \neq X$ can be given in one space dimension. Let $X=\RR$, $\ZZ$, $I_l=\{0, \dots, l-1\}$, $\cF =[0,1]$  and consequently $\Lambda_l=[0,l]$. One can use this sequence of sets to define the IDS of random Schr\"o\-dinger operators although $\cup_l \Lambda_l = [0,\infty[$.
A non-trivial example where the sets $\Lambda_l$ do not exhaust $X$ can be found in \cite{Sznitman-1989,Sznitman-1990}. There Sznitman considers random Schr\"o\-dinger operators in \index{hyperbolic space}hyperbolic spaces, in which setting the approach presented here does not work due to lack of amenability. However, it is interesting that Sznitman obtains the IDS by choosing a sequence $\Lambda_l$ which converges to a horosphere which is properly contained in the hyperbolic space.

\end{rem}
\smallskip

Thus an admissible exhaustion always exists in our setting. By
Lemma 2.4 in \cite{PeyerimhoffV-2002} every admissible exhaustion
satisfies the \index{van Hove property}van Hove property. Inequality \eqref{e-volest}
implies that for a sequence with the van Hove property
   \[
   \lim_{l \to \infty} \frac{\vol_\omega( \partial_h \Lambda_l )}{\vol_\omega(\Lambda_l)} = 0
   \text{ for all $h > 0$}
   \]
holds for all $\omega\in \Omega$.
Let us remark that one could require for the sets $\Lambda_l$ in the exhaustion sequence to have
smooth boundary, see Definition 2.1 in \cite{PeyerimhoffV-2002}. Such sequences also exist for any $X$ with amenable deck-transformation group $\Gamma$. This may be of interest, if one wants to study Laplacians with Neumann boundary conditions.
For groups of polynomial growth it is possible to construct analoga of admissible exhaustions
by taking metric open balls $B_{r_l}(o)$ around a fixed point $o\in X$ with increasing radii $r_1, \dots, r_n, \dots \to \infty$, cf.~Theorem 1.5 in \cite{PeyerimhoffV-2002}.

We denote by $H_\omega^l$ the Dirichlet restriction of $H_\omega$ to $\Lambda_l$, cf.~Remark \ref{r-QFsa}, and define the finite volume IDS by the formula
\[
N_\omega^l(\lambda):= \vol_\omega(\Lambda_l)^{-1} \# \{ n \mid \lambda_n(H_\omega^l)<\lambda    \}
\]

Now we are able to state the result on the existence of a self-averaging IDS.

\begin{thm}
\label{t-selfaverIDS}
Let $\{H_\omega\}_\omega$ be a random Schr\"o\-dinger operator and $\Gamma$ an ame\-na\-ble  group.
For any \index{exhaustion!admissible ---|)}admissible exhaustion $\{\Lambda_l\}_l$  there exists a set $\Omega'\subset \Omega$ of full measure
such that
\be
\label{e-selfaverIDS}
\lim_{l\to\infty} N_\omega^{l}(\lambda) = N_H(\lambda),
\ee
for every $\omega \in \Omega'$ and every continuity point $\lambda\in \RR$ of $N_H$.
\end{thm}

\begin{dfn}
The limit in \eqref{e-selfaverIDS} is called \index{integrated density of states}\emph{integrated density of states}.
\end{dfn}

Thus all properties (i)--(iii) on page \pageref{i-IDSconv} can be established for the model under study. 
In particular, formula \eqref{e-selfaverIDS} is a variant of the \index{Pastur-\v Subin trace formula}Pastur-\v Subin trace formula in the context of manifolds.
The theorem
is proven in \S\S~\ref{ss-meas}--\ref{ss-EtLt}. It recovers in particular the result of Adachi and Sunada \cite{AdachiS-1993} on the existence of the IDS of \index{periodic!--- operator on manifold}periodic Schr\"o\-dinger operators on manifolds.

\begin{rem}
\label{r-amen}
Let us motivate  for readers acquainted only with Euclidean space why it is natural that the amenability requirement enters in the theorem. In the theory of random operators and in statistical mechanics one often considers a sequence of sets $\Lambda_l, l \in \NN$  which tends to the whole space.
Even in Euclidean geometry it is known that the exhaustion sequence $\Lambda_l, l \in \NN$ needs to tend to $\RR^d$ in an appropriate way, e.g.~in the sense of Van Hove or \mindex{Fisher sequence}Fisher \cite{Ruelle-69a}. Convergence in the sense of Van Hove \cite{VanHove-49} means that
\be
\label{e-isop2}
\lim_{l\to \infty}\frac{|\partial_\epsilon \Lambda_l|}{|\Lambda_l|} = 0
\ee
for all positive $\epsilon$.

If one chooses the sequence $\Lambda_l, l \in \NN$, badly, one cannot expect the convergence of the finite volume IDS' $N_\omega^l, l \in \NN,$ to a limit. In a \index{amenable group}non-amenable geometry, any exhaustion sequence is bad, since \eqref{e-isop2} cannot be satisfied, cf.~Proposition 1.1 in \cite{AdachiS-1993}.
\end{rem}

\begin{rem}
We have assumed the potentials $V_\omega$ to be nonnegative and some of our proofs will rely on this fact.

However, the \emph{statements} of Theorem \ref{t-nrSpec} on the non-randomness of the spectrum and Theorem \ref{t-selfaverIDS} on the existence of the IDS carry over to  $V_\omega$ which are uniformly bounded below by a constant $C$ not depending on $\omega\in \Omega$.
Indeed, in this case our results directly apply to the shifted operator family $\{H_\omega
-C\}_{\omega\in\Omega}$. This implies immediately the same statements for the original operators, since the spectral properties we are considering are invariant under shifts of the spectrum.
\end{rem}

\subsection{Measurability}
\label{ss-meas}

Since we want to study the operators $H_\omega^\Lambda$ as random variables we need a notion of \index{measurability}measurability.
To this aim, we extend the definition introduced by Kirsch and Martinelli \cite{KirschM-82a} for random operators on a fixed Hilbert space to families of operators where the spaces and domains of definition vary with $\omega \in \Omega$.

To distinguish between the scalar products of the different $L^2$-spaces we denote by
$\langle \cdot,\cdot\ra_0$ the scalar product  on $L^2(\Lambda,\vol_0)$ and
by $\Vert \cdot \Vert_0$ the corresponding norm. Similarly,  $\langle\cdot, \cdot\rangle_\omega$
and  $\Vert \cdot \Vert_\omega$ are the scaler product and the norm, respectively, of
$L^2(\Lambda,\vol_\omega)$.

\begin{dfn}
\label{d-measurableH}
Consider a family of selfadjoint operators $\{H_\omega\}_\omega$, where the
domain of $H_\omega$ is a dense subspace $\cD_\omega$ of $L^2(\Lambda, \vol_\omega)$.
The family $\{H_\omega\}_\omega$ is called a \index{measurable family of operators}\emph{measurable family of operators} if
\begin{equation}
\label{e-weakmeas}
\omega \mapsto \langle f_\omega, F(H_\omega) f_\omega\rangle_\omega
\end{equation}
is measurable for all  measurable and bounded $F\colon \RR\rightarrow \CC$
and all measurable functions $f \colon \Omega\times \Lambda\rightarrow \RR$ with $f(\omega,\cdot)=f_\omega \in L^2(\Lambda,\vol_\omega)$ for every $\omega\in \Omega$.
\end{dfn}

\begin{thm}
\label{t-measur}
A random Schr\"o\-dinger operator $\{H_\omega\}_{\omega\in\Omega}$ as in Definition \ref{d-randomoperator} is a measurable
family of operators. The same applies to the Dirichlet restrictions $\{H_\omega^\Lambda\}_{\omega\in\Omega}$ to any open subset $\Lambda$ of $X$.
\end{thm}
For the proof of this theorem we need some preliminary considerations.

Assumption \eqref{M2} in our setting implies that it is sufficient to show the weak measurability
\eqref{e-weakmeas} for functions $f$ which are constant in $\omega$.
Note that $L^2(\Lambda, \vol_0)$ and $L^2(\Lambda, \vol_\omega)$ coincide
as sets for all $\omega \in \Omega$, though not in their scalar products.
Thus it makes sense to speak about $f_\omega \equiv f \in L^2(\Lambda, \vol_\omega) \, "=" \, L^2(\Lambda, \vol_0)$.

\begin{lem}
A random Schr\"o\-dinger operator $\{H_\omega\}_\omega$ is measurable if and only if
\begin{equation}
\label{e-wenigermeas}
\omega\mapsto \langle f , F(H_\omega) f \rangle_\omega \text{ is measurable }
\end{equation}
for all  measurable and bounded $F\colon \RR\rightarrow \CC $ and all $f \in
L^2(\Lambda, \vol_0)$.
\end{lem}
\begin{proof}
To see this, note that \eqref{e-wenigermeas} implies the same statement
if we  replace $f(x)$ by $h(\omega,x)= g(\omega) f(x)$ where $g
\in L^2(\Omega)$ and $f \in L^2(\Lambda, \vol_0)$. Such functions form a
total set in $L^2(\Omega\times \Lambda, \PP\circ \vol)$.

\begin{sloppy}
Now, consider a measurable $h \colon \Omega\times \Lambda\rightarrow \RR$
such that $h_\omega:= h(\omega, \cdot) \in L^2(\Lambda,\vol_\omega)$ for
every $\omega\in \Omega$.
Then $h^n(\omega,x):= \chi_{h,n}(\omega)\, h(\omega,x)$ is in
\mbox{$L^2(\Omega\times \Lambda, \PP\circ \vol)$}
where $\chi_{h,n}$ denotes the characteristic function of the set
$\{\omega| \, \|h_\omega\|_{L^2(\Lambda,\vol_\omega)} \le n \} \subset \Omega $. Since $\chi_{h,n} \to
1$ pointwise on $\Omega$ for $n \to \infty$ we obtain
\[
\langle  h^n_\omega, F(H_\omega) h^n_\omega\rangle_\omega
\to \langle  h_\omega, F(H_\omega) h_\omega \rangle_\omega
\]
which shows that $\{H_\omega\}_\omega$ is a measurable family of operators.
\end{sloppy}
\end{proof}

To prove Theorem \ref{t-measur} we will  pull all operators $H_\omega^\Lambda$ onto the same Hilbert space
using the unitary transformation $S_\omega$ induced by the density $\rho_\omega$
\[
S_\omega\colon L^2(\Lambda,\vol_0) \to L^2(\Lambda,\vol_\omega), \quad
(S_\omega f)(x) = \rho_\omega^{1/2}(x) f (x)
\]
The transformed operators are
\begin{gather}
A_\omega := -S_\omega^{-1} \, \Delta_\omega^\Lambda \, S_\omega
\\      \nn
A_\omega \colon S_\omega^{-1} \, \cD(\Delta_\omega^\Lambda) \subset L^2 (\Lambda,\vol_0)
\longrightarrow L^2(\Lambda,\vol_0)
\end{gather}
The domain of definition $S_\omega^{-1} \, \cD(\Delta_\omega^\Lambda)$ is dense in  $L^2 (\Lambda,\vol_0)$ and contains
all smooth functions of compact support in $\Lambda$.

The first fact we infer for the operators $A_\omega, \omega \in \Omega$ is that
they are uniformly bounded with respect to each other, at least in the sense of quadratic forms.
This is the content of Proposition 3.4 in \cite{LenzPV-2002?} which we quote without proof.

\begin{prp}
\label{p-glmBProp}
Let $Q_0, Q_\omega$ be the quadratic forms associated to the operators
$-\Delta_0^\Lambda$ and $A_\omega$, and $\cD \subset L^2(\Lambda,\vol_0)$ the closure of $C_c^\infty(\Lambda)$
with respect to the norm $\left( Q_0(f,f) + \Vert f \Vert_0^2 \right)^{1/2}$.
Then
\[
\cD=\cD (Q_0) = \cD (Q_\omega)
\]
and there exists a constant $C_A$ such that
\begin {equation} \label{QomQ0}
C_A^{-1} \left( Q_0(f,f) + \Vert f \Vert_0^2 \right)
\le  Q_\omega(f,f) + \Vert f \Vert_0^2
\le  C_A \left( Q_0(f,f) + \Vert f \Vert_0^2 \right).
\end{equation}
for all $f \in \cD$ and $\omega\in\Omega$.
\end{prp}
Here $\cD (Q)$ denotes the \index{quadratic form!--- domain}\emph{domain of definition of the quadratic form} $Q$.
In the proof of this proposition the bound \eqref{M3} --- more precisely \eqref{e-sqrtrho} --- on the gradient of the density $\rho_\omega$ is needed. It seems to be a technical assumption and in fact dispensable by using a trick from \cite{Davies-1990a}, at least if $\Lambda$ is precompact or of finite volume.
\smallskip

Since we are dealing now with a family of operators on a fixed Hilbert space,
we are in the position to apply the theory developed in \cite{KirschM-82a}.
The following result is an extension of Proposition 3 there.
It suits our purposes and shows that our notion
of measurability is compatible with the one in \cite{KirschM-82a}.

Let $\cH$ be a Hilbert space, $\cD \subset \cH$ a (fixed) dense subset and
$B_\omega\colon \cD \to \cH, \omega\in \Omega$ nonnegative operators.
Denote by $\tilde\Sigma= \overline{\bigcup_\omega \sigma(B_\omega)}$ the closure of all spectra,
and by $\tilde\Sigma^c$ its complement. To establish the measurability of the family $\{B_\omega\}_\omega$
one can use one of the following classes of test functions:

\begin{itemize}
\item
$\cF_1= \{\chi_{]-\infty,\lambda[} | \, \lambda \ge 0 \}$,
\item
$\cF_2= \{ x\mapsto e^{itx} | \, t \in \RR \}$,
\item
$\cF_3= \{ x \mapsto e^{-tx} |\, t \ge 0 \}$,
\item
$\cF_4= \{ x \mapsto (z-x)^{-1} | \, z\in\CC\setminus\tilde\Sigma \}$,
\item
$\cF_5=\cF_4(z_0)= \{ x \mapsto(z_0-x)^{-1} \}$ for a fixed $z_0\in \CC\setminus\tilde\Sigma $,
\item
$\cF_6= C_b =\{ f\colon \RR \to \CC | \, f \text{ bounded, continuous} \}$,
\item
$\cF_7 =\{ f\colon \RR \to \CC | \, f \text{ bounded, measurable} \}$.
\end{itemize}

The following proposition says, that it does not matter which of the above sets of functions one chooses for
testing the  measurability of $\{B_\omega\}_\omega$.

\begin{prp}
\label{equivProp}
For $ i =1,\dots,7$ the following statements are equivalent:
\begin{equation*}
({{\mathbf F}_i}) \hspace{4em} \omega \mapsto \langle f, F(B_\omega)h
\ra_{\, \cH} \text{ is measurable for all } f,h \in \cH \text{ and
} F \in \cF_i,
\end{equation*}
\end{prp}

\begin{proof}
It is obvious that (\textbf{F}$_4$) $\Rightarrow$ (\textbf{F}$_5$),
(\textbf{F}$_7$) $\Rightarrow$ (\textbf{F}$_6$), and (\textbf{F}$_6$)
$\Rightarrow$ (\textbf{F}$_3$).  The equivalence of (\textbf{F}$_1$),
(\textbf{F}$_2$) and (\textbf{F}$_4$) can be found in
\cite{KirschM-82a}.

To show (\textbf{F}$_5$) $\Rightarrow$ (\textbf{F}$_4$), consider the
set
\[
Z:= \{z\in \tilde\Sigma^c| \, \omega\mapsto (z-H_\omega)^{-1} \text{ is weakly measurable }\}
\]
in the topological space $\tilde\Sigma^c$. It is closed, since $z_n
\to z$ implies the convergence of the resolvents, see
e.g.~\cite[Theorem VI.5]{ReedS-80}. A similar argument using the
resolvent equation and a Neumann series expansion shows that $z\in Z$
implies $B_\delta(z) \subset Z$ where $\delta :=
d(z,\tilde\Sigma)$. Since $\tilde\Sigma^c$ is connected,
$Z=\tilde\Sigma^c$ follows.

(\textbf{F}$_3$) $\Rightarrow$ (\textbf{F}$_1$): By the
\mindex{Stone-Weierstrass Theorem}Stone-Weierstrass Theorem, see e.g.~\cite[Thm.~IV.9]{ReedS-80},
applied to $C([0,\infty])$ it follows that $\cF_3$ is dense in the set of Functions
$\{f\in C([0,\infty]) \, | \, f(\infty)=0\}=C_\infty ([0,\infty[\, )$.
We may approximate any
$\chi_{]-\infty,\lambda[}$ pointwise by a monotone increasing sequence
$0 \le f_n, n\in\NN$ in $C_\infty (\RR)$. Polarisation, the spectral
theorem, and the monotone convergence theorem for integrals imply that
$\chi_{]-\infty,\lambda[}(H_\omega)$ is weakly measurable.
An analogous argument shows (\textbf{F}$_1$) $\Rightarrow$
(\textbf{F}$_7$), since any non-negative $f\in \cF_7$ can be
approximated monotonously pointwise by non-negative step functions $f_n, n \in
\NN$.
\end{proof}

We use the following proposition taken from \cite{Stollmann-01} (Prop.~1.2.6.)
to show that $\{A_\omega\}_\omega$ is a measurable family of  operators.

\begin{prp}
\label{p-stoll}
Let $B_\omega, \omega\in\Omega$ and $B_0$ be nonnegative operators on a Hilbert space $\cH$. Let
$Q_\omega$, $\omega \in \Omega$ and $ Q_0$ be the associated closed \index{quadratic form}quadratic forms with the following properties:
\stepcounter{equation}
\begin{enumerate}
\item[{\rm (\theequation)} \label{P1}]
$Q_\omega$, $\omega \in \Omega$ and $Q_0$ are defined on the same dense subset $\cD\subset \cH$.
\item[\refstepcounter{equation} {\rm (\theequation)} \label{P2}]
There is a constant $C > 0$ such that
\[
C^{-1} \left( Q_0(f,f) + \Vert f \Vert_0^2 \right) \le
Q_\omega(f,f) + \Vert f \Vert_0^2 \le C \left( Q_0(f,f) + \Vert f
\Vert_0^2 \right)
\]
for all $\omega \in \Omega$ and $f \in \cD$.
\item[\refstepcounter{equation} {\rm (\theequation)} \label{P3}]
For every $f \in \cD$ the map $\omega \mapsto Q_\omega(f,f)$ is measurable.
\end{enumerate}
Then the family $\{B_\omega\}_\omega$ of operators satisfies the equivalent properties of Proposition
\ref{equivProp}.
\end{prp}
By property (\textbf{F}$_7$), this implies that $\{B_\omega\}_\omega$ is
a measurable family of operators.

We apply the proposition to  $B_\omega= A_\omega$, where $\{A_\omega\}_\omega$ is defined
in Proposition \ref{p-glmBProp}. To do so we check that the properties \eqref{P1}--\eqref{P3}
are satisfied:
Since $C^\infty_c (\Lambda)$ is dense in $\cD(Q_\omega)$ for all $\omega$, the closures of
$C^\infty_c (\Lambda)$ with respect to any of the equivalent norms in \eqref{QomQ0} coincide,
which shows assumption \eqref{P1}.
Property \eqref{P2} is just \eqref{QomQ0}, property  \eqref{P3} is obvious for $f\in C^\infty_c (\Lambda)$ and
follows by approximation for all $f\in \cD$.

\begin{proof}[Proof of Theorem \ref{t-measur}]
We already know that the transformed 'kinetic' part $A_\omega$,
$\omega\in \Omega$ of the Hamiltonian is measurable. To deal with
the singular potential we introduce the cut off
\[
V_\omega^n (x) :=  \min\{n, V_\omega (x)\} \text{ for $n\in \NN$ and $\omega \in \Omega$ }
\]
The auxiliary potential $V_\omega^n $ is bounded and in particular its domain of definition
is the whole Hilbert space $L^2(\Lambda,\vol_0)$. Thus the operator sum
\[
A_\omega^n := A_\omega + V_\omega^n , \quad \omega \in \Omega
\]
is well defined and \cite[Prop.~4]{KirschM-82a} implies that it forms a measurable family of operators.
To recover the unbounded potential $V_\omega$, we consider the semigroups\index{semigroup}
$\omega\mapsto \exp(- t A_\omega^n), t>0$ which are weakly measurable.

 The quadratic forms of $A_\omega^n$ converge monotonously
 to the form of $A_\omega^\infty:= A_\omega +V_\omega$. Now Theorems VIII.3.13a and IX.2.16 in
 \cite{Kato-80} imply that the  semigroups of $A_\omega^n$ converge weakly towards the one of
 $ A_\omega^\infty$ for $n\to \infty$.
 Thus $\exp(-t A_\omega^\infty)$ is weakly measurable, which implies the
 measurability of the family $A_\omega^\infty$.

 Finally, since $S_\omega$ is multiplication with the measurable function $(x,\omega) \mapsto
 \rho_\omega (x)$, this implies the measurability of the family  $H_\omega =
 S_\omega A_\omega^{\infty} S_\omega^{-1}, \, \omega \in \Omega$.
\end{proof}

For later use let us note that the \mindex{trace}trace of measurable operators is measurable. More precisely
we will need the fact that the mappings
\be
\label{e-measTr}
\omega \mapsto \Tr (\chi_{\Lambda} e^{- t H_\omega})
\quad \text{ and } \quad
\omega \mapsto \Tr (e^{- t H_\omega^\Lambda})
\ee
are measurable. Note that one can chose an orthonormal basis for $L^2(\Lambda,\vol_\omega)$ which depends in a measurable way on $\omega$, cf.~for instance Lemma II.2.1 in \cite{Dixmier-69}. Thus \eqref{e-measTr} follows immediately from the Definition \ref{d-measurableH} of measurable operators.

\subsection{Bounds on the heat kernels uniform in $\omega$}
\label{ss-unifB}

This paragraph is devoted to \index{heat equation!kernel@--- kernel}heat kernel estimates of the Schr\"o\-dinger operators $H_\omega$.
It consists of four parts. Firstly we discuss existence of $L^2$-kernels of $e^{-t H_\omega}, t >0$
and derive rough upper bounds relying on results in \cite{Davies-1989}. Secondly, we infer Gau\ss ian
\index{decay!--- of heat kernel}\index{heat equation!kernel@--- kernel!off-diagonal decay of ---}off-diagonal decay estimates 
of the kernels using estimates derived in \cite{LiY-1986}. We then present an idea of H.~Weyl to derive the \index{principle of not feeling the boundary}\emph{principle of not feeling the boundary}, and finally we state a proposition which summarises the information on the heat kernel needed in the next section.

We have to control the dependence on the metric and potential of all these estimates since those quantities vary with the random parameter $\omega\in \Omega$.
\bigskip

As $H_\omega$ is non-negative the \index{semigroup}semigroup $e^{-t H_\omega}, t >0$ consists of \index{semigroup!contraction ---}contractions. Moreover, the semigroup satisfies some nice properties formulated in the following definition which enable us to derive estimates on the corresponding heat kernel.
\begin{dfn}
Let $\Lambda\subset X$ be open and $\mu$ a $\sigma$-finite Borel measure on $\Lambda$.
Let $A$ be a real, non-negative, selfadjoint operator on the Hilbert space $L^2(\Lambda, \mu)$.
The semigroup $e^{-t A}, t >0$ is called \index{semigroup!positivity preserving ---}\emph{positivity preserving} if $e^{-t A}f \ge 0$ for any $0\le f\in L^2(\Lambda, \mu)$ and $t >0$.
Furthermore, $e^{-t A}, t >0$ is called a \index{semigroup!Markov ---}\emph{Markov semigroup}, if it is well defined on $L^\infty(\Lambda, \mu)$ and the two following properties hold
\begin{align}
\label{e-BDi}
e^{-tA} &\colon L^{2}(\Lambda,\mu) \longrightarrow
L^{2}(\Lambda,\mu) && \text{ is positivity preserving for every $t>0$}
\\
\label{e-BDii}
e^{-tA} &\colon L^{\infty}(\Lambda,\mu) \rightarrow
L^{\infty}(\Lambda,\mu)  && \text{ is a contraction for every $t>0$}
\end{align}
In this case $A$ is called a \index{Dirichlet form}\emph{Dirichlet form}.

A Markov semigroup $e^{-tA}$ is called \index{semigroup!ultracontractive ---}\emph{ultracontractive} if
\be
\label{e-HC}
e^{-tA} \colon L^2 (\Lambda,\mu) \rightarrow L^{\infty} (\Lambda,\mu) \text{ is bounded for all $t>0$}
\ee
\end{dfn}
The above \eqref{e-BDi} and \eqref{e-BDii} are called \emph{Beurling-Deny}  conditions \cite{BeurlingD-58,BeurlingD-59}.
\bigskip

We infer from \cite{Davies-1989} the following facts:
A Markov semigroup is a contraction on $L^p(\Lambda, \mu)$ for all
$1 \le p\le \infty$ (and all $t>0$). For all $\omega \in \Omega$ the Schr\"o\-dinger operator $H_\omega^\Lambda$ on
$L^2(\Lambda, \vol_\omega)$ is a Dirichlet form, \cite[Thm.~1.3.5]{Davies-1989}.
There the proof is given for $X=\RR^d$, but it applies to manifolds, too.
By Sobolev embedding estimates and the spectral theorem $ e^{t\Delta_\omega^\Lambda} $ is ultracontractive.
Thus by Lemma 2.1.2 in \cite{Davies-1989} each $e^{t \Delta_\omega^\Lambda}$ has a 
\index{semigroup!kernel@--- kernel}kernel, which we denote by
$k_\omega^\Lambda$, such that for almost all $x,y \in \Lambda$
\begin {equation}
\label{e-ctfirst}
0\leq k_{\omega}^\Lambda (t,x,y) \leq \Vert e^{t \Delta_\omega^\Lambda}
\Vert_{1,\infty} =: C_\omega^\Lambda(t)
\end{equation}
Here $\|B\|_{1,\infty}$ denotes the norm of $B \colon L^1\rightarrow L^\infty$.
For $\Lambda=X$ we use the abbreviation $k_\omega^X=k_\omega $.

To derive an analogous estimate to \eqref{e-ctfirst} for the full Schr\"o\-dinger operator with potential
we make use of the \index{Feynman-Kac formula}\emph{Feynman-Kac formula}.
Using the symbol $\Erw$ for the expectation with respect to the \index{Brownian motion}\emph{Brownian motion} $b_t$ starting in $x\in X$ the formula reads
\[
(e^{-t H_\omega}f)(x) = \Erw \left ( e^{-\int_0^t V_\omega(b_s) \, ds} \, f(b_t) \right )
\]
For a \mindex{stochastically complete}stochastically complete manifold $X$ and bounded, continuous $V_\omega$
the formula is proven, for instance, in Theorem IX.7A in \cite{Elworthy-82}. It extends to general non-negative potentials which are in $L^1_{\loc}$ using semigroup and integral convergence theorems similarly as in the proof of Theorem X.68 in \cite{ReedS-75}. Since we consider \mindex{geodesically complete}(geodesically) complete manifolds whose Ricci curvature is bounded below, they are all stochastically complete, cf.~for instance \cite{Grigoryan-1986} or Theorem 4.2.4 in \cite{Hsu-02}.

Since the potential is non-negative, the Feynman-Kac formula implies for non-negative $f\in L^1(\Lambda,\vol_\omega)$
\[
0\leq \big (e^{-t H_\omega^\Lambda} f\big )(x) \leq \big (e^{t\Delta_\omega^\Lambda} f\big ) (x) \le C_\omega^\Lambda(t) \, \|f\|_{L^1}
\]
for almost every $x\in \Lambda$. Thus $e^{-t H_\omega^\Lambda} \colon L^1 (\Lambda,\vol_\omega) \rightarrow L^\infty (\Lambda,\vol_\omega) $ has the same bound $C_\omega^\Lambda(t)$ as the semigroup where the potential is absent.
This yields the pointwise estimate on the kernel $k_{H_\omega}^\Lambda $ of $e^{-t H_\omega^\Lambda} $:
\begin {equation}
\label{e-bounds}
0\leq k_{H_\omega}^\Lambda (t,x,y) \leq C_\omega^\Lambda(t) \quad\text{ for almost every $x,y\in X$.}
\end{equation}
\medskip

In the following we derive sharper upper bounds on the kernels which imply their decay in the distance between the two space arguments $x$ and $y$. Such estimates have been proven by Li and Yau
\cite{LiY-1986} for \index{heat equation!fundamental solution of ---}\emph{fundamental solutions} of the heat equation.
One would naturally expect that the fundamental solution and the 
\index{heat equation!kernel@--- kernel}$L^2$-heat kernel of the semigroup coincide under some regularity assumptions. This is actually the case as has been proven for instance in \cite{Dodziuk-1983} for vanishing, and in \cite{LenzPV-2002?} for smooth, non-negative potentials. The proof in the last cited source uses that
$H_\omega$ is a Dirichlet form.

To formulate the results of Li and Yau \cite{LiY-1986} which we will be using,
we denote by $d_\omega\colon X \times X \to [0,\infty[$ the \mindex{Riemannian distance}Riemannian distance function on $X$
with respect to $g_\omega$. Note that the following proposition concerns the heat kernel of the pure Laplacian.

\begin{prp}
\label{p-LiYau}
For every $t>0$ there exist constants $C(t)>0$, $\alpha_t>0$ such that
\begin{equation}
\label{e-LiYau}
k_{\omega}(t,x,y) \leq C(t) \, \exp \big(-\alpha_t \,d^2_0 (x,y)\big)
\end{equation}
for all $\omega \in \Omega$ and $x,y \in X$.
\end{prp}
\begin{proof}
For a fixed Schr\"o\-dinger operator the estimate (with $d_0$ replaced by $d_\omega$) is contained in Corollary  3.1 in \cite{LiY-1986}. The properties \eqref{M2}, \eqref{e-rhoest} and
\[
C_g^{-1} d_0(x,y) \le d_\omega(x,y) \le C_g d_0(x,y)
\]
ensure that the constants $C(t)$ and $\alpha_t$ in \eqref{e-LiYau} may be chosen uniformly in $\omega$.
Moreover, for measuring the distance between the points $x$ and $y$ we may always replace $d_\omega$ by $d_0$
by increasing $\alpha_t$.
\end{proof}
\smallskip

Let us collect various consequences of  Proposition \ref{p-LiYau} which will be useful later on.
\begin{enumerate}[\rm (i)]
\item
The pointwise kernel bound on the left hand side  of \eqref{e-bounds} can be chosen uniformly in $\omega\in \Omega$.
\item \label{i-Vnachtraeglich}
We stated Proposition \ref{p-LiYau} for the pure Laplacian, although Li and Yau
treat the case of a Schr\"o\-dinger operator with potential. The reason for this is that we want to avoid
the regularity assumptions on the potential imposed in \cite{LiY-1986}.

To recover from \eqref{e-LiYau} the case where a (non-negative) potential is present we use again the Feynman-Kac formula. We need now  a local version of the argument leading to \eqref{e-bounds}, more precisely we consider $e^{-tH_\omega}$ as an operator form $L^1(B_\epsilon(y))$ to $L^\infty(B_\epsilon(x))$ for small $\epsilon>0$.
Thus we obtain
\begin{equation*}
0 \le k_{H_\omega}(t,x,y) \le C(t) \, \exp \big(-\alpha_t \,d^2_0 (x,y)\big)
\end{equation*}
\item
The estimates derived so far immediately carry over to the case where the entire manifold is replaced by an open subset $\Lambda \subset X$.
\[
0\leq k_{H_\omega}^\Lambda(t,x,y) \leq k_{H_\omega} (t,x,y)
\]
This is due to \index{domain monotonicity}\emph{domain monotonicity}, see for example \cite[Thm.~2.1.6]{Davies-1989}.
\item
The \index{Bishop's theorem|see{volume comparison theorem}}\index{volume!--- comparison theorem}\emph{Bishop volume comparison} theorem controls the growth of the volume of balls with radius $r$,
see for instance \cite{BishopC-01}, \cite[Thm.~III.6]{Chavel-84} or \cite{Burstall-99}.
It tells us that the lower bound  \eqref{M4} on the Ricci curvature is sufficient to bound the growth of the volume of balls as $r$ increases. The volume of the ball can be estimated by the volume of a ball with the same radius in a space with constant curvature $K$. The latter volume grows at most exponentially in the radius.
For our purposes it is necessary to have an $\omega$-uniform version of the volume growth estimate. Using Properties \eqref{M2}, \eqref{M4} and \eqref{e-rhoest} we obtain the uniform bound
\[
\vol_\omega \big ( \{y| \, d_\omega(x,y)<r \} \big ) \le C_1 \, e^{C_2 r}  \quad \text{ for all } x\in X
\]
where $C_1, C_2$ do not depend on $x$ and $\omega$.
This implies that for all exponents $p> 0$, there exists a $M_p(t)<\infty $ such that the 
\mindex{moment estimate}moment estimate
\[
\int_\Lambda [k_{H_\omega}^\Lambda(t,x,y)]^{p} \, d\vol_\omega (y) \leq M_p(t)
\]
holds uniformly in $\Lambda \subset X$ open, in $x\in \Lambda$ and $\omega\in \Omega$. We set $M(t) := M_1(t)$.
\item
The heat kernel estimates imply a uniform bound on the \mindex{trace}traces of the semigroup localised in space.
Let $\Lambda\subset X$ be a (fixed) open set of finite volume.
There exists a constant $C_\Tr=C_\Tr(\Lambda,t)>0$ such that for all $\omega\in \Omega$
\[
\Tr \big (\chi_{\Lambda} \, e^{- t H_\omega} \big ) \le C_\Tr
\]
Intuitively this is the same as saying that $\int_\Lambda k_{H_\omega}(t,x,x) \, d \vol_\omega (x)$ is uniformly bounded.
However, since the diagonal $\{(x,x)|\, x \in \Lambda\}$  is a set of measure zero, the integral does not make sense as long as we consider $k_{H_\omega}$ as an $L^2$-function. We do not want here to address the question of continuity of the kernel. Instead we use the \index{semigroup}semigroup property $e^{-2t H_\omega} = e^{- tH_\omega} e^{-t H_\omega}, t>0$ and selfadjointness to express the trace as
\begin{equation}
 \label{e-HGsa}
\Tr \big (\chi_{\Lambda} \, e^{- t H_\omega} \big )
=\int_{\Lambda} \int_{\Lambda} [k_{H_\omega} (t/2,x,y)]^2 \, d\vol_\omega(x) \, d\vol_\omega(y)
\le M_2(t/2) \, \vol_\omega(\Lambda)
\end{equation}
By \eqref{e-volest} this is bounded uniformly  in $\omega \in \Omega$.
Applying domain monotonicity once more, we obtain
\begin{equation}
\label{e-HGsal}
\Tr \big ( e^{- t H_\omega^\Lambda} \big ) \le M_2(t/2) \, \vol_\omega(\Lambda)
\end{equation}
\end{enumerate}
\bigskip

The following lemma is a \index{maximum principle}maximum principle for Schr\"o\-dinger operators with non-negative potentials.
Combined with the \mindex{off-diagonal decay}off-diagonal decay estimates  in Proposition \ref{p-LiYau} it will give us a proof of the
principle of not feeling the boundary.

\begin{lem}[Maximum principle for heat equation with nonnegative potential]
\label{l-Max}
Let $\Lambda \subset X$ be open with compact closure, $V$ be a non-negative function,
and $u \in C ( [0, T[ \times \overline{\Lambda}) \, \cap \, C^2 ( ]0, T[
\times \Lambda)$ be a solution of the \index{heat equation}heat equation
$\frac{\partial}{\partial t} u + (-\Delta + V)u = 0$ on $ ]0, T[ \times
\Lambda$ with nonnegative supremum $s = \sup\{ u(t,x) \mid (t,x) \in [0,T[
\times \overline{\Lambda} \}$. Then,
\[
s = \max \left \{ \max_{x \in \overline{\Lambda}} u (0,x) , \sup_{[0,
T[ \times \partial \Lambda} u (t,x) \right \}
\]
\end{lem}
Note that regularity of $V$ is not assumed explicitly, but implicitly by the requirements on $u$.
They are e.g.~satisfied if $V$ is smooth. Indeed, in that case the heat kernel is smooth, as can be seen
following the proof of \cite[Thm.~5.2.1]{Davies-1989}.
\smallskip

Now we are in the position to state the second refined estimate on the heat kernels, the \emph{principle of not feeling the boundary}. It is a formulation of the fact that the heat kernel of the Laplacian restricted to a large open set $\Lambda$ should not differ much from the heat kernel associated to the Laplacian on the whole manifold, as long as one stays away from the boundary of $\Lambda$. As before, we derive this estimate first for the pure Laplacian and then show that it carries over to Schr\"o\-dinger operators with non-negative potential.

\begin{prp}
\label{p-freierfall}
For any fixed $t, \epsilon>0$, there exists an $h=h(t,\epsilon)>0$ such
that for every open set $\Lambda\subset X$ and all $\omega \in \Omega$
\[
0\leq k_{\omega}(t,x,y) - k_{\omega}^\Lambda (t,x,y)\leq
\epsilon,
\]
for all $x\in \Lambda, y \in {\Lambda_h}$.
\end{prp}

\begin{proof} The first inequality is a consequence of domain monotonicity. So we just have to prove the second one.

Fix $\omega \in \Omega$ and  $t,\epsilon >0$. Choose  $h > 0$ such that
\[
C(t) \exp \Big ( -\alpha_t \big(h/2\big)^2\Big )  \le \epsilon
\]
Note that the choice is independent of $\omega$. For any $y \in \Lambda_h$ and $0 < \delta < h/2$
denote by $B_\delta (y)$ the open $d_0$-ball around $y$ with radius $\delta$.
Let $f_\delta\in C_0^\infty (B_\delta (y))$ be a non-negative approximation of the $\delta$-distribution at $y$.

We consider now the time evolution of the initial value $f$ under the two semigroups generated by $ \Delta_\omega$ and
$ \Delta_\omega^\Lambda$, respectively.
\begin{align*}
u_1 (t,x)&:= \int_X k_\omega (t,x,z) f_\delta(z) d\vol_\omega(z) = \int_\Lambda
k_\omega (t,x,z) f_\delta(z) d\vol_\omega(z).
\\
u_2 (t,x)&:= \int_\Lambda k_\omega^\Lambda(t,x,z) f_\delta(z) d\vol_\omega(z).
\end{align*}
The difference $u_1(t,x)-u_2(t,x)$ solves the heat equation $ \frac{\partial }{\partial t} u = \Delta_\omega  u$
and satisfies the initial condition $u_1(0,x)-u_2(0,x)= f_\delta(x)-f_\delta(x)=0$ for all $x \in \Lambda$.
Now, by domain monotonicity we know $k_\omega (t,x,z)-k_\omega^\Lambda(t,x,z) \ge 0$, thus
\bea
u_1(t,x)-u_2(t,x)= \int_\Lambda \big [ k_\omega (t,x,z) -k_\omega^\Lambda (t,x,z) \big] f_\delta(z) \,
d\vol_\omega(z) \ge 0
\eea
for all $t >0$ and $x \in \Lambda$. The application of the maximum principle yields
\begin{equation} \label{Max-am-Werk}
u_1(t,x)-u_2(t,x) \le \max_{]0,t] \times \partial \Lambda} \left \{
u_1(s,w)-u_2(s,w) \right \}.
\end{equation}
The right hand side can be further estimated by:
\begin{align*}
u_1(s,w)-u_2(s,w) \le \int_\Lambda k_\omega (s,w,z) f_\delta(z) \, d\vol_\omega(z)
= \int_{\Lambda_{h/2}} k_\omega (s,w,z) f_\delta(z) \, d\vol_\omega(z).
\end{align*}
Since $w \in \partial \Lambda$ and $z \in \Lambda_{h/2}$, we conclude using Proposition \ref{p-LiYau}:
\[
\int_{\Lambda_{h/2}} k_\omega (s,w,z) f_\delta(z) d\vol_\omega(z) \le C(t)
\exp\big(-\alpha_t  (h/2)^2\big) \le \epsilon
\]
Since the bound is independent of $\delta$ we may take the limit $\delta \to 0$ which concludes the proof.
\end{proof}
One can prove the principle of not feeling the boundary by other means too, see for instance \cite{LueckS-1999,DodziukM-97,PeyerimhoffV-2002}.
This alternative approach uses information on the behaviour of solutions of the wave equation. Unlike the solutions of the heat equation,
they do not have the unphysical property that their support spreads instantaneously to infinity. Actually, the solutions of the wave equation have \mindex{finite propagation speed}finite propagation speed \cite{Taylor-96a}.
Fourier transforms and the spectral theorem turn this information into estimates on the difference of the solutions of the free and restricted heat equation. Sobolev estimates lead then to the principle of not feeling the boundary.

See also Section 7 in \cite{RayS-1971}.

\begin{rem}
Similarly as in Lemma \ref{l-Max}, one can prove the proposition if a potential is present.
More precisely, Proposition \ref{p-freierfall} is valid  for Schr\"o\-dinger operators with potentials $V$ such that for continuous initial and boundary values the solution of the heat equation $ \frac{\partial }{\partial t} u = -(-\Delta_\omega +V) u$ is in $ C ( [0, T[ \times \overline{\Lambda}) \, \cap \, C^2 ( ]0, T[\times \Lambda)$.
However, Proposition \ref{p-freierfall} implies an analogous estimate for the case where a non-negative
potential is present, similarly as in \eqref{i-Vnachtraeglich} on page \pageref{i-Vnachtraeglich}.
\end{rem}
\smallskip

Consider $e^{-tH_\omega} -e^{-tH_\omega^\Lambda} $  as an operator from $L^1(\Lambda_h)$ to $L^\infty(\Lambda)$,
and denote by $\tau_x^\Lambda$ the \mindex{exit time}\emph{first exit time} from $\Lambda$ for a Brownian motion starting in $x$.
By the Feynman-Kac formula, we have for $0 \le f \in L^1(\Lambda_h)$
\begin{multline*}
[(e^{-tH_\omega} -e^{-tH_\omega^\Lambda})f](x)
 = \Erw \Big ( e^{-\int_0^tds V(b_s)} f(b_t) \chi_{\{b| \, \tau_x^\Lambda \le t\}} \Big )
 \\
 \le
 \Erw \Big ( f(b_s) \chi_{\{b| \, \tau_x^\Lambda \le t\}}\Big )
 =\int [k_\omega(t,x,y)-k_\omega^\Lambda(t,x,y)]f(y) \, d\vol_\omega
 \le \epsilon \int f(y) \, d\vol_\omega
\end{multline*}
if we chose $h$ as in Proposition \ref{p-freierfall}. Thus for almost all $x\in \Lambda,y \in \Lambda_h$
\be
\label{e-PNFBmitv}
k_{H_\omega}(t,x,y)-k_{H_\omega}    ^\Lambda(t,x,y)
\le \|e^{-tH_\omega} -e^{-tH_\omega} \|_{L^1(\Lambda_h) \to L^\infty(\Lambda)}
\le \epsilon
\ee
\bigskip

The upper bounds on the heat kernel and the principle of not feeling the boundary enable us to prove
a result on the traces of localised \index{semigroup!heat ---}heat-semigroups.

In the macroscopic limit, as $\Lambda$ tends (in a nice way) to the whole of $X$, the two quantities
\[
\Tr (\chi_{\Lambda} e^{-tH_\omega})
\quad \text{ and } \quad
\Tr (e^{- t H_\omega^\Lambda})
\]
are approximately the same.
The precise statement is contained in the following

\begin{prp} \label{p-heatkernel}
Let $\{ \Lambda_l \}_{l\in \NN}$, be an sequence of subsets of $X$
which satisfies the \index{van Hove property}van Hove property \ref{e-isop} and let
$\{H_\omega\}_\omega$ be a random Schr\"o\-dinger operator. Then
\[
\lim_{l\to\infty}
\sup_{\omega\in \Omega} \ \frac{1}{\vol_\omega(\Lambda_l)}
\left|  \Tr (\chi_{\Lambda_l} e^{-t H_\omega}) - \Tr( e^{-t H_\omega^{l}})\right|
=0
\]
\end{prp}

\begin{proof} We consider first a fixed $l\in \NN$ and abbreviate $\Lambda=\Lambda_l$.
For the operator $e^{-t H_\omega^\Lambda}$ we may write the trace in the same way
as in \eqref{e-HGsa} to obtain
\begin{equation}
 \label{e-HGsaL}
\Tr(e^{-t H_\omega^\Lambda})
= \int_{\Lambda} \int_{\Lambda} [k_{H_\omega}^\Lambda(t/2,x,y)]^2 d\vol_\omega(x) d\vol_\omega(y)
\end{equation}
We express the difference of \eqref{e-HGsa} and \eqref{e-HGsaL} using
\[
(k_{H_\omega})^2 - (k_{H_\omega}^\Lambda)^2
=(k_{H_\omega} - k_{H_\omega}^\Lambda) (k_{H_\omega} + k_{H_\omega}^\Lambda)
\]
Next we chose $h=h(t/2,\epsilon)>0$ as in Proposition \ref{p-freierfall} and
decompose the integration domain according to
\[
\Lambda \times \Lambda = (\Lambda \times \Lambda_h) \cup (\Lambda \times \partial_h \Lambda)
\]
The difference of the traces can be now estimated as
\begin{multline}
\label{e-diffSGT}
\Tr(\chi_\Lambda \, e^{-t H_\omega}) - \Tr(e^{-t H_\omega^\Lambda})
 \\
 = \int_{\Lambda} \int_{\Lambda_h} \big [k_{H_\omega}\big(\frac{t}{2},x,y\big) - k_{H_\omega}^\Lambda\big(\frac{t}{2},x,y\big)\big ] \, \big [k_{H_\omega}\big(\frac{t}{2},x,y\big)
 + k_{H_\omega}^\Lambda\big(\frac{t}{2},x,y\big)\big ] \, d\vol_\omega(x,y)
 \\
  + \int_{\Lambda} \int_{\partial_h \Lambda} \big [k_{H_\omega}\big(\frac{t}{2},x,y\big) - k_{H_\omega}^\Lambda\big(\frac{t}{2},x,y\big)\big ] \, \big [k_{H_\omega}\big(\frac{t}{2},x,y\big)
 + k_{H_\omega}^\Lambda\big(\frac{t}{2},x,y\big)\big ] \, d\vol_\omega(x,y)
\end{multline}
The first term is bounded by $2 M(t/2) \,  \epsilon  \, \vol_\omega (\Lambda)$
and the second  by
\[
2  M(t/2)  \, C (t/2)  \,\vol_\omega (\partial_h \Lambda)
\]
It follows that
\[
0 \le\frac{1}{\vol_\omega(\Lambda)} \Big (\Tr(\chi_\Lambda \, e^{-t H_\omega}) - \Tr(e^{-t H_\omega^\Lambda})\Big )
\le
2 M(t/2) \epsilon + 2 M(t/2) C (t/2) \frac{\vol_\omega (\partial_h \Lambda)}{\vol_\omega(\Lambda)}
\]
Now, we let $l$ go to infinity. Since the sequence $\Lambda_l$
satisfies the van Hove property \eqref{e-isop} and since our
bounds are uniform in $\omega$, the proposition is proven.
\end{proof}

\subsection{Laplace transforms and Ergodic Theorem}
\label{ss-EtLt}
This paragraph completes the proof of Theorem \ref{t-selfaverIDS}. It relies, apart from the results results established in the previous paragraphs \ref{ss-meas}--\ref{ss-unifB}, on a general \index{ergodic!--- theorem}ergodic theorem and a criterion for the convergence of \index{distribution!function@--- function}distribution functions.

Lindenstrauss proved in \cite{Lindenstrauss-2001,Lindenstrauss-1999} an ergodic theorem which
applies to locally compact, second countable \index{amenable group}amenable groups. It includes as a special case the following statement for discrete groups.

\begin{thm}
\label{ergthm}
Let $\Gamma$ be an amenable discrete group and $(\Omega,\cB_\Omega,\PP)$ be a probability space.
Assume that $\Gamma$ acts ergodically on $\Omega$ by measure preserving transformations.
Let $\{ I_l \}_l $ be a tempered {F\oe lner} sequence in $\Gamma$.
Then for every $f \in L^1(\Omega)$
\begin {equation}
\label{aver}
\lim_{j \to \infty} \frac{1}{\vert I_l \vert} \sum_{\gamma \in I_l} f(\gamma \omega)
= \EE(f)
\end{equation}
for almost all $\omega \in \Omega$.
\end{thm}
In the application we have in mind $f\in L^{\infty}$, so the convergence holds in the $L^1$-topology, too.
\smallskip

\begin{rem}
\label{r-ergthr}
Some background on previous results can be found for instance in Section 6.6 of Krengel' s book \cite{Krengel-1985}, in Tempelman's works \cite{Tempelman-67,Tempelman-1972,Tempelman-1992} or some other sources \cite{EmersonG-1967,Aribaud-70,EmersonG-1974,OrnsteinW-83}. The book \cite{Tempelman-1992} gives in \S~5.6 a survey of Shulman's results \cite{Shulman-1988}.
Mean ergodic theorems hold in more general circumstances, see for instance \cite[\S~6.4]{Krengel-1985}
or \cite[Ch.~6]{Tempelman-1992}.
\end{rem}
\smallskip

We will apply the ergodic theorem above not to the normalised eigenvalue
counting functions $N_\omega^l$, but to their \index{Laplace transform}Laplace transforms $\cL_\omega^l$.
The reason is, that the $\cL_\omega^l$  are bounded, while the original $N_\omega^l$ are not.
The following criterion of Pastur and \v  Subin \cite{Pastur-1971,Shubin-1979}
says that it is actually sufficient to test the convergence of the Laplace transforms.

\begin{lem}[Pastur-\v Subin convergence criterion] \index{Pastur-\v Subin convergence criterion}
  Let $N_n $ be a sequence of distribution functions
  such that
  \begin{enumerate} [\rm (i)]
  \item
  there exists a $\lambda_0 \in \RR$ such that $N_l (\lambda ) =0$ for all $\lambda \le \lambda_0$ and $l \in \NN$,
  \item
  there exists a function $C: \RR^+ \to \RR$ such that
  $\cL^l(t) := \int e^{-\lambda t} dN_l (\lambda) \le C(t)$ for all $l \in \NN$ and $t > 0 $,
  \item
  $\lim_{l \to \infty} \cL^l(t) =: \cL(t)$  exists for all $t > 0$.
  \end{enumerate}
  Then $ \cL$ is the Laplace transform of a distribution function $N$
  and for all continuity points $\lambda$ of $N$ we have
  \[
  N(\lambda) := \lim_{l \to \infty} N_l (\lambda)
  \]
\end{lem}

Finally, we present the proof of Theorem \ref{t-selfaverIDS} on the existence of a self-averaging IDS:

\begin{proof}[Proof of Theorem \ref{t-selfaverIDS}]
We have to check the conditions in the previous lemma for the normalised eigenvalue counting functions $N_\omega^l$.
The first one is clearly satisfied for $\lambda_0=0$, since all operators we are dealing with are non-negative.
To see (ii), express the Laplace transform by the trace of the heat semigroup\index{semigroup}
\[
\cL_\omega^l(t)= \frac{1}{\vol_\omega(\Lambda)} \sum_{n, \, \lambda_n\in \sigma} e^{-t\lambda_n}
=\frac{1}{\vol_\omega(\Lambda)} \,  \Tr (e^{-tH_\omega^l})
\]
The sum extends over all eigenvalues $\lambda_n$ of $H_\omega^l$, counting multiplicities.
Now, \eqref{e-HGsal} implies condition (ii) of the Pastur-\v Subin criterion.

To prove (iii) we will show for all $t > 0$ the convergence
\[
\lim_{j \to\infty} \cL_\omega^l(t) = \int_\RR e^{-t\lambda } dN_H(\lambda)
\]
in ($L^1$ and) $\PP$ almost sure-sense.

Introduce for two sequences of random variables $a_l(\omega),b_l(\omega), l \in \NN$ the equivalence relation $a_l\stackrel{j\to \infty}{\sim} b_l$ if they satisfy $a_l - b_l \to 0$, as $l \to \infty$, in $L^1$ and
$\PP$-almost surely.

For technical reasons we will deal separately with the convergence of the enumerator and denominator in
\[
\cL_\omega^l(t) = \vol_\omega(\Lambda_l)^{-1} \, \Tr(e^{-tH_\omega^l})
\]
However, we need \emph{some} normalisation, to avoid divergences. Consider first
the enumerator with an auxiliary normalisation
\be
\label{e-auxnorm}
\vert I_l \vert^{-1} \, \Tr(e^{-t H_\omega^l})
\ee
By Proposition \ref{p-heatkernel}, the equivariance, and Lindenstrauss' ergodic theorem \ref{ergthm}
\begin{eqnarray*}
\vert I_l \vert^{-1}  \, \Tr(e^{-t H_\omega^l})
&\stackrel{j\to \infty}{\sim}& \vert I_l \vert^{-1}  \, \Tr (\chi_{\Lambda_l} \, e^{-t H_\omega})
= \vert I_l\vert^{-1} \sum_{\gamma \in I_l^{-1}} \Tr( \chi_{\gamma \cF} \, e^{-t H_\omega})
\\
&=& \vert I_l\vert^{-1} \sum_{\gamma \in I_l} \Tr(\chi_{\cF} \, e^{-t H_{\gamma \omega}})
 \,  \stackrel{j\to \infty}{\sim} \, \EE \big \{ \Tr (\chi_{\cF}  \, e^{-t H_\bullet})\big \}
\end{eqnarray*}
Similarly we infer for the normalised denominator
\begin{equation*}
\vert I_l\vert^{-1} \vol_\omega(\Lambda_l)
= \vert I_l\vert^{-1} \sum_{\gamma \in I_l^{-1}} \vol_\omega(\gamma \cF)
= \vert I_l\vert^{-1} \sum_{\gamma \in I_l} \vol_{\gamma \omega}(\cF)
\stackrel{j\to \infty}{\sim} \EE \left \{\vol_\bullet( \cF)\right \}
\end{equation*}
Note that by \eqref{e-volest} all terms in the above line are bounded form above and below uniformly in $\omega$.
By taking quotients we obtain
\begin{equation*}
\cL_\omega^l(t)
= \frac{\vert I_l\vert^{-1}  \, \Tr (e^{-tH_\omega^l})}{\vert I_l\vert^{-1} \, \vol_\omega(\Lambda_l)}
\stackrel{j\to \infty}{\sim}
\frac{\EE\left \{ \Tr(\chi_\cF e^{-tH_\bullet})\right \}}{\EE \left \{\vol_\bullet(\cF)\right \}}
\end{equation*}
The right hand side is the Laplace transform of $N_H$, see the proof of Theorem 6.1 of \cite{LenzPV-2002}
for a detailed calculation.
\end{proof}

\subsection{Approach using Dirichlet-Neumann bracketing}  \index{Dirichlet-Neumann bracketing}
\label{ss-DNb}

We sketch briefly an alternative proof of the existence of the IDS due to Kirsch and Martinelli \cite{KirschM-82c}.
It applies to random Schr\"o\-dinger operators on $\RR^d$.

It relies on a ergodic theorem for superadditive processes by Akcoglu and Krengel \cite{AkcogluK-1981}
and estimates on the number of bound states essentially implied by the \index{Weyl asymptotics}Weyl asymptotics.
\smallskip

Let us explain the notion of an superadditive process in our context. Denote by $Z$ the set of all multi-dimensional intervals or boxes $\Lambda$ in $\RR^d$  such that $\Lambda =\{x | \,  a_j< x_j<b_j , \text{ for } j = 1,\dots, d\}$
for some $a, b \in \ZZ^d$ with $a_j<b_j$  for all $j = 1,\dots, d$.
The restriction of $H_\omega$ to an $\Lambda \in Z$ with Dirichlet boundary conditions is denoted by $H_\omega^\Lambda$ and with Neumann boundary conditions by $H_\omega^{\Lambda,N}$.
Consider a group  $\{T_k\}_{k\in\ZZ^d}$
(or semigroup  $\{T_k\}_{k\in\NN_0^d}$) of measure preserving transformations on the probability space $(\Omega,\cB_\Omega,\PP)$.

\begin{dfn}
A set function $F\colon Z \to L^1(\Omega)$ is called a (discrete) \index{superadditive process}\emph{superadditive process}
(with respect to  $\{T_k\}_{k}$) if the following conditions are satisfied
\begin{align}
&   \label{i-Fequi}
F_\Lambda \circ T_k = F_{\Lambda+k} \text{ for all } k \in \ZZ^d \text{ (or $\NN_0^d$) }, \Lambda \in Z
\\
&   \label{i-Fsupadd}
\text{if $\Lambda_1, \dots,\Lambda_n\in Z$ such that }
\Lambda := \inter \Big (\bigcup_{k=1}^n \overline{\Lambda}_k \Big ) \in Z
\text{ then \ } F_\Lambda \ge \sum_{k=1}^n F_{\Lambda_k}
\\
&   \label{i-Fbeschr}
\gamma :=\gamma(F) :=\sup_{\Lambda \in Z} |\Lambda|^{-1} \EE \{ F_\Lambda\}  < \infty
\end{align}
$F$ is called \index{subadditive process}\emph{subadditive} if $-F$ is superadditive.
\end{dfn}
Similarly one can define continuous superadditive processes with respect to an action of $\RR^d$ on $\Omega$.

We formulate the main result of \cite{AkcogluK-1981} in the way it suits our needs.

\begin{thm}
Let $F$ be a discrete superadditive process and
$\Lambda_l:=[-l/2,l/2]^d$, $l \in \NN$. Then the limit \,
$\lim_{l\to \infty} \, l^{-d} \, F_{\Lambda_l}$ exists for almost
all $\omega\in \Omega$.
\end{thm}

More generally, one can replace the $\Lambda_l, l\in \NN$ by a so called \emph{regular sequence}, cf.~\cite{Tempelman-1972,AkcogluK-1981,KirschM-82c} or \S~6.2 in \cite{Krengel-1985}.
\smallskip

To apply the superadditive \index{ergodic!--- theorem}ergodic theorem we consider for arbitrary, fixed $\lambda\in \RR$ the processes given by the eigenvalue counting functions of the Dirichlet and Neumann Laplacian
\begin{align*}
F_\Lambda^D:= F_\Lambda^D(\lambda,\omega)&:= \#\{n| \, \lambda_n(H_\omega^\Lambda) < \lambda\}, \quad \Lambda \in Z
\\
F_\Lambda^N := F_\Lambda^N(\lambda,\omega)&:= \#\{n| \, \lambda_n(H_\omega^{\Lambda,N}) < \lambda\}, \quad \Lambda \in Z
\end{align*}
where $H_\omega$ is a random operator as in Definition \ref{d-ergO}.
Obviously for $\Lambda=\Lambda_l=[-l/2,l/2]^d$ we have $F_\Lambda^D(\lambda)= l^d N_\omega^l(\lambda)$.
We will show that $F_\Lambda^D, \Lambda \in Z$ is a superadditive process, which is also true for $-F_\Lambda^N, \Lambda \in Z$.  Property \eqref{i-Fequi} follows from the equivariance of $\{H_\omega\}_\omega$, while
\eqref{i-Fsupadd} and \eqref{i-Fbeschr} are implied by the following

\begin{lem}
\label{l-DNbWa}
Let $H_\omega$ be a random operator as in Definition \ref{d-ergO} and $\lambda$ a fixed energy value.
\begin{enumerate}[\rm (i)]
\item
For two cubes $\Lambda_1\subset \Lambda_2$ we have
$F_{\Lambda_2}^D \ge F_{\Lambda_1}^D$ and $F_{\Lambda_1}^N \ge F_{\Lambda_1}^D$.
\item
If $\Lambda_1, \Lambda_2\in Z$ are disjoint such that $\Lambda= \Lambda_1 \cup \Lambda_2 \cup M\in Z$ where $M\subset \RR^d$ is a set of measure zero, then
\begin{align*}
F_\Lambda^D \ge F_{\Lambda_1}^D + F_{\Lambda_2}^D
\\
F_\Lambda^N \le F_{\Lambda_1}^N + F_{\Lambda_2}^N
\end{align*}
\item
There exists an $C_\lambda\in \RR$ such that for all $\Lambda \in Z$ and $\omega \in \Omega$ we have $F_\Lambda^D (\omega)\le C_\lambda \, |\Lambda|$.
\end{enumerate}
\end{lem}
\begin{proof}
The first two statements are known as Dirichlet-Neumann bracketing and are stated e.g.~in Proposition XIII.15.4 in \cite{ReedS-78}. See also Section I.5 in \cite{Chavel-84} for analogous results on manifolds.

To prove (iii) we consider first the case where the potential is identically equal to zero and infer from the Weyl asymptotics that there is a constant $C(\lambda)\in \RR$ such that
\[
\#\{n| \, \lambda_n (-\Delta^{\Lambda}) < \lambda\} \le C(\lambda) \, |\Lambda|
\]
Here $\Delta^{\Lambda}$ denotes the Laplace operator on $\Lambda$ with Dirichlet boundary conditions.
Since the potentials we consider are infinitesimally bounded with respect to the Laplacian, uniformly in $\omega$,
the \mindex{Min-Max principle}Min-Max principle for eigenvalues implies the same estimate for the full Schr\"o\-dinger operator, with an suitably changed constant. See Sections XIII.3, 15 and 16 in \cite{ReedS-78} for more details and, for explicit estimates,
the works \cite{Rozenbljum-72,LiebT-76,Lieb-76,Cwikel-77} deriving the Lieb-Thirring and Cwikel-Lieb-Rozenblum bounds
on the number of bound states.
\end{proof}

Now we can state the main result of \cite{KirschM-82c}.

\begin{thm}
There exists a set $\Omega'\subset \Omega$ of full measure such that
\be
\label{t-KMIDS}
N(\lambda) :=\lim_{l\to\infty} N_\omega^{l}(\lambda)
\ee
exists for every $\omega \in \Omega'$ and every continuity point $\lambda\in \RR$ of $N$.
\end{thm}

\begin{proof}
For a fixed $\lambda \in \RR$ one applies the ergodic theorem of \cite{AkcogluK-1981} to the superadditive process $F^D=F^D(\lambda,\omega)$. Since in our case the transformation group is ergodic, the limit $N(\lambda)$ equals $\gamma(F^D(\lambda))$, in particular it is independent of $\omega$. Almost sure convergence means that there is a  set $\Omega_\lambda$ of measure one for which the convergence holds. Let $S\subset \RR$ be dense countable set. Then $\Omega'=\cap_{\lambda\in S}\Omega_\lambda$ still has full measure and \eqref{t-KMIDS} holds for all $\lambda \in S $ and $\omega \in \Omega'$. Since $S$ was dense, this shows the
convergence \eqref{t-KMIDS} at all continuity points of the limit function. Afterwards one modifies the limit function $N$ to be left continuous. See \cite{KirschM-82c,Kirsch-83} for details.
\end{proof}

For models which satisfy both the conditions of the previous theorem and of \ref{t-selfaverIDS}  the two  IDS's coincide.

Under certain regularity conditions the theorem remains true if Neumann boundary conditions are used to define the IDS,
 and also for $\RR^d$-ergodic potentials, cf.~for instance \cite{KirschM-82c,HupferLMW-2001b}.

\subsection{Independence of the choice of boundary conditions}
\index{boundary!conditions@--- conditions}
Consider again the more general setting of Schr\"o\-dinger operators on a Riemannian covering manifold $X$.
If the open subset $\Lambda \subset X$ of finite volume is sufficiently regular, the Neumann Laplacian $H_\omega^{\Lambda,N}$ on $\Lambda$ has discrete spectrum. One condition which ensures this is the 
\index{extension!property@--- property}\emph{extension property} of the domain  $\Lambda$, see e.g.~\cite{Davies-1989}, which is in turn satisfied if the boundary $\partial \Lambda$ is piecewise smooth. Minimal conditions which ensure the extension property are discussed in \S~VI.3 of \cite{Stein-70}.
Thus it is possible to define the normalised eigenvalue counting function
    \[
    N_\omega^{\Lambda,N}(\lambda)
    := \frac{1}{|\Lambda|} \, \#\{ n \in \NN | \, \lambda_n(H_\omega^{\Lambda,N})<\lambda \}
    \]
Let $\Lambda_l$ be an admissible exhaustion $\Lambda_l \subset X , l \in \NN$ of sets which all have the extension property.
Consider the sequence of distribution functions $N_\omega^{l,N}:=N_\omega^{\Lambda_l,N}$. It is natural to ask whether it converges almost surely, and, moreover, whether its limit coincides with $N$ as defined in \eqref{t-KMIDS}.
If this is true, the IDS is independent of the choice of Dirichlet or Neumann boundary conditions used for its construction. This indicates that boundary effects are negligible in the macroscopic limit.

However, this turns out not to be true for all geometric situations.
Sznitman studied in \cite{Sznitman-1989,Sznitman-1990} the IDS of a random Schr\"o\-dinger operator on hyperbolic space with potential generated by a Poissonian field.
He showed that the IDS does depend on the choice of boundary condition used for its construction. Actually, he computes the \index{Lifshitz tails}Lifshitz asymptotics of the IDS at energies near the bottom of the spectrum and shows that it is different for Dirichlet and Neumann boundary conditions.

In contrast, in the case of Euclidean geometry $X=\RR^d$, the question of 
\index{boundary!conditions@--- conditions!--- independence}boundary condition independence has been answered positively already some decades ago \cite{BenderskiiP-70,KirschM-82c,Shubin-1982,DroeseK-86} for a large class of $\ZZ^d$ or $\RR^d$-ergodic random potentials. Recently, there has been interest in the same question if a magnetic field is included in the Hamiltonian, see also \S~\ref{ss-MagF}.
In this case the coincidence of the IDS defined by the use of Dirichlet and Neumann boundary conditions was established for bounded potentials in \cite{Nakamura-2001}, for non-negative potentials in \cite{DoiIM-2001}, and for certain potentials assuming both arbitrarily large positive and negative values in \cite{HupferLMW-2001b} and \cite{HundertmarkS-02?}. The last mentioned approach seems to be extensible to non-Euclidean geometries which is matter of current research.

\section{Wegner estimate}
\index{Wegner estimate}

In 1981 Wegner \cite{Wegner-81} proved on a physical level of rigour a lower and upper bound on the 
\index{density of states}\emph{density of states} (DOS) of the (discrete) \index{Anderson model}Anderson model and similar lattice Hamiltonians. The density of states --- if it exists --- is the density function of the IDS.
Wegner's  argument did in particular not rely on any information about the type of the spectrum in the considered energy interval.
This was important since before Wegner's result there where various conjectures in the physics community how the
DOS should behave at the \index{mobility edge}\emph{mobility edge}, if it exists. This is the name for the critical energy which is supposed to form the boundary between an interval with pure point spectrum and another one with continuous spectrum. Note however that there is so far no rigorous proof of the existence of continuous spectrum for ergodic random Schr\"o\-dinger operators.

There were arguments suggesting that the DOS should diverge to infinity at the mobility edge, others that it should vanish. Wegner's estimate discarded this misconceptions. In a sense it is a negative result: you cannot recognize the  borderline of different spectral types by looking at the IDS.

In the sequel we concentrate on \index{alloy type!--- model}alloy type models as defined in \ref{d-AM} (and Remark \ref{r-AM}).
We will be concerned here with upper bounds on the DOS only.
It is derived by considering its analoga on finite boxes.
So what we are speaking about in this section is an estimate on
      \bea
      \EE \left \{ \Tr P_\omega^l(I) \right \} =  |\Lambda_l|   \EE \left \{ N^l_\omega(E_2) -N^l_\omega(E_1)  \right \}
      \eea
where for the moment for notational convenience we only consider half open energy intervals $I=[E_1, E_2[$. 
By the \mindex{Cebysev-inequality@\v Ceby\v sev-inequality}\v Ceby\v sev-inequality one sees
    \begin {equation}
    \PP \left \{ \sigma(H_\omega^l) \modcap I \neq \emptyset \right \}
    = \PP \left \{ \Tr P_\omega^l(I) \neq 0  \right \}
    \label{use-ceb}
     \le  \EE \big \{  \Tr P_\omega^l(I) \big \}
    \end{equation}
This means that a bound on the averaged trace of the projection gives immediately a bound on the probability to find an eigenvalue in the considered energy interval. Actually, in the literature on Anderson localisation often the (weaker) bound on $\PP \left \{ \Tr P_\omega^l(I) \neq 0  \right \}$ is called Wegner estimate, since it is sufficient for the purposes of multiscale analysis, see \S~\ref{ss-MSA}.
\medskip

In the following we will adopt the following notation:
$\Lambda_l:= ]-l/2,l/2[^d\subset \RR^d$ denotes the cube of side length $l$ centred at zero.
Occasionally we suppress the dependence on the size and just write $\Lambda$. A cube centred at $x\in \RR^d$ is denoted by $\Lambda_l+x=\{y+x|\, y \in \Lambda_l\}$ or $\Lambda_l(x)$. The characteristic function of the unit cube $\Lambda_1+j$ is abbreviated by $\chi_j$.
For $l \in \NN$ the symbol $\tL_l$ denotes the lattice points in $\ZZ^d$ such that
 \[
 \inter \Big (\bigcup_{j \in \tL_l} \overline{\Lambda_1(j)}\Big)= \Lambda_l
 \]
More explicitly: $\tL_l=\Lambda_l \cap \ZZ^d$.

\subsection{Continuity of the IDS}
\label{ss-ContIDS}

The estimates on the expected number of energy levels in $I$, which most authors derive or use (for localisation proofs) are 'polynomial', more precisely
    \be
    \label{e-polyWE}
     \EE \big \{  \Tr P_\omega^l(I) \big\} \le C_W \, |I|^a \, |\Lambda_l|^b
    \ee
Here $|I|$ and $|\Lambda_l|$ denote the ($1$-dimensional, respectively $d$-dimensional) Lebesgue measure of the energy interval $I$, or the set $\Lambda_l$, respectively. The \index{Wegner constant}\emph{Wegner constant} $C_W$ depends on the various parameters of the model and for continuum Hamiltonians on the supremum of $I$. Actually, $C_W$ can be assumed to be a monotone non-decreasing function of $|\sup I|$. However, once $\sup I$ is fixed, $C_W$ is independent both of the energy interval length and the volume. It is obvious that the energy and volume exponents must satisfy $a\in ]0,1] ,b\in [1,\infty[$. As far as the exponents are concerned, the Wegner estimate is optimal if the dependence on the volume and energy length is linear, i.e.~$a=b=1$.

For, if $b=1$, the bound \eqref{e-polyWE} carries over to the infinite volume IDS:
    \be
    \lim_{l \to \infty}  \EE \left \{   N^l_\omega(E_2) -N^l_\omega(E_2)   \right \}
    =  \lim_{l \to \infty} \frac{\EE \left \{  \Tr P_\omega^l([E_1, E_2[) \right \}}{|\Lambda_l|} \le C_W \, |E_2-E_1|^a
    \ee
Since we know from Theorem \ref{t-selfaverIDS} and dominated convergence that for $E_1,E_2$ in a dense set of energies
    \[
    \lim_{l \to \infty} \EE \left \{ N_\omega^l(E_2) -N_\omega^l(E_1) \right \} = N(E_2) -N(E_1)
    \]
it follows
    \[
    N(E_2) -N(E_1)\le C_W \, |E_2-E_1|^a
    \]
where $C_W=C_W(E_2)$. Now the monotonicity of the IDS implies its \index{H\"older continuity!--- of IDS}H\"older continuity. Moreover, if the estimate \eqref{e-polyWE} is linear in the energy, i.e.~$a=1$, the IDS is even 
\index{Lipschitz continuity}Lipschitz continuous. Thus its derivative, the \index{density of states}\emph{density of states},
    \be
    n(E) := \frac{dN(E)}{dE}
    \ee
exists for almost every energy $E \in \RR$ and is locally bounded
    \[
    n(E)\le C_W(E_2)  \quad \text{ for all $E\le E_2$}
    \]
So the Wegner constant turns out to be a locally uniform bound on the DOS.

\begin{rem}
\label{r-subexpWE}
\begin{sloppy}
For certain models the bounds derived on $\PP \left \{ \sigma(H_\omega^l) \modcap I \neq \emptyset \right \}$ are not polynomial in the volume. This is the case for one-dimensional Anderson or alloy type models where the \index{coupling constants!Bernoulli distributed ---}coupling constants $\omega_j, j \in \ZZ$ are distributed according to the \index{distribution!Bernoulli ---}Bernoulli distribution: for some $p\in ]0,1[$ the random variable $\omega_0$ takes on the value $1$ with probability equal to $p$ and the value $0$ with probability $1-p$. Since this disorder regime is highly 
\mindex{singular disorder}singular, the 'usual' proofs of the Wegner estimate, see Sections \ref{s-WK} and \ref{s-KSCH}, fail. The ones that do work, yield somewhat weaker results. Namely, it is proven in \cite{CarmonaKM-87} (cf.~also \cite{ShubinVW-98}) for the discrete Anderson model  and in \cite{DamanikSS-02}  for the continuum alloy type model,
that for a fixed compact energy interval $I$ and all $\beta \in ]0,1[,\gamma >0$ there exist $l_0 \in \NN$ and $\alpha >0$ such that
\end{sloppy}
    \be
    \label{e-expWE}
    \PP \{d (\sigma(H_\omega^l),E  )\le e^{-\gamma l^\beta}    \} \le e^{-\alpha l^\beta}
    \ee
for all $E\in I$ and all $l \ge l_0$. Here in the case of the continuum model it has to be assumed that $I$ is disjoint from a discrete set of exceptional energies.

This estimate obviously does not imply a continuity  estimate for the infinite volume IDS.
Interestingly, for these models, the H\"older continuity of the IDS is established using other techniques which are specifically tailored for the one dimensional case, see \cite{lePage-84}, \cite[App.~to \S~5]{CarmonaKM-87} and \cite[Thm.~4.1]{DamanikSS-02}. Subsequently, this regularity result is used to derive the finite volume estimate  \eqref{e-expWE}.  In higher dimensions, as we have discussed above, one proceeds in the other direction,
carrying over finite volume estimates to the macroscopic limit.

The bound \eqref{e-expWE} is still sufficient as an input for the \index{multiscale analysis}multiscale analysis which proves localisation, cf.~e.g.~\cite{DreifusK-89} or our discussion in \S~\ref{ss-MSA}.  In the discussion of \mindex{Spencer's example}Spencer's example in \S~\ref{ss-MSA} we will obtain an insight why subexponentially small \mindex{eigenvalue splitting}eigenvalue splittings are effective enough to prevent \mindex{resonance}resonances.
\end{rem}

\begin{rem}[Continuity of the IDS on the lattice and in one dimension]
\label{r-DS-CS}
In \cite{DelyonS-84} Delyon and Souillard showed by a very simple argument that the IDS of the \index{Anderson model}Anderson model
on the lattice $\ZZ^d$ is continuous, regardless of the continuity of the 
\index{distribution!of potential values@--- of potential values}distribution of the potential values.
The potential may even be correlated over long distances, as long as it is an ergodic stochastic field.
Delyon and Souillard use the \index{unique continuation property}unique continuation property of the discrete Schr\"o\-dinger equation, to prove that no eigenvalue can be sufficiently degenerated to produce a jump of the IDS.
At the end of Remark \ref{r-cIDSgeo} we contrast their theorem  with the situation in \mindex{graph}graphs other than the lattice.
In \cite{CraigS-83a} a quantitative version of the continuity is proven: for random, ergodic lattice Hamiltonians the IDS is actually \index{log-H\"older continuity}log-H\"older continuous. See also \cite{Lueck-94c} and \cite{Farber-98,DodziukM-98,Clair-99,MathaiY-02,DodziukMY} for a related result for \mindex{graph Hamiltonian}graph Hamiltonians.

Similarly, the IDS is continuous for one dimensional Hamiltonians, both on $\ZZ$ and on $\RR$, \cite{Pastur-80,JohnsonM-82,AvronS-83}. Again, this result can be strengthened to log-H\"older continuity, cf.~\cite{CraigS-83b}.
\end{rem}

\begin{rem}[Continuity of the IDS and geometry]
\label{r-cIDSgeo}
So far we have only mentioned proven or expected assertions on the continuity of the IDS. One might ask whether there are interesting models which exhibit a \index{discontinuous IDS}discontinuous IDS. It turns out that this phenomenon may occur, if the configuration space has a more complicated geometry than $\ZZ^d$ or $\RR^d$. Another example would be the IDS of the Landau Hamiltonian, cf.~e.g.~the references in \S~\ref{ss-MagF}, in particular \cite{Nakamura-2001}.

Maybe the simplest example to illustrate the difference between Euclidean and more general geometry is provided by periodic Schr\"o\-dinger operators. Under mild assumptions on the $\ZZ^d$-periodic potential $V_\per$ the IDS of the Schr\"o\-dinger operator $H_0=-\Delta +V_\per$ on $\RR^d$ is \index{absolutely continuous!--- IDS}absolutely continuous, cf.~e.g.~\cite[Problem 145]{ReedS-78},\cite{Thomas-73}. In particular the IDS cannot have jumps. However, precisely this can occur for \index{Laplace-Beltrami operator}Laplace-Beltrami operators (even without potential) on a Riemannian covering manifold $X$, as it was mentioned in \cite[App.~2]{Sunada-1988} and is a subject of current research \cite{LenzPPV,LenzPV-03?}. This phenomenon can be deduced from the fact that Laplacians on covering manifolds
may have eigenvalues, as has been shown in \cite{KobayashiOS-1989}. Furthermore, the size of the jumps of the IDS is related to certain geometric invariants. Examples of such invariants are the order of the torsion subgroup of the deck transformation group $\Gamma$ of $X$ and the
\index{Betti number}$L^2$-Betti numbers of $X$, which can be expressed in terms of the \index{Gamma-trace@$\Gamma$-trace}$\Gamma$-trace on a certain von Neumann algebra, establishing the connection to the representation of the IDS discussed before Theorem \ref{t-bigH}.
\cite{Dodziuk-77,DodziukM-97,DodziukM-98,Schick-01a,DodziukMY}. Related material can be found in \cite{Atiyah-76,Sunada-1990,Lueck-94c,Shubin-96,Schick-00,GrigorchukLSZ-00,Schick-01b,MathaiY-02,DodziukLMSY-03,LueckS-02a,LueckS-02b}.

Some of the Wegner estimates we present in Sections \ref{s-WK} and \ref{s-KSCH} extend to alloy type models on manifolds. A particularly interesting phenomenon occurs if one considers a periodic Laplace-Beltrami operator with discontinuous IDS, and perturbs it randomly such that the IDS of the perturbed  operator is continuous. This happens if either an appropriate alloy type potential is added to the Hamiltonian or if the metric is multiplied by an appropriate  alloy type perturbation, see \cite{LenzPPV,LenzPV-03?}.

A discontinuous IDS may also occur for models with a random geometry. This is the case for the tight-binding Hamiltonian defined on Delone sets studied in \cite{LenzS,KlassertLS-02}. Ideas related to the ones in \cite{KlassertLS-02} have been used in \cite{SchenkerA-00}  and in \S~2 of  \cite{MathaiY-02}. The paper \cite{SchenkerA-00} is devoted to the proof of the existence of spectral gaps for certain graph Hamiltonians.

We will discuss a different example, the \index{quantum!--- percolation model}\emph{quantum percolation model}, in some more detail, since it fits readily in the class of models which we have described so far.
This model has been studied amongst others in  \cite{deGennesLM-59a,deGennesLM-59b,KirkpatrickE-72,ShapirAH-82,ChayesCFJS-86,Veselic-QP}.
We sketch the site \index{percolation}percolation problem on $\ZZ^d$ with probability parameter $p$ above the percolation threshold:
let $v_k\colon \Omega\to \{0, \infty\}, k\in \ZZ^d$ be a sequence of independent, identically distributed random variables which take on the value $0$ with probability $p$ and the value $\infty$ with probability $1-p$. Define $X_\omega$ to be the infinite component of the set of \emph{active sites} $\{k \in \ZZ^d| \, v_k(\omega)=0\}$. The graph  $X_\omega$ is called the (active) \emph{infinite cluster}. For $p$ above a certain critical value $p_c$ it is known that almost surely    an infinite cluster exists \cite{Kesten-82,Grimmett-99}  and is unique \cite{AizenmanKN-87a,GandolfiGR-88}.

One defines the Laplacian $h_\omega$ on $X_\omega$ as the restriction of the finite difference operator onto $l^2(X_\omega)$. For a cube $\Lambda_l$ one defines $X_\omega^l$ to be those active sites in $\Lambda_l \cap \ZZ^d$ which are connected to the boundary $\partial \Lambda_l$ by a chain of active sites. The finite volume
Laplacian $h_\omega^l$ is the usual finite difference operator restricted to $l^2(X_\omega^l)$.

Although the finite active clusters, which would obviously give rise to bound states, are not taken into consideration, it turns out that $h_\omega$ has bound states. This was  as first observed in \cite{KirkpatrickE-72}.
Eigenstates with finite support in the infinite cluster are called \emph{molecular states}.
The existence of such states affects the properties of the IDS of $h_\omega$,
which is defined in the following way. For each $l\in \NN$  the normalised eigenvalue counting function of the Hamiltonian $h_\omega^l$ is defined as
    \[
    N_\omega^l(E):= \frac{1}{\#(\Lambda_l \cap \ZZ^d)} \, \# \{n | \, \lambda_n(h_\omega^l) < E \}
    \]
which converges for $l \to \infty$ to an non-random limit almost surely \cite{ChayesCFJS-86,Veselic-QP}. A simple construction shows that there are locally supported eigenfunctions, which depend only on the pattern of $X_\omega$ in a bounded region. Consequently the patterns and the associated localised eigenfunctions occur with a non-zero density along the infinite cluster and thus produce jumps of the IDS at the corresponding energy. In \cite{ChayesCFJS-86} it is shown by physical arguments that the discontinuities of $N$ constitute a dense set of energies.

Actually, uniqueness of the infinite cluster is not used in the arguments of \cite{ChayesCFJS-86} and a similar argument for constructing finitely supported eigenfunctions does work on the Bethe lattice as well, although there the infinite cluster is not unique. For the quantum percolation model on the Bethe lattice locally supported eigenfunctions have been
constructed in \cite[\S~7]{CarmonaKM-87}.

A mathematically rigorous study of the quantum percolation model
on amenable graphs is undertaken in \cite{Veselic-QP}. There the
discontinuities of the IDS are explained in terms of the breakdown
of the \index{unique continuation property}unique continuation property of eigenfunctions of the
adjacency operator, see also Remark \ref{r-DS-CS}. Moreover, the set of these energies is characterised in the case $X=\ZZ^d$.
From a wider perspective, the properties of this set are related to the Atiyah conjecture,
cf.~
\cite{DodziukLMSY-03}.
\end{rem}

\begin{rem}
While the continuity of the IDS has clearly to do with the
\index{distribution!of eigenvalues@--- of eigenvalues}distribution of eigenvalues of the random Hamiltonian, it only
captures a part of the properties of this distribution. The theory
of \index{level statistics}\emph{level statistics} is concerned with the finer structure of the
fluctuations of eigenvalues. It can be studied by an appropriate
scaling procedure. This has been carried out for certain
one-dimensional and discrete models in
\cite{Molchanov-78a,Molchanov-81,Reznikova-81a,Reznikova-81b,Minami-1996,Giere-98}.
\end{rem}

\subsection{Application to Anderson localisation} \index{localisation!Anderson ---}
\label{ss-MSA}

In the last paragraph the implications of Wegner estimates for the IDS were presented. Now we focus on the second main application of those bounds, namely \emph{Anderson localisation}.

As we discussed earlier in \S~\ref{ss-transport}, this phenomenon tells us that a random family of Schr\"o\-dinger operators exhibits in a certain energy interval dense \index{pure point spectrum}pure point spectrum, almost surely. Moreover, the eigenfunctions of the eigenvalues lying in this interval decay exponentially. Even a stronger property, namely \emph{dynamical localisation}, holds. See \S~\ref{ss-transport} for more details and references.
\smallskip

For multi-dimensional configuration space there are two techniques at disposal to prove localisation: the \index{fractional moment method}\emph{fractional moment method} and the \emph{multiscale analysis} (MSA). The first one is also called \index{Aizenman-Molchanov technique|see{fractional moment method}}\emph{Aizenman-Molchanov} technique and was introduced in \cite{AizenmanM-93,Aizenman-94a,Aizenman-94b,Graf-94,AizenmanG-98}.
It was so far applicable only to lattice Hamiltonians, up to the recent extension to continuum configuration space \cite{AizenmanENSS}. For discrete models it has in fact been proven \cite{AizenmanSFH-00,AizenmanSFH-01} that in the energy regime where the MSA applies, the Aizenman-Molchanov method works, too.

However, we will discuss in a little more detail only the MSA, since
it has found applications to a variety of models
and since the Wegner estimate is a key ingredient in the MSA.
We first sketch the basic ideas of the MSA, and then discuss shortly its history.
\smallskip

To carry through the MSA one needs two a priori estimates: the \index{initial scale estimate}\emph{initial scale estimate} and the \index{Wegner estimate}\emph{Wegner estimate}.
These two conditions essentially determine for which \index{single site!--- potential}single site potentials $u$, 
\index{single site!--- distribution}single site distribution measures $\mu$ and which energy intervals localisation can be derived.
Note that $u$ and $\mu$ are  parameters which determine our alloy type potential, see Definition \ref{d-AM}.

In the literature one can find multiscale analyses which are adapted to
operators describing the propagation of classical waves
or to abstract families of  differential operators, see among others
\cite{FigotinK-1996,FigotinK-1997a,CombesHT-1999,DamanikS-2001,Stollmann-01,GerminetK-2001a,KleinK-?}.
In this context one has also to make sure that certain other conditions are satisfied, like the  
\mindex{geometric resolvent inequality}geometric resolvent inequality, the \mindex{generalised eigenfunction expansion}generalised eigenfunction expansion, a rough upper estimate on the trace of spectral projections of finite box operators (obtained e.g.~by the \index{Weyl asymptotics}Weyl asymptotics), etc. However, since we discuss here only (random) Schr\"o\-dinger operators, these conditions are automatically satisfied, cf.~\cite[\S~ 3.2]{Stollmann-01} or \cite[App.~A]{GerminetK-2001b}.
\smallskip

The multiscale analysis is an induction argument over a sequence of increasing length scales $l_k, k\in \NN$.
Each scale $l_{k+1}$ is a power $l_k^\alpha$ of the preceding one, where $\alpha\in ]1,2[$.
Actually, for technical reasons one truncates the scales so that  all $l_k$ lie in $6\NN$.

One considers the restriction of the alloy type model $H_\omega$ to the open cube $\Lambda^{(k)}:=\Lambda_{l_k}(0)$ of side length $l_k$.
The corresponding restricted operator is denoted by $H_\omega^{(k)}$, where Dirichlet, Neumann or periodic boundary conditions ensure its selfadjointness. One wants to study decay properties of the Green's function of $H_\omega^{(k)}$,
i.e.~the integral kernel of the resolvent operator $R_\omega^{(k)}(z) =(H_\omega^{(k)}-z)^{-1}$, where $z$ is taken from the resolvent set $\CC \setminus \sigma(H_\omega^{(k)})$. Since we are not interested in pointwise properties of the kernel of  $R_\omega^{(k)}(z)$, and since they tend to be unpleasant near the diagonal, we may consider instead the sandwiched resolvent
    \[
    \chi^{out} (H_\omega^l-E)^{-1} \chi^{in}
    \]
Here $\chi^{out}$ denotes the characteristic function of the boundary belt \mbox{$\Lambda_{l-1}\setminus\Lambda_{l-3}$}, and $\chi^{in}$ the characteristic function of the interior box $\Lambda_{l/3}$.

The initial scale estimate is stated in terms of the notion of \emph{regular cubes}.
A box $\Lambda_l=\Lambda_l(0)$ is called \emph{$(E,\gamma)$-regular} if $l \in 6 \NN$, $E \not \in \sigma(H_\omega^l)$, and
\be
\label{e-regular}
\| \chi^{out} (H_\omega^l-E)^{-1} \chi^{in} \| \le e^{-\gamma l}
\ee
The exponent has to satisfy $\gamma \ge l^\beta$ for some $\beta > -1$. So regularity describes quantitatively how fast the Green's function on a finite box decays. The exponent $\beta$ must be greater than $-1$ otherwise the rhs of \eqref{e-regular} could be just one.

The \index{initial scale estimate}\emph{initial scale estimate} is satisfied if there exist a scale $l_1 < \infty$
such that for some $\xi >0$
\beq
\label{e-inise}
\PP\{\omega| \,\forall E \in I : \, \Lambda_l \text{ is $(E,\gamma)$-regular for } \omega  \} \ge 1- l^{-\xi}
\eeq
for any $l \ge l_1$.
\smallskip

The \emph{weak form} of the \index{Wegner estimate}\emph{Wegner estimate} as it is needed for the MSA is:
    \be
    \PP \{d (\sigma(H_\omega^l),E  )\le e^{-l^\theta}    \} \le l^{-q}  \text{ for all } l \in 6\NN
    \ee
where $\theta <1/2$ and $q>d$.
\smallskip

The initials scale estimate \eqref{e-inise}
serves as the induction anchor of the MSA.
The induction step uses the Wegner estimate and proves that the exponential \index{decay!--- of Green function}decay property holds on the subsequent scale $l_2$ with even higher probability, and that the decay exponent $\gamma$ essentially does not change. As one repeats the procedure on the scales $l_1, l_2, \dots$ one proves that the decay of the Green's functions $\chi^{out} (H_\omega^{(k)}-E)^{-1} \chi^{in}$ holds with probability which tends to one, with error bounded polynomially
in $l_k^{-1}$.

Thus one establishes the exponential decay of the sandwiched resolvent on arbitrary large cubes, where the decay exponent $\gamma$ is bounded away from zero uniformly in the scales.
Now one uses polynomial bounds on the growth of eigenfunctions and subsolution estimates to prove spectral localisation, cf.~for instance \cite[\S~3.3]{Stollmann-01}.
To prove dynamical localisation one has to do  more work, see e.g.~\cite[\S~3.4]{Stollmann-01} or \cite{GerminetDB-98b,DamanikS-2001,GerminetK-2001a}.

The assumptions for the MSA depend on several parameters, and so do the various versions of localisation which may be obtained by it.
Recently Germinet and Klein showed in \cite{GerminetK-2001a} how to optimize the dependence of the MSA on the various parameters, i.e.~how to obtain with the weakest assumptions in the input the strongest conclusions.


\medskip

    The MSA was introduced by Fr\"ohlich and Spencer in \cite{FroehlichS-1983}. The method applied to the Anderson
    model on the lattice and experienced various improvements and applications
    \cite{MartinelliS-1985a,MartinelliS-1985b,FroehlichMSS-85,SimonW-86,DelyonLS-85a}.

    Based on results from \cite{Dreifus-87} and \cite{Spencer-88} von Dreifus and Klein presented in \cite{DreifusK-89}
    a streamlined version of
    the MSA. Although results on localisation for continuum Hamiltonians existed
    earlier \cite{MartinelliH-84,KotaniS-87},
    it was this simplification of the MSA, which made alloy type Schr\"o\-dinger operators more accessible to systematic
    research.

There was a series of articles which proved various variants of the MSA for continuum models
\cite{Klopp-93,CombesH-94b,Klopp-95a,KirschSS-1998a,GerminetDB-98b,FischerLM-00,DamanikS-2001,Stollmann-01,GerminetK-2001a}.
Other works concentrated onto identifying energy/disorder regimes where one can prove the assumptions needed for the MSA to work     \cite{Klopp-93,Klopp-95b,Klopp-95a,BarbarouxCH-97b,KirschSS-1998a,KirschSS-1998b,GerminetK-2001a,Stollmann-2000b,
 Veselic-2002b,Veselic-2002a,HislopK-02,Zenk-02,Klopp-02c}.
\smallskip

\begin{rem}
\label{r-1dim}
One dimensional models play a special role in the game of localisation. Namely, for $d=1$ there are some specific techniques available which do not exist in higher dimensions, or are not as effective. Some examples are: the
transfer-matrix method, study of the Ljapunov exponent, ODE techniques, Pr\"ufer transformation.

Consequently, in one space dimension localisation has been proven for some models
even before the MSA technique was available. See
    \cite{GoldsheidMP-77,Molchanov-78a,Goldsheid-81,Carmona-82,Royer-82,KotaniS-87}
for various  models on $\ZZ$ or $\RR$.
Furthermore there are certain models which even now can be treated only in one dimension.
This applies to the following random Schr\"o\-dinger operators: \mindex{random displacement model}random displacement model \cite{BuschmannS-2001,SimsS-00}, potentials generated by a \mindex{Poissonian field}Poissonian field \cite{BuschmannS-2001}, alloy type potentials with \index{single site!--- potential!--- of changing sign}changing sign (at all energies) \cite{Stolz-2000}, discrete and continuous models with \index{Bernoulli!--- disorder}Bernoulli disorder \cite{CarmonaKM-87,ShubinVW-98,DamanikSS-02}.
This restriction to $d=1$ is partially due to the fact, that there is no Wegner estimate at disposal in these cases.
\end{rem}
\smallskip

The following paragraph gives an idea where the Wegner estimate is used in the MSA.

\subsection{Resonances of Hamiltonians on disjoint regions}

Rather than describing precisely how the Wegner estimate enters in the induction step of the MSA we will
confine ourselves to present an illuminative example due to Spencer \cite{Spencer-86}. It was originally formulated for lattice Hamiltonians, but can also be considered in the continuum case, as we have learned from P.~M\"uller.
\smallskip

As we mentioned earlier \eqref{e-polyWE} implies for $0\le\delta<1$
    \be
    \label{e-polyWWE}
    \PP\{\omega| \, d(\sigma(H_\omega^l),E) \le \delta  \}   \le C_W(E) \, \delta^a \, |\Lambda_l|^b
    \ee
This inequality implies that with respect to the parameter $\omega$ the eigenvalues of $H_\omega^l$ do not cluster on the energy axis. To give a more precise meaning to this statement, consider two box Hamiltonians $H^1=H_\omega^{\Lambda_l(x)}$ and $H^2=H_\omega^{\Lambda_l(y)}$. Assume that the boxes $\Lambda_l(x)$ and $\Lambda_l(x)$ are sufficiently far apart such  that $H^1$
and $H^2$ are independent.  Let $I$ be a bounded interval and consider the event
\[
\Omega(\sigma_1,\sigma_2)
:= \{\omega | \, B_\delta( \sigma_1) \modcap B_\delta( \sigma_2) \neq \emptyset \}
\]
where $\sigma_i$ stands for $\sigma(H^i) \modcap I$. Let $\lambda_1,\dots,\lambda_N$ be the eigenvalues of $H^1$ in $I$.
By \index{Weyl asymptotics}Weyl's law we know that $N\le C_I |\Lambda_l|$, where $C_I$ is independent of $\omega$.
Since
\bea
\Omega(\sigma_1,\sigma_2) \subset \bigcup_{n=1}^N \Omega(n ,\sigma_2)
\quad \text{ where }
\Omega(n ,\sigma_2) := \{\omega| \, B_{2\delta}(\lambda_n(\omega)) \modcap \sigma_2 \neq \emptyset \}
\eea
we may use \eqref{e-polyWWE} to conclude
    \be
    \PP\{\Omega (\sigma_1,\sigma_1) \}   \le C  \delta^a \, |\Lambda_l|^{b+1}
    \ee
This means that \index{resonance}resonances of $H^1$ and $H^2$, i.e.~the occurrence of approximately the same eigenvalues for both operators are very unlikely.
\smallskip

The feature which is common to Spencers example and the MSA is the effect of resonances between Hamiltonians which are localised to disjoint cubes. As we mentioned earlier, in the induction step of the MSA one puts together boxes $\Lambda_l$ of side length $l$ to form a larger cube $\Lambda_L$ of side length $L$. Assume that one knows already that the Green's functions of the operators $H_\omega^l$ living on any one of the small cubes $\Lambda_l$ decays exponentially.

The Schr\"o\-dinger operator $H_\omega^L$ on $\Lambda_L$ is obtained when we remove the boundary conditions which separate the smaller boxes $\Lambda_l$. The question  is whether the Green's function on $\Lambda_L$ will still decay (with approximately the same rate). To answer this question affirmatively it is not enough to know the exponential decay of the individual Green's functions on the small boxes $\Lambda_l$, but it has to be ensured that they are not in resonance with each other.

\index{resonance}Resonance means in this context that the spectra of two restriction $H^1, H^2$ of $H_\omega$ to disjoint cubes are very close to each other, and can be formulated quantitatively in terms of
    \be
    \label{dist-Spektrum}
    d (\sigma_1, \sigma_2) := \inf \{d(\lambda_1, \lambda_2)| \, \lambda_1 \in \sigma_1,\lambda_2 \in \sigma_2\}
    \ee
as we will see below.

The model situation we are about to consider is easier than the one occurring in the MSA because we do not introduce boundary conditions but confine ourselves to the analysis of the discrete spectrum below zero.

\begin{exm}[Spencer's example \protect{\cite[p.~903--904]{Spencer-86}}]
\label{ex-Spencer}
Consider two smooth potentials $V_1, V_2 \le 0$ with compact support and set
    \be
    \label{e-dist-Vi}
    \cV_i := \supp V_i \modsubset B_r(a_i) , r > 0, a_i \in \RR^d, i=1,2
    \ee
It follows $d(\cV_1,\cV_2 ) \ge | a_1- a_2| -2r =: \varrho$. Consider furthermore the operators
    \begin{align}
    \label{e-Hi}
    H &:= H_0+V_1+V_2, \qquad H_0 :=-\Delta
    \\
    H_i &:= -\Delta +V_i , \quad i =1,2
    \end{align}

Denote by $\sigma_i := \sigma (H_i) \modcap ]-\infty, 0[ , \, i =1,2$ the negative spectra, which are purely discrete.
We are interested in the localisation and decay properties of the corresponding eigenfunctions.

We look at two cases, where the first has a special symmetry and the second corresponds to the situation one expects to occur in a random medium.
\\[1em]
\textbf{Case (A):}\\
Consider first the \emph{exceptional} case in which $V_2$ is obtained from $V_1$ by a reflection along an axis of symmetry.
Without loss of generality
    \be
    \label{e-Spiegelung}
    V_2(x_1, x_2, \dots, x_d) = V_1(-x_1, x_2, \dots, x_d)
    \ee
Thus $H$ commutes with the reflection operator
    \be
    \label{e-pi-Operator}
    \Pi \colon L^2(\RR^2) \to L^2(\RR^2), \quad (\Pi f ) (x_1, x_2, \dots, x_d) = f  (-x_1, x_2, \dots, x_d)
    \ee
In particular, for every eigenfunction $\psi$
    \[
    H\psi = \lambda \psi  , \quad \lambda < 0
    \]
the reflected function $\Pi \psi$ is an eigenvector of $H$ as well. If $\psi$ is localised
around $a_1$ $\Pi \psi$ will be localised around $a_2$. Thus a typical vector form
$\Span \{ \psi, \Pi \psi\}$ will have non negligible amplitudes both at $a_1$ and $a_2$, even for large distances $\varrho$. For short, eigenfunctions of $H$ do not need to have just one centre of localisation.

We are dealing with a resonance between the two disjoint regions $ \cV_1 , \cV_2 $,
or more precisely between the spectra of $H_1$ and $H_2$.
Actually, we encountered the extreme case where the the spectra $\sigma_1$ and $\sigma_2$
are not only close to each other but identical.
\smallskip

The example we just considered exhibited a special symmetry, namely  \mbox{$[H,\Pi]=0$}.
For random potentials we expect generically that such symmetries are absent and that the spectra $\sigma_1$ and $\sigma_2$ have positive distance. This situation is considered in
\pagebreak[3]\\[1em]
\textbf{Case (B):}\\ \nopagebreak[3] \noindent
We give a condition on $d(\sigma_1,\sigma_2)$ which ensures that the eigenfunctions of
$H$ (defined in \eqref{e-Hi}) are localised at \emph{only one} of
the potential wells $V_1,V_2$. Namely, assume that
    \be
    \label{e-epsilon}
    d(\sigma_1, \sigma_2) \ge e^{-\sqrt{ \varrho}} =: \epsilon
    \ee
For an eigenvalue $\lambda \le -\varrho^{-1}$, with corresponding equation $H\psi =\lambda\psi$, we have
either
    \begin{align}
    \label{e-weg-von-E}
    | \lambda_1^j - \lambda | \ge \epsilon /2 , \quad \forall  \lambda_1^j \in \sigma_1
    \qquad  \text{ or } \qquad
    | \lambda_2^j - \lambda | \ge \epsilon /2 , \quad \forall  \lambda_2^j \in \sigma_2
    \end{align}
Assume without loss of generality the first case. The eigenfunction equation implies
    \[
    - \psi = (H_1 -\lambda )^{-1} V_2 \psi
    \]
Applying twice the resolvent equation we obtain
    \be
    \label{e-2x-res}
    - \psi = [ R_0-R_0 V_1 R_0 + R_0 V_1 R_1 V_1 R_0 ] \  V_2 \psi ,
    \ee
where  $R_i := (H_i -\lambda)^{-1} , \, i =0,1$ denotes the resolvents. We show that the amplitude of $\psi$ on $\cV_1$ is exponentially small in the parameter $\varrho$.
Denote with $\chi^i$ the characteristic function of $\cV_i$ for $i=1,2$ and multiply (\ref{e-2x-res}) with $\chi^1$
    \begin{multline}
    \label{auf-V1}
    - \chi^1\psi
    = \chi^1 R_0 \chi^2 V_2 \psi - \chi^1 R_0 V_1 \chi^1 R_0 \chi^2 V_2 \psi
    + \chi^1 R_0 V_1 R_1 V_1 \chi^1 R_0 \chi^2  V_2 \psi .
    \end{multline}
The free resolvent decays exponentially, see e.g.~\cite{Agmon-1982} or  \cite[IX.30]{ReedS-75},
    \[
    R^{1,2} := \| \chi^1 R_0 \chi^2 \|  \lesssim e^{-\varrho \sqrt{-\lambda}} ,
    \]
the terms $ \| V_2 \psi \|, \| \chi^1 R_0 V_1 \| $ are bounded uniformly in $\varrho$,
and (\ref{e-weg-von-E}) implies
    \[
    \| R_1 \| \le \frac{2}{\epsilon} = 2 e^{\sqrt \varrho}
    \]
Consequently
    \begin{multline*}
    \| \chi^1 \psi \|
     \le R^{1,2} \ \| V_2 \psi \|
                            + \| \chi^1 R_0 V_1 \|  \ R^{1,2}  \ \| V_2 \psi \|
    + \| \chi^1 R_0 V_1 \|  \ \| R_1 V_1 \| \  R^{1,2}  \ \| V_2 \psi \|
    \end{multline*}
is bounded by a constant times  $\exp ( - \sqrt \varrho) \, \|\psi\|$, since $\varrho^{-1} \le -\lambda$.
\end{exm}
\smallskip

Let us finish this section  by discussing some aspects and contrasts of the two cases considered in the example.
\begin{enumerate}[\rm (i)]
\item
In general, the spectrum alone describes only general properties of the solution of the eigenvalue equation.
In our example it is the additional information contained in \eqref{e-Spiegelung} and \eqref{e-epsilon},
respectively, which allows us to analyse the eigenfunctions more precisely.

\item
Obviously, in Case (A), the Green's function  decays in space, too. However, this decay is not yet felt at the scale $\varrho$, since $  |\psi(a_1)\psi(a_2)|$
converges to a positive constant for $\varrho\to \infty$. On the contrary, in Case (B), the amplitude is small either
at $a_1$ or $a_2$, so
    \[
    |\psi(a_1)\psi(a_2)| \lesssim e^{-\sqrt{ \varrho} } .
    \]
\item
A semiclassical analysis of double well potentials is carried out, for instance, in \cite{HislopS-1996}.
\end{enumerate}

\section{Wegner's original idea. Rigorous implementation}
\label{s-WK}
\setcounter{thm}{0}

In this section we present a proof of Wegner's estimate following his original ideas in \cite{Wegner-81}. His proof was originally formulated for the discrete Anderson model. In the meantime, it has been cast into mathematically rigorous form and adapted for continuum Hamiltonians. We follow mostly the arguments of Kirsch \cite{Kirsch-96}.
There are proofs of Wegner's estimate by other authors, which make use of the ideas in \cite{Wegner-81}.
Let us mention \cite{MartinelliH-84,Maier-87,Klopp-93,Klopp-95b,FischerLM-96,CombesHKN-2002,CombesHK-03}. \nocite{Fischer-96}

The theorem to be proven is

\begin{thm}
\label{t-WK}
Let $H_\omega$ be as in Definition \ref{d-AM} and assume additionally that there exists an $\kappa >0$ such that
\be
\label{e-posdefu}
u \ge \kappa \chi_{[-1/2,1/2]^d}
\ee
Then for all $E_0 \in \RR$ there exists a constant $C_W=C_W(E_0)$ such that for all $l \in \NN$, $E\le E_0$
and all $\epsilon \in [0,1]$
    \be
    \label{e-WK}
    \EE \big \{  \Tr \big [ P_\omega^l(\left [E -\epsilon,E+\epsilon \right ]) \big ] \big \} \le C_W \ \epsilon \ l^{2d}
    \ee
\end{thm}

This theorem is proven in the next section. Its bound with respect to the volume term $l^d$ is quadratic and does not yield a continuity statement for the IDS.
Subsequently we show how this estimate was improved in \cite{CombesHN-2001}.
Denote by $\omega_+$ and $\omega_-$ the largest, respectively the smallest value a coupling constant may take.

\subsection{Spectral averaging of the trace of the spectral projection} \index{spectral!--- averaging}
\label{ss-SAtraceSP}
We show that the expectation over the randomness smears out the eigenvalues of $H_\omega^l$ and thus regularises the trace $P_\omega^l(I)$.
\\

By definition the \index{spectral!--- projection}spectral projection $P_\omega^l(I)=\chi_I (H_\omega^l)$ is the characteristic function of
$H_\omega^l$. For certain purposes it will be necessary to differentiate this function with respect to the energy parameter, which motivates the introduction of the following smooth 'switch function'.

Let  $\rho$ be a smooth, non-decreasing function such that on $]-\infty, -\epsilon]$ it is identically equal to $-1$, on $[\epsilon, \infty[$ it is identically equal to zero and $\|\rho'\|_\infty \le 1/\epsilon$. Then
\[
\chi_{]E-\epsilon,E+\epsilon [} (x)
\le \rho(x-E+2\epsilon) -\rho(x-E-2\epsilon)
=
\int_{-2\epsilon}^{2\epsilon} dt \,\rho'(x-E+t)
\]
Thus by the spectral theorem
\[
P_\omega^l(]E-\epsilon,E+\epsilon [) \le  \int_{-2\epsilon}^{2\epsilon} dt \,\rho'(H_\omega^l-E+t)
\]
in the sense of quadratic forms. Since $B_\epsilon(E) =]E-\epsilon,E+\epsilon [$ is bounded and $ \sigma(H_\omega^l)$
discrete, the above operators are \index{trace!--- class}trace class and we may estimate:
\[
\Tr \Big [ P_\omega^l(B_\epsilon(E)) \Big ]
\le
\Tr \Big [ \int_{-2\epsilon}^{2\epsilon} dt \,\rho'(H_\omega^l-E+t)  \Big ]
=
\sum_{n\in \NN} \int_{-2\epsilon}^{2\epsilon} dt \,\rho'(\lambda_n^l(\omega)-E+t)
\]
where $\lambda_n^l(\omega)$ denotes the eigenvalues of $H_\omega^l$ enumerated in non-decreasing order and counting multiplicities. Only a finite number of terms in the sum are non-zero. More generally the above arguments prove the following

\begin{lem}
Let $H$ be an operator with purely discrete spectrum. Denote by $\lambda_1 \le \lambda_2 \le \dots $ the eigenvalues of $H$. Then for $E\in \RR$ and $\epsilon>0 $
\[
\Tr \Big [ \chi_{B_\epsilon(E)}(H) \Big ]
\le
\sum_{n\in \NN} \int_{-2\epsilon}^{2\epsilon} dt \,\rho'(\lambda_n-E+t)
\]
\end{lem}
\smallskip

In the following we analyse the behaviour of the spectrum of the Schr\"o\-dinger operator
under the perturbation $\omega_j \, u(\cdot-j)$. Fix a box-size $l \in \NN$, a lattice site $j \in \tL$
and a configuration of coupling constants $\omega \in \Omega$ and consider the one-parameter family of operators
\[
t \mapsto H_t:=H+ tU , \text{ where $ H= H_\omega^l$ and $U=u(\cdot-j)$}
\]
By the arguments in \S\ref{ss-MN} the single site potential is infinitesimally bounded with respect to $H$, thus
$H_t$ forms a \index{holomorphic family of operators}holomorphic family of type (A) in the sense of Kato \cite{Kato-80} for $t$ in a neighbourhood of the real line, cf.~e.g.~XII.\S~2 in \cite{ReedS-78}. Moreover, $H_t$ has compact resolvent by XII.\S~14 in \cite{ReedS-78}.  Hence one may apply a theorem of Rellich \cite{Rellich-42}, see also Theorem VII.\S3.9 in \cite{Kato-80}. It says that the eigenvalues and eigenvectors of $H_t$ can be chosen to be real analytic on $\RR$.
Actually, each eigenvalue is holomorphic on a neighbourhood of $\RR$ in the complex plane, but their intersection may contain only $\RR$.

If $\lambda_n(t)$ is a non-degenerate eigenvalue of $H_t$, first order perturbation theory tells us that there exists a normalised eigenfunction $\psi_n(t)$ such that
\be
\label{e-HF}
\frac{d \lambda_n}{dt}(t_0)= \la \psi_n(t_0), U  \psi_n(t_0)\ra
\ee

\begin{rem}
This is sometimes called \index{Hellmann-Feynman formula}\emph{Hellmann-Feynman formula}, and it holds true also if the eigenvalue $\lambda_n$ happens to be degenerate at $t=t_0$, cf.~for instance \cite{IsmailZ-88}.
One has however to chose the enumeration of the eigenvalues $\lambda_n$ and eigenvectors $\psi_n$ in such a way that
the pair $\lambda_n(t), \psi_n(t), t <t_0$ continues holomorphically into $\lambda_n(t), \psi_n(t), t >t_0$.
Note that this is actually not the case with the enumeration we chose earlier, where $\lambda_n(t)$ denotes the $n$-th eigenvalue of $H_t$.
There are two possibilities to solve the problem: either one chooses a somewhat unintuitive enumeration of eigenvalues which makes them --- together with the eigenvectors --- holomorphic functions of $t$. Or one sums over the eigenvalues.
Namely, formula \eqref{e-HF} remains true if we sum over all eigenvalues which correspond to a degeneracy.
More precisely, for a degenerate eigenvalue $\lambda_n(t_0)$ denote by $l,k\in \NN$ the largest numbers such that
$\lambda_{n-l}(t_0) = \dots =\lambda_n(t_0)= \dots =\lambda_{n+k}(t_0)$ and set $S(t)= \sum_{m=n-l}^{n+k}\lambda_m(t)$.  Then we have
\[
\frac{dS}{dt}(t_0)= \sum_{m=n-l}^{n+k}   \la \psi_m(t_0), U\psi_m(t_0) \ra
\]
In the application of \eqref{e-HF} in the next proposition we will be considering all eigenvalues below a certain energy. Thus if we consider one eigenvalue participating in a degeneracy we will actually take into account all participating eigenvalues.
\end{rem}
\smallskip

The results for one parameter families of operators carry over to the multi-parameter family $\omega \mapsto H_\omega^l$. Thus we have
\[
\sum_{j \in \tL} \frac{\partial\lambda_n(H_\omega^l)}{\partial\omega_j}
=\sum_{j \in \tL}  \la \psi_n, u(\cdot-j)  \psi_n\ra
\]
where $\psi_n$ are normalised eigenvectors corresponding to $\lambda_n(H_\omega^l)$. By  assumption \eqref{e-posdefu}
we have
\be
\label{e-upartUnity}
\sum_{j \in \tL}  \la \psi_n, u(\cdot-j)  \psi_n \ra \ge \kappa >0
\ee
Now the chain rule
\begin{eqnarray*}
\sum_{j \in \tL}
\frac{ \partial \rho (\lambda_n(H_\omega^l) -E + t) }{ \partial \omega_k }
=
\rho'(\lambda_n(H_\omega^l) -E + t)
\sum_{j \in \tL} \frac{\partial \lambda_n(H_\omega^l)}{\partial \omega_k}
\end{eqnarray*}
implies
\begin{eqnarray}
\label{e-HFlowerbound}
\rho'(\lambda_n(H_\omega^l) -E + t)
\le
\kappa^{-1}  \sum_{j \in \tL}
\frac{ \partial \rho (\lambda_n(H_\omega^l) -E + t) }{ \partial \omega_j } .
\end{eqnarray}
Due to monotonicity, integrating over one coupling constant  we obtain
\begin{multline*}
\int  d\omega_j \, f(\omega_j) \, \frac{\partial  \rho(\lambda_n(H_\omega^l)-E+t)}{\partial \omega_j}
\le \| f \|_\infty \, \int  d\omega_j \, \frac{\partial  \rho(\lambda_n(H_\omega^l)-E+t)}{\partial \omega_j}
\\
=
\|f\|_\infty \big [ \rho(\lambda_n(\omega,j=\max)-E+t)  -\rho(\lambda_n(\omega,j=\min)-E+t) \big]
\end{multline*}
where $\lambda_n(\omega,j=\max)$ denotes the $n$-th eigenvalue of the operator
\[
H_\omega^l(j=\max):= H_\omega^l +(\omega_+-\omega_j) \, u(x-j)
\]
where $\omega_j$ takes its maximal value. Analogously we use the notation $\lambda_n(\omega,j=\min)$. This proves
\begin{prp}
\label{p-forboth}
\begin{multline*}
\EE \big \{  \Tr \big [ P_\omega^l(\left [E -\epsilon,E \right ]) \big ] \big \}
\\
\le
\frac{\|f\|_\infty}{\kappa}  \sum_{n\in \NN} \int_{-2\epsilon}^{2\epsilon}dt \,
\sum_{j \in \tL} \EE
\Big \{\rho[\lambda_n(\omega, j=\max)-E+t]- \rho[\lambda_n(\omega, j=\min)-E+t] \Big \}
\end{multline*}
\end{prp}
\smallskip

The upper bound can be also written as
\begin{multline*}
\frac{\|f\|_\infty}{\kappa}   \int_{-2\epsilon}^{2\epsilon}dt \,
\sum_{j \in \tL} \EE
\Big \{ \Tr
\big [ \rho[H_\omega^l(j=\max)-E+t]- \rho[H_\omega^l( j=\min)-E+t] \big ]
\Big \}
\end{multline*}

Since $\rho \le 0$

\begin{gather}
\nn
\sum_{n\in \NN}\rho[\lambda_n(\omega, j=\max)-E+t]- \rho[\lambda_n(\omega, j=\min)-E+t]
\\
\label{e-rough}
\le  - \sum_{n\in \NN} \rho[\lambda_n(\omega, j=\min)-E+t]
\le C_{E+3\epsilon} \ l^d
\le C_{E_0+3} \ l^d
\end{gather}
by bound (iii) in Lemma \ref{l-DNbWa}. This proves Theorem \ref{t-WK}.
\begin{rem}
\label{r-rough}
The suboptimality of the volume bound in Theorem \ref{t-WK} is due to the rough estimate \eqref{e-rough}.
The right hand side of the inequality is the net increase of the number of eigenvalues in the energy interval $]E-t- \epsilon, E-t+\epsilon[$ due to the decrease of the $j$-th coupling constant from its maximal to its minimal value.
This quantity is expected to be independent of $\Lambda$. However, in \eqref{e-rough} we estimated it by the total number of eigenvalues below the energy $E+3\epsilon$, which is by \index{Weyl asymptotics}Weyl's law proportional to the volume of $\Lambda$.
Thus we get an extra volume factor in the upper bound of the Wegner estimate.
\end{rem}

\subsection{Improved volume estimate}
\label{ss-CHN}
In \cite{CombesHN-2001}  Combes, Hislop and Nakamura obtained a Wegner estimate analogous to the one in the last section, but with upper bound linear in the volume. The Wegner estimate they proof is somewhat weaker in the energy parameter.

More precisely the main result in \cite{CombesHN-2001} may be formulated as follows.
Let $I_G:=]E_-, E_+[, I_G \cap \sigma(H_0)=\emptyset$ be a spectral gap of $H_0$.
Denote by $\tilde H_\omega^l$ the operator $H_0+\sum_{k\in \tL_l}   \omega_k u(\cdot-k)$ and by $\tilde P_\omega^l$ the corresponding spectral projector.

\begin{thm}
\label{t-CHN}
Let $H_\omega$ be as in Definition \ref{d-AM} and assume additionally that the periodic potential $V_{\per}$ is bounded below, $0 \le u \in C_c(\RR^d)$ and $u$ is not identically equal to zero.
Let  $E\in I_G$, $ 0< \epsilon_0 := \frac{1}{2} d(E, I_G^c)$ and $\alpha<1$ then there exists a finite constant $C$ such that
    \be
    \label{e-CHN}
    \EE \big \{  \Tr \big [ \tilde P_\omega^l(B_\epsilon(E)) \big ] \big \}
    \le C \, \epsilon^\alpha \ l^{d}
    \quad \text{ for all $ 0\le \epsilon < \epsilon_0$ and $l\in \NN$}
    \ee
\end{thm}
Note that $\tilde H_\omega^l$ is \emph{not} an operator restricted to a bounded domain. This is the reason, why the theorem does not apply to energies inside the unperturbed spectrum $\sigma(H_0)$.
Estimate \eqref{e-CHN} implies the \index{H\"older continuity!--- of IDS}H\"older continuity of the IDS outside the spectrum of the unperturbed operator $H_0$.
A sharper bound, linear in $\epsilon$, was obtained in \cite{BarbarouxCH-97b}.
However, its proof is based on a different technique, cf.~Section \ref{s-KSCH}.
We will prove here a slightly simpler fact than the above, namely
\begin{thm}
\label{t-KS}
Let $H_\omega$ be as in Definition \ref{d-AM} and assume additionally that the periodic potential $V_{\per}$ is bounded below, $u \in L^p_c(\RR^d), p=p(d)$ and for some $\kappa >0$
\[
u \ge \kappa \chi_{[0,1]^d}
\]
Then, there exists for any $E_0\in \RR$ and $\alpha<1$ a finite constant $C$ such that
\[
\EE \left \{ \Tr \big [ P_\omega^l(B_\epsilon(E) \big ]  \right \}
\le C  \, \epsilon^\alpha \ l^{d}
\]
for all $ \epsilon \in [0,1]$, $E\le E_0$ and $l\in \NN$.
\end{thm}
This result is proven in Section 3 of \cite{KostrykinS-01b} using partially different methods. Note that the earlier papers
\cite{KotaniS-87,CombesH-94b} contain sharper estimates, cf.~Section \ref{s-KSCH}.

The fundamental contribution of \cite{CombesHN-2001} was that it replaced  the Weyl-type volume estimate \eqref{e-rough}, whose drawback we explained in Remark \ref{r-rough}. We present now the improved volume estimate of \cite{CombesHN-2001}. Together with the argument from the previous \S~\ref{ss-SAtraceSP} it proves Theorem \ref{t-KS}.

The trace
\be
\label{e-SSFtr}
\Tr \big [ \rho[H_\omega^l(\omega, j=\max)-E+t]- \rho[H_\omega^l(\omega, j=\min)-E+t] \big ]
\ee
can be expressed using the \index{spectral!--- shift function}\emph{spectral shift function} $\xi=\xi(\cdot,H+U,H)$, abbreviated \index{SSF|see{spectral shift function}}SSF, of the pair of operators
\[
H:= H_\omega^l(\omega, j=\min) \text{ and } H+U, \text{ where } U= (\omega_+ - \omega_-) \, u(x-j)
\]
The necessary estimates on the SSF are collected in Appendix \ref{a-SSF}. Since the difference of our operators is not \index{trace!--- class}trace class, we have to use the indirect definition of the SSF by the \emph{invariance principle}, cf.~\eqref{e-IP}.
Let $C_0 \in \RR$ be such that $H_\omega, H_0 \ge C_0$ for all $\omega$ and $g(x)=(x-C_0+1)^{-k}$ for some $k\in \NN$.

For $k > \frac{d+4}{2}, k \in \NN $ the operator $g(H+U)-g(H)$ is trace class and the invariance principle implies
\begin{gather}
\nn
\Tr \big (\rho(H+U-E-t)  - \rho(H-E-t) \big)
=
-\int \rho'(\lambda) \xi\big(g(\lambda), g(H+U), g(H)\big) \, d\lambda
\\
\le
-\big ( {\textstyle\int} \rho'(\lambda)^q d\lambda\big)^{1/q} \,
\left ( \int_{\supp \rho'} \xi\big (g(\lambda), g(H+U), g(H)\big)^p  \, d\lambda \right)^{1/p}
\end{gather}
In the last line we used the H\"older inequality  and $p,q \ge 1$ are conjugate exponents $\frac{1}{p}+\frac{1}{q}=1$.
Remember that we choose $\rho$ depending on $\epsilon$. Thus, its derivative is bounded by $\epsilon^{-1}$ times a constant and
\[
\big ( {\textstyle\int} \rho'(\lambda)^q d\lambda\big)^{1/q}
\le
C \big ( 1/\epsilon \big)^{\frac{q-1}{q}} \big ( {\textstyle\int} \rho'(\lambda) d\lambda\big)^{1/q}
=C \, \epsilon^{\frac{1}{q}-1} =C \, \epsilon^{\frac{1}{p}}
\]
Since on the support of $\rho'$ the function $g$ is uniformly bounded away from zero
a transformation of variables gives
\begin{multline*}
\hspace*{-1em}- \left (\int_{\supp \rho'} \xi(g(\lambda), g(H+U), g(H))^p \, d\lambda \right)^{1/p}
\le
C  \left (\int_{\RR} \xi(\lambda, g(H+U), g(H))^p \, d\lambda \right)^{1/p}
\\
=
C \, \|g(H+U)- g(H)  \|_{J_{1/p}}^{1/p}
\le
\tilde C
\end{multline*}
where we used in the last line Theorems \ref{t-LpSSF} and \ref{t-NakPain} from the Appendix. There one can find the definition of the \index{super-trace class}\emph{super-trace class ideal} $J_{1/p}$ and its norm. The constant $\tilde C$ is independent of the box $\Lambda_l$, the lattice site $j$, the configuration of the coupling constants $\omega_k, k\neq j$, and of $\epsilon$.

Hence we have a volume independent bound for \eqref{e-SSFtr}, in contrast to the estimate discussed in Remark
\ref{r-rough}. The bound is H\"older continuous in the energy parameter. In view of Proposition \ref{p-forboth} the proof of Theorem \ref{t-CHN} is finished.
\bigskip

\begin{rem}
\label{r-HKV}
Recently Hundertmark, Killip and the author \cite{HundertmarkKV-?} found a different, shorter way to prove the super-trace class estimates
and to apply them to bound the SSF. The basic observation is, that one can control the singular values of the
difference of two Schr\"o\-dinger 
\index{Schr\"odinger operator!semigroup generated by ---}\index{semigroup!Schr\"odinger ---}semigroups. 
In fact the \index{singular values}singular values decay almost exponentially, and the semigroup difference is therefore in any super-trace class ideal.

We state without proofs their result on the decay of the singular values and the estimate on the SSF it implies.
For simplicity we consider here the case where the magnetic vector potential is absent.
\end{rem}

\begin{thm}
\label{t-singval}
Let $H_1=-\Delta+V$ and $H_2=H_1+u$, with $V, u \in L_{\loc}^1(\RR^d), V, u
\ge -\frac{1}{2}C_0 $.
Denote by $H_1^l,H_2^l$ the corresponding Dirichlet restrictions to the cube $\Lambda_l=[-l/2,l/2]^d$.
Set  $V_{\eff}^l:=e^{-H_1^l}- e^{-H_2^l}$.
There are finite positive constants  $c_1, c_2$  such that the singular values $\mu_n$ of the operator $V_{\eff}^l$ obey \be
\label{e-singval}
\mu_n \le c_1 \, e^{-c_2 \,  n^{1/d}}
\ee
The constants  depend only on $d$, $C_0$ and the diameter of the support of $u$.
\end{thm}

Let $\rho$ be a switch function as above.
\begin{thm}
\label{c-SSF}
There is a constant depending only on $d$, $C_0$, $\diam \, \supp \, u$ and $E+\epsilon$ such that
\[
\Tr \left [ \rho (H_2^l-E)-\rho (H_1^l-E)  \right  ]
\le const  \, \log(1 + 1/\epsilon)^d
\]
for all $\epsilon>0$ and $l\in \NN$.
\end{thm}

\subsection{Sparse potentials}\mindex{sparse potential}
Form the physical point of view there are some interesting models which
have a potential
\be
\label{e-genAP}
V_\omega(x) = \sum_{k \in \Gamma} \omega_k u(x-k)
\ee
resembling the alloy type model. However, the set $\Gamma$ may be much more general than the lattice $\ZZ^d$.
A class of particular interest are \mindex{surface model}surface models where $\Gamma= \{0\} \times \ZZ^\nu$ and $\nu<d$ is the dimension  of a hyperplane in whose neighbourhood the potential is concentrated. The literature on such models includes
\cite{Chahrour-99a,KostrykinS-00a,KostrykinS-01b,BoutetdeMonvelS-?}, see also \cite{EnglischKSS-1988,EnglischKSS-90}. The results in this paragraph are taken from \cite{KirschV-2002a}.

Here we will consider arbitrary sets $\Gamma$, which are \emph{uniformly discrete} in the following sense
\[
\sup_{x\in \RR^d }\# \{\Gamma \modcap B_1(x)\} < \infty
\]
For uniformly discrete $\Gamma$ the number of points of $\Gamma$ contained in the cube $\Lambda_l(x)$
can be bounded linearly in the volume of the cube and independently of its centre $x$.

Consider a background Schr\"o\-dinger operator $H_0=-\Delta+ V_{\per}$ with a periodic potential $V_{\per}\in L_{\unif, \loc}^p(\RR^d)$ where $p=p(d)$ is as in \eqref{e-defp(d)}. By adding a constant we may assume that $\inf\sigma(H_0)=0$.
Let $H_\omega=H_0+V_\omega$ be an random operator with an \mindex{generalised alloy type potential}\emph{generalised alloy type potential} $V_\omega$ as in \eqref{e-genAP}.

As before $H_\omega^l$ stands for the restriction of $H_\omega$ to the cube $\Lambda_l$ with Dirichlet boundary conditions (we may as well use Neumann or periodic ones), and $P_\omega^l $ denotes the corresponding spectral projection.
\begin{thm}
Assume that the single site potential $u\in L_c^\infty(\RR^d)$ is non positive, and that the single site distribution $\mu$ has a density $f\in L_c^\infty([0,\infty[)$.

Then, for any $\alpha<1$ and $-E' <0$ there exists a finite $C$ such that for any $E\in \RR,\epsilon\ge 0$ satisfying $E+3\epsilon \le -E'$:
\[
\EE \left[ \Tr P_\omega^l(B_\epsilon(E))   \right ] \le  C\, \epsilon^\alpha \, l^{d}
\]
\end{thm}

The proof of the theorem follows from the arguments of \S~\ref{ss-SAtraceSP} and \S~\ref{ss-CHN}, once
a replacement for the estimate \eqref{e-HFlowerbound} has been established. This is provided by
the following lemma. Denote by $\Lambda_l^+ = \{ k \in \Gamma | \, \supp \, u(\cdot-x) \cap \Lambda_l \neq
\emptyset\}$ the set of indices whose coupling constants influence the
value of the potential in the cube $\Lambda_l$.
Recall that the supremum of the support of $f$ is denoted by $\omega_+$.

\begin{lem}
\label{l-lower}
Assume that the $n$-th eigenvalue of the operator $H_\omega^{l}$ satisfies \\
$\lambda_n^l(\omega) \le -E' <0$.
Then
\bea
\rho'(\lambda_n^l(\omega) -E + t )
\le
\frac{\omega_+}{E'} \left [- \sum_{k \in \Lambda^+}\frac{\partial\rho(\lambda_n^l(\omega)
-E + t )}{\partial \omega_k } \right ]
\eea
\end{lem}

\begin{proof}
Let $\psi_n$ be the normalised eigenfunction corresponding  to $\lambda_n^l(\omega)$.
Then $\psi_n$ satisfies by definition $\la \psi_n , (H_0^{l} -
\lambda_n^l(\omega) ) \psi_n \ra = -\la \psi_n, V_\omega \psi_n\ra$.
We have
\bea
\sum_{k \in \Lambda^+} \,  \omega_k  \, \la \psi_n , -u_k(\cdot-k)
\psi_n\ra
= - \la \psi_n, V_\omega  \psi_n \ra = \la \psi_n, (H_0^{l} -
\lambda_n^l(\omega) ) \psi_n\ra \ge E'
\eea
Now we have by the \index{Hellmann-Feynman formula}Hellmann-Feynman theorem
\begin{multline*}
\hspace*{-.7em}
-\sum_{k \in \Lambda^+} \frac{\partial \lambda_n^l(\omega)}{\partial \omega_k}
 =
\sum_{k \in \Lambda^+} \la \psi_n , -u_k(\cdot-k)  \psi_n\ra
 \ge
\omega_+^{-1} \sum_{k \in \Lambda^+} \, \omega_k \, \la \psi_n , -u_k(\cdot-k) \psi_n\ra
 \ge
  \frac{E'}{\omega_+}
\end{multline*}
This gives
\begin{align}
\nn
\rho'(\lambda_n^l(\omega) -E + t )
&=
-
\left [-\sum_{k \in \Lambda^+} \frac{\partial \lambda_n^l(\omega)}{\partial \omega_k }
\right ]^{-1} \sum_{k \in \Lambda^+}\frac{\partial\rho(\lambda_n^l(\omega) -E + t )}{\partial \omega_k }
\\
\label{e-rho'up}
& \le
\frac{\omega_+}{E'}
\left [- \sum_{k \in \Lambda^+}\frac{\partial\rho(\lambda_n^l(\omega) -E + t )}{\partial \omega_k } \right ]
\end{align}
\end{proof}
Note that since $\rho$ is monotone increasing and $u $ is
non-positive, \eqref{e-rho'up} is a non-negative real.

\smallskip

Related models and results as presented in this paragraph are discussed in Section 3.1 of \cite{CombesHKN-2002}.
In \cite{KirschV-2002a} two more classes of generalised alloy type models are analysed.
Firstly, the case where the number of points in $\Gamma \modcap B_1(x)$ is not uniformly bounded, but grows at a controlled rate as $x$ goes to infinity. Secondly, the case where $\Gamma$ is itself a random point process, for example of Poissonian type, cf.~also \cite{CombesH-94b}.

\subsection{Locally continuous coupling constants} \index{coupling constants!locally continuous ---}

In this paragraph we present a Wegner estimate which requires the coupling constants $\omega_k$ to have a continuous distribution merely in a neighbourhood of their extremal value $\omega_+=\sup \supp \, f$. Both the result and its proof are taken from \cite{KirschV-2002a}.

Consider a background Schr\"o\-dinger operator $H_0=-\Delta+ V_{\per}$ with a periodic potential $V_{\per}\in L_{\unif, \loc}^p(\RR^d)$.
Let $H_\omega=H_0+V_\omega$ be an random operator with an alloy type potential $V_\omega$.
Assume that the coupling constants $\omega_k, k \in \ZZ^d$ take values in the bounded interval $[\omega_-, \omega_+]$.
 By modifying the periodic background potential we may consider only the case  that the coupling constants are non-negative.
For a value $\omega_c\in [0, \omega_+]$ introduce the auxiliary periodic potential $V_c = \omega_c\sum_{k\in\ZZ^d}u(x-k)$
and the threshold energy $E_c=\inf \sigma(H_0+V_c)$.

\begin{thm}
Assume that the single site potential $u\in L_c^\infty(\RR^d)$ is non positive, and that the restriction of the single site distribution $\mu_c:=\mu|_{]\omega_c,\omega_+]}$ has a density $f\in L^\infty$.

Then, for any $\alpha<1$ and $E' <E_c$ there exists a $C$ such that for any $E\in \RR,\epsilon\ge 0$ satisfying $E+3\epsilon \le E'$:
\[
\EE \left[ \Tr P_\omega^l(B_\epsilon(E))   \right ] \le  C\, \epsilon^\alpha \, l^{d}
\]
\end{thm}

\begin{proof}
The value  $\omega_c$ is a critical one for the random variable $\omega_k$ in the sense that for
$\omega_k >\omega_c$ we know that it is continuously distributed, while for smaller
values we do not know anything.
We introduce a corresponding decomposition of the 'probability' space
$\Omega_l:=\times_{k \in \Lambda^+} \RR \cong \RR^L$.
This is the part of the randomness on which the restricted Hamiltonian $H_\omega^l$ depends.
For a given configuration of coupling constants $\{\omega_k\}_{k\in \Lambda^+}$ set
\[
\Lambda^{ac} (\omega)= \{ k \in \Lambda^+| \, \omega_k > \omega_c\}
\]
This defines an equivalence relation on $\Omega_l$ by setting for any $A \subset \Lambda^+$
\[
\Omega(A):= \{\omega | \, \Lambda^{ac}(\omega) =A\} \
\]
Consequently
\be
\label{e-deco}
 \sum_{A \subset \Lambda^+} \int_{\RR^L} \prod_{k \in \Lambda^+} d\mu_k(\omega_k)\,
\chi_{\Omega(A)}  (\omega) =1.
\ee
Split the potential now into two parts, a singular and an absolutely continuous one. The singular one
\begin{align*}
V_\omega^s(x) &:= \sum_{k \in \Lp, \omega_k \le \omega_c}
\omega_k u_k(x-k)  + \sum_{k \in \Lp, \omega_k > \omega_c} \omega_c u_k(x-k)
\ge V_c(x)
\intertext{will be considered as part of the background operator, while the absolutely continuous one}
 V_\omega^{ac}(x) &:= \sum_{k \in \Lp, \omega_k > \omega_c} r_k u_k(x-k)
=\sum_{k \in \Lambda^{ac}} r_k u_k(x-k)  , \quad\text{ with } r_k = \omega_k
-\omega_c> 0
\end{align*}
will be used for spectral averaging.

Consider an eigenvalue $\lambda_n^l \le E' <E_c$ and an eigenfunction $ H_\omega^l \psi_n = \lambda_n^l \psi_n$ and set $\delta = E_c-E'$.
We have
\bea
 -\la \psi_n , V_\omega^{ac}  \psi_n\ra
 = \la \psi_n , (H_0^l + V_\omega^s -\lambda_n^l) \psi_n\ra \ge \la \psi_n , (H_0^l + V_c -\lambda_n^l) \psi_n\ra \ge
\delta
\eea
which implies similarly as in Lemma \ref{l-lower}
\begin{multline*}
\hspace*{-.9em} - \sum_{j \in \Lambda^{ac}} \frac{\partial \lambda_n(\omega)}{\partial \omega_j}
\ge
 \frac{1}{\omega_+-\omega_c} \sum_{j \in \Lambda^{ac}} r_j \, \la \psi_n , -u_j(\cdot-j)  \psi_n\ra
=
- \frac{\la \psi_n , V_\omega^{ac}  \psi_n\ra }{\omega_+-\omega_c} \ge  \frac{\delta}{\omega_+-\omega_c}
\end{multline*}
 Consider first the case $\emptyset \neq A\subset \Lp$ and estimate
\begin{multline*}
\int_{\RR^L} \prod_{k \in\Lp} d\mu(\omega_k) \chi_{\Omega(A)}
(\omega)  \sum_{n \in \NN} \int_{-2 t }^{2 t } \rho'(\lambda_n(\omega) -E+ t )
\\
\le
\frac{\omega_+-\omega_c}{\delta}
\int_{\RR^L} \prod_{k \in\Lp} d\mu(\omega_k) \chi_{\Omega(A)}  (\omega) \sum_{n \in \NN}
\int_{-2 t }^{2 t } d t \left [- \sum_{j \in \Lambda^{ac}}
\frac{\partial\rho(\lambda_n(\omega) -E + t)}{\partial \omega_j } \right]
\end{multline*}
As we know that all sites $j \in \Lambda^{ac}$ correspond to coupling constants $\omega_j$
with values in the absolutely continuous region of the conditional density $f$
we may estimate as in  \S~\ref{ss-SAtraceSP}:
\bea
\lefteqn{
 -\sum_{n \in \NN}\int_\RR d\mu( \omega_j) \,  \chi_{\Omega(A)} (\omega)  \,
\frac{\partial\rho(\lambda_n(\omega)-E+ t )}{\partial \omega_j}
}
\\
&=&
-\sum_{n \in \NN}\int_{\omega_c}^{\omega_+} f(\omega_j) d \omega_j  \,
\frac{\partial\rho(\lambda_n(\omega)-E+ t)}{\partial \omega_j}
\\
& \le  &
\|f\|_\infty  \sum_{n \in \NN}\left [ \rho (\lambda_n(\omega,\omega_j=\omega_c)-E+ t )- \rho( (\lambda_n(\omega,\omega_j=\omega_+)-E+ t ) \right]
\eea
which can be estimated as in \S~\ref{ss-CHN}.
We have to say something how we deal with the special case $A=\emptyset$. In
this situation $V_\omega^{ac} \equiv 0$ and $H_\omega = H_0+V_\omega^s \ge
H_0+V_c\ge E_c$. Thus there are no eigenvalues in the considered energy
interval for this potential configuration.

Finally we use the decomposition \eqref{e-deco} to finish the proof:
\bea
\lefteqn{\EE \left (\Tr P_\omega^l([E-\epsilon,E+\epsilon]) \right ) }
\\
& \le & \sum_{A \subset \Lambda^+} \int_{\RR^L} \prod_{k \in \Lambda^+} d\mu(\omega_k)\,
\chi_{\Omega(A)}   (\omega) \sum_{j \in A}
\frac{(\omega_+-\omega_c)}{\delta} \, 4\epsilon \|f\|_\infty
C(\alpha) \epsilon^{\alpha-1}
\\
& \le &
4 C(\alpha) \frac{(\omega_+-\omega_c)}{\delta} \|f\|_\infty    \epsilon^{\alpha} l^d
\eea
\end{proof}

\subsection{Potentials with small support}\mindex{small support}
In \S~\ref{ss-SAtraceSP} we used in a crucial step in the derivation of the Wegner estimate that the single site potentials were lower bounded by a partition of unity
\[
\sum_{k\in \ZZ^d} u(x-k) \ge \kappa \quad \text{ on } \RR^d
\]
It is of natural interest, whether a Wegner estimate holds if this condition is relaxed. In this paragraph we consider the case that $u$ is of fixed sign, but has small support. More precisely, we assume throughout this paragraph merely that there is an open set $\cO\subset \RR^d$ and a positive $\kappa$ such that
\be
\label{e-smsu}
u(x) \ge \kappa \chi_{\cO}
\ee
The first Wegner estimates under this relaxed condition on the single site potential were derived for spectral boundaries, i.e.~for energies either near the bottom of the spectrum, or near an internal spectral boundary. The case of the infimum of the spectrum was treated e.g.~in \cite{Klopp-95a,Kirsch-96}, and internal spectral boundaries in \cite{KirschSS-1998a}. These works derived a Wegner estimate where the volume dependence of the bound was growing faster than linearly. Thus they were not sufficient to derive a result on the regularity of the IDS, cf.~our discussion in \S~\ref{ss-ContIDS}. A linear bound for the same energy regimes was found in \cite{BarbarouxCH-97b,CombesHN-2001}.

By now there are Wegner estimates which under the relaxed condition \eqref{e-smsu} derive Wegner estimates
valid for any bounded interval on the energy axis. We consider first the one-dimensional case where the result is particularly clear and the proof simple. We follow \cite{KirschV-02b} in the presentation, see \cite{Giere-02,CombesHK-03} for other proofs.

Assume that the single site potential $u$ and the periodic potential $V_{\per}$ are bounded.

\begin{thm}
\label{t-smsud1}
Assume that $u$ is compactly supported and obeys \eqref{e-smsu}. For any $E_0 \in \RR$ there exist a constant $C$ such that
\begin{equation}
\label{e-smsud1}
\EE \left [ \Tr P_\omega^l( B_\epsilon(E)) \right ]
\le C \, \epsilon \, l , \quad \forall \, \epsilon \in [0,1] , E \le E_0, l \in \NN
\end{equation}
\end{thm}
Thus the IDS is \index{Lipschitz continuity}Lipschitz-continuous.

\begin{proof}
First we show how to replace \eqref{e-upartUnity} in the case of small support. By shifting the origin of  $\RR^d$ we may assume without loss of generality that there is a  $s>0$ such that $\Lambda_s(0)\subset\cO $. Likewise, we may assume $\kappa=1$ by rescaling the single site potential and the coupling constants.

We set $S= \bigcup_{k\in \tL} \Lambda_s(k)$. The \index{Hellmann-Feynman formula}Hellmann-Feynman theorem gives us
\begin{eqnarray*}
\sum_{k \in \tL} \frac{\partial \lambda_n^l(\omega)}{\partial \omega_k}
=
\sum_{k \in \tL} \la \psi_n, u(\cdot-k)\psi_n \ra
\ge
\int_{S} |\psi_n|^2.
\end{eqnarray*}
where $\psi_n$ is a normalised eigenfunction corresponding to $\lambda_n^l(\omega)$.

If the integral on the right hand side would extend over the whole of $\Lambda_l$ it would be equal to $1$ due to the normalisation of $\psi_n$.  A priori the integral over $S$ could be arbitrary close to zero, but the following Lemma shows that this is not the case.

\begin{lem}
\label{l-uc}
Let $I$ be a bounded interval and $s>0$. There exists a constant $c>0$ such that
\begin{eqnarray*}
\int_{\Lambda_s(k)} |\psi|^2 \ge  c\int_{\Lambda_1(k)} |\psi|^2
\end{eqnarray*}
for  all $l\in \NN$, all $k\in \tilde\Lambda_l$ and for any eigenfunction $\psi$ corresponding to an eigenvalue $E\in I$ of $H_\omega^l$.
\end{lem}

\begin{proof}[Proof of the Lemma]
For
\begin{equation*}
 \phi ( y) := \int_{\Lambda_s(k+y)} dx \, |\psi(x)|^2  = \int_{\Lambda_s(k)} dx \, |\psi(x-y)|^2
\end{equation*}
 one has
\begin{eqnarray*}
\left | \frac{\partial}{\partial y}  \phi ( y)   \right |
& = &
\left |  \int_{\Lambda_s(k)} dx \, \left [  \frac{\partial}{\partial y} \psi (x-y)  \right ] \overline{ \psi(x-y) }
        + \int_{\Lambda_s(k)}  dx  \, \psi(x-y) \, \frac{\partial}{\partial y} \overline{ \psi(x-y) } \right |
\\
& \leq & 2 \left \|  \psi \right \|_{ L^2( \Lambda_s(k+y) ) } \left \| \psi'\right \|_{ L^2( \Lambda_s(k+y) ) }.\end{eqnarray*}
\mindex{Sobolev estimates}Sobolev norm estimates (e.g.~Theorems 7.25 and 7.27 in \cite{GilbargT-83}) imply
\[
\|  \psi' \|_{L^2( \Lambda_s(k+y) )} \le  C_5 \,  \|  \psi \|_{L^2( \Lambda_s(k+y) )} +\|  \psi'' \|_{L^2( \Lambda_s(k+y) )}
\]
By the eigenvalue equation  we have
\begin{equation}
\left | \frac{\partial}{\partial y}  \phi ( y)   \right |
\le C_6 \  \|  \psi  \|_{ L^2( \Lambda_s(k+y) ) }^2 = C_6 \, \phi ( y), \qquad C_6 =C_6(\|V_{\per}+V_\omega-E\|_\infty)
\end{equation}
\index{Gronwall's Lemma}Gronwall's Lemma  implies  $ \phi(y)  \le  \exp(C_6 |y| ) \, \phi(0)$ and thus
\begin{eqnarray*}
\int_{\Lambda_1(k)} |\psi|^2 \le  e^{C_6}    \ s^{-1} \int_{\Lambda_s(k)} |\psi|^2
\end{eqnarray*}
\end{proof}
Thus $\int_{S} |\psi|^2 \ge  c\int_{\Lambda_l} |\psi|^2$ with the same constant as in Lemma \ref{l-uc}.
\medskip

It remains to estimate the \index{spectral!--- shift function}spectral shift
\begin{equation}
\label{e-TrEst}
\sum_{n \in \NN}
\left [ \rho (\lambda_n^l(\omega,j=\max) -E + t)- \rho (\lambda_n^l(\omega,j=\min) -E + t)\right ]
\end{equation}
We may assume without loss of generality that the single site potential $u$ is supported in the interval $[-R,R]$.
Introduce now the operator $H_\omega^{l,D}(j=\max)$ which coincides with $H_\omega^{l}(j=\max)$ up to additional Dirichlet boundary conditions at the points $j-R$ and $j+R$. Likewise, $H_\omega^{l,N}(j=\min)$ coincides with $H_\omega^{l}(j=\min)$ up to additional Neumann boundary conditions at the same points.
Their eigenvalues are $\lambda_n^{l,D}(\omega,j=\max)$ and $\lambda_n^{l,N}(\omega,j=\min)$, respectively. By \index{Dirichlet-Neumann bracketing}Dirichlet-Neumann bracketing, the square brackets in \eqref{e-TrEst} are bounded by
\beq
\label{DNb}
\rho (\lambda_n^{l,D}(\omega,j=\max) -E + t)- \rho (\lambda_n^{l,N}(\omega,j=\min) -E + t)
\eeq
Since for both  $*\!=\!D,N$ the Hamiltonian  $H_\omega^{l,*} $ is a direct sum of an operator $H_\omega^{j,*}$ acting on $L^2(j\!-\!R,j\!+\!R)$ and another one $H_\omega^{c,*}$ acting on  $L^2(\Lambda_l \setminus [j\!-\!R,j\!+\!R])$ the sum over the terms in \eqref{DNb} can be separated:
\beq
\label{e-cOp}
&&\sum_n \rho (\lambda_n^{c,D}(\omega ) -E + t)- \rho (\lambda_n^{c,N}(\omega ) -E + t)
\\
\label{e-jOp}
&+&
\sum_n \rho (\lambda_n^{j,D}(\omega,j=\max) -E + t)- \rho (\lambda_n^{j,N}(\omega,j=\min) -E + t)
\eeq
Note that the eigenvalues in \eqref{e-cOp} are independent of $\omega_j u(\cdot-j)$.
Since the difference in the boundary conditions is a rank two perturbation in resolvent sense (see e.g.~\cite{Simon-95}), the \index{interlacing theorem}interlacing theorem says that
\[
\rho (\lambda_n^{c,D}(\omega,j=\max) -E + t) \le \rho (E_{n+2}^{c,N}(\omega,j=\max) -E + t)
\]
A telescoping argument bounds the whole sum in \eqref{e-cOp} by twice the total
variation of $\rho$, which is equal to one. The sum in \eqref{e-jOp} we estimate by
\begin{multline*}
\Tr\left [\chi_{[E-3\epsilon ,\infty[} (H_\omega^{j,D}(j= \max))- \chi_{]E+3\epsilon,\infty[} (H_\omega^{j,N}(j=\min))\right ]
\\
\le
2 +
\Tr\left [\chi_{[E-3\epsilon ,\infty[} (H_\omega^{j,D}(j=\min)+\|u_j\|_\infty)- \chi_{]E+3\epsilon,\infty[} (H_\omega^{j,D}(j=\min))\right ]
\end{multline*}
which is bounded by a constant, that is independent of $\Lambda_l$, $j \in \tL_l$ and $\epsilon >0$ (on which $\rho$ depends).
\end{proof}
\bigskip

In the remainder of this section we give an overview of various Wegner estimates which are based or related to techniques presented at the beginning of Section \ref{s-WK}. However, we refrain from giving the proofs of this results but refer to the original articles.
\bigskip

In \cite{CombesHK-03} Combes, Hislop and Klopp study multi-dimensional alloy type models with single site potentials of small support, and establish the \index{H\"older continuity!--- of IDS}H\"older continuity of the IDS at all energies.
They consider the case where the single site potential $u\in L^\infty_c(\RR^d)$ is non-negative and not identically equal to zero, and treat three different situations. In all of them the unperturbed background operator $H_0=(-i\nabla-A)^2+V_{0}$ may include a \index{magnetic potential}magnetic vector potential $A$ and a (scalar) electric potential $V_0$.
They have to satisfy some regularity conditions such that $H_0$ is selfadjoint and has $C_0^\infty(\RR^d)$ as an operator core. The coupling constants are distributed according to a bounded, compactly supported density.
\begin{enumerate}[\rm (i)]
\item
\label{i-IDS0Hc}
The background operator $H_0$  has an IDS $N_0$, which is H\"older continuous
\[
|N_0(E_2)-N_0(E_1)| \le C_0 |E_2-E_1|^{\tilde \alpha}
\]
with H\"older exponent $\tilde \alpha\in ]0,1]$. The constant $C_0=C_0(I)$ can be chosen uniformly for $E_2,E_1$ in a given compact interval $I$.
\item
The background operator $H_0$ is periodic with respect to the lattice $\ZZ^d$ and has the 
\index{unique continuation property}unique continuation property, cf.~for instance \cite{Wolff-95}. The set $\{x| \, u(x) >0\}$ contains an open subset of $\RR^d$.
\item
\label{i-LandauHc}
Let the space dimension be $d=2$. Let $H_0=(-i\nabla-A)^2+V_{\per}$ be a Landau Hamiltonian with vector potential
$A(x_1,x_2)=\frac{B}{2} (-x_2,x_1) $ where $B>0$ is the (constant) magnetic field strength. The magnetic flux trough a unit cell satisfies the rationality condition
\be
\label{e-ratB}
B\in 2\pi \, \QQ
\ee
The scalar potential $V_{\per}$ is a $\ZZ^d$-periodic function in $L_{\loc}^2(\RR^d)$.
\end{enumerate}
In case \eqref{i-IDS0Hc} set $\alpha_c = \frac{\tilde\alpha}{\tilde\alpha +2}$, otherwise $\alpha_c = 1$.

In a follow up work \cite{CombesHKR-03} on the Landau Hamiltonian in collaboration with Raikov condition \eqref{e-ratB} has been removed.
\begin{thm}
Let $H_\omega$ be an alloy type model satisfying either one of the above conditions \eqref{i-IDS0Hc}--\eqref{i-LandauHc}.
Then, for each $\alpha\in]0,\alpha_c[$, the IDS of $H_\omega$ is H\"older continuous at all energies, with H\"older exponent $\alpha$.
\end{thm}

\subsection{H\"older continuous coupling constants} \index{coupling constants!H\"older continuous ---}
\index{H\"older continuity!--- of coupling constants}
There is special interest to extend the known Wegner estimates to coupling constants with singular distribution.
The reason is the intuitive interpretation of the coupling consonants as nuclear charge numbers modulating the strength of atomic potentials. In this case their distribution would correspond to a pure point measure.

So far the best result in this direction for multi-dimensional alloy type models was obtained by Stollmann in \cite{Stollmann-2000b}.
In Remarks \ref{r-subexpWE} and \ref{r-1dim} we mentioned already results for one-dimensional models with singular randomness.

Stollmann's result applies to a single site measure $\mu$ which has compact support $[\omega_-, \omega_+]$ and which is merely H\"older continuous. For $\epsilon\ge 0$ denote
\[
s(\epsilon):=s(\mu,\epsilon):= \sup\{\mu([a,b])| \, b-a \le \epsilon \}
\]


\begin{thm}
Let $H_\omega$ be an alloy type model as in
Definition \ref{d-AM}, but let the single site measure be merely H\"older continuous.
Assume additionally that the single site potential obeys
$u\ge \chi_{[0,1]^d}$.
Then for any $E\in \RR$ there exists a constant $C$ such that for any open interval $I\subset ]-\infty, E[$ and any $l\in 2\NN$
\[
\PP\{\omega| \, \sigma(H_\omega^l) \modcap I \neq \emptyset \} \le C \, s(|I|) \, l^{2d}
\]
holds.
\end{thm}

\subsection{Single site potentials with changing sign} \index{single site!--- potential!--- of changing sign} 

\label{ss-KHK}
First Wegner estimates for indefinite alloy type potentials were derived in \cite{Klopp-95a}.
In \cite{HislopK-02} Hislop and Klopp combine the techniques from \cite{Klopp-95a} and  \cite{CombesHN-2001} to prove a Wegner estimate valid for general indefinite single site potentials and for  energy intervals at edges of $\sigma(H_\omega)$. They assume the single site potential $u\in C_c (\RR^d)$ satisfies $u(0)\neq 0$. The density $f\in L_c^\infty$ of the random variable $\omega_0$ (which may be in fact the conditional density with respect to $\omega^{\bot 0} := (\omega_k)_{k \not= 0}$) is assumed to be piecewise absolutely continuous. For any $\alpha < 1$ and any compact energy interval $I$ strictly below the spectrum of the unperturbed operator $H_0$ they prove
\begin{eqnarray*}
\PP \{\sigma (H_\omega^l) \cap I \not= \emptyset \} \le C \, |I|^\alpha \, l^d
\end{eqnarray*}
where the constant $C$ depends only on $\alpha, d$ and the distance between the interval $I$ and $\sigma(H_0)$. With a sufficiently small global coupling constant $\lambda$ the same result holds for the operator $H_0 + \lambda V_\omega$ for $I$ in an internal spectral gap of $H_0$. The results of \cite{HislopK-02} extend to more general models including certain operators with random magnetic field.

In \S~\ref{ss-indefWE} we discuss in more detail an alternative technique to obtain a Wegner estimate valid for single site potentials which change sign. It applies to a more restricted class of potentials but yields stronger results. In particular, it proves the Lipschitz continuity of the IDS at all energies.

\subsection{Uniform Wegner estimates for long range potentials} \index{long range!--- single site potentials}
\index{single site!--- potential!--- of long range}
Kirsch, Stollmann and Stolz proved in \cite{KirschSS-1998b} a Wegner estimate for single site potentials which do not need to have compact support, but merely need to decay sufficiently fast. They consider $u$ of polynomial decay
\be
\label{e-uploydec}
|u(x)|\le C (1+|x|^2)^{-m/2}
\ee
where $m>0$ is required to be larger than $3d$. For certain applications they can also deal with the case where $m$ is only larger than $2d$, cf.~\cite{KirschSS-97,Zenk-02}.

For such single site potentials
the restrictions of the alloy type potential to two finite cubes may be correlated,  even if the cubes are far apart.
This makes it necessary to use a enhanced version of the \index{multiscale analysis}multiscale analysis for the proof of localisation. Among others, this requires a \emph{uniform Wegner estimate}. By this we mean a Wegner estimate for the Hamiltonian $H_\omega^l$ restricted to the cube $\Lambda_l$ which is uniform in the coupling constants $\omega_k$ with index $\|k\|_\infty > r$
where $r$ is a function of $l$.

To formulate the Wegner estimate from  \cite{KirschSS-1998b} let us first introduce some notation.
For any cube $\Lambda\subset \RR^d$ and $\tL = \Lambda \modcap \ZZ^d$ we denote by $\Pi_\Lambda$ the projection
\[
\Pi_\Lambda \colon \Omega \mapsto \medtimes_{\tL} \supp \, \mu   \qquad
\Pi_\Lambda (\omega) := \{\omega_k\}_{k \in \tL}
\]
For a measurable set $A\subset \Omega$ we denote by $A_\Lambda^*$ the cylinder set
\[
A_\Lambda^* := \Pi_\Lambda^{-1}(\Pi_\Lambda A)
=\{\omega \in \Omega| \, \exists \omega' \in A \text{ such that } \Pi_\Lambda (\omega') = \Pi_\Lambda (\omega)\}
\]
The following observation plays a crucial role in the enhanced multiscale analysis.
\begin{lem}
For two disjoint cubes $\Lambda, \Lambda'$ and two events $A,B \in \Omega$, the induced events $A_{\Lambda}^*$ and $B_{\Lambda'}^*$ are independent.
\end{lem}

The following lemma allows one to turn a 'usual' Wegner estimate, as we have it discussed before, into a uniform Wegner estimate. It relies on the polynomial decay of the single site potential \eqref{e-uploydec}.
Let $I$ be a compact interval, $E\in I$ and  $\epsilon \in ]0,1]$.
We denote by $A(E,\epsilon, l )$ the event $\{\omega | \, d(E,\sigma(H_\omega^l))< \epsilon\}$ and use the abbreviations $\Pi_l :=\Pi_{\Lambda_l},A_l^*:=A_{\Lambda_l}^*$.

\begin{lem}
Under the above assumptions there exists a finite constant $c$, independent of $\omega\in \Omega$, $l,r\in \NN$ and $\epsilon \le 1 $ such that
\[
\PP\{A(E,\epsilon, l )_{l+r}^* \}
\le
\PP\{A(E,\epsilon + c r^{-(m-d)}, l ) \}
\]
\end{lem}

\begin{proof}
By definition, for an $\omega \in A_{l+r}^*$ there exists an $\omega' \in A$ such that
\[
\Pi_{l+r} \omega' = \Pi_{l+r} \omega
\]
Thus, the coupling constants of  $\omega$ and $\omega'$ with index $k$ within the cube of size $l+r$ coincide and we have for $x\in \Lambda_l$
\bea
|V_\omega(x)-V_{\omega'}(x)| \le \sum_{|k|_\infty > l+r} |\omega_k - \omega_k'| \, u(x-k)
\le
c' \sum_{|k|_\infty > l+r} |x-k|^{-m} \le c r^{-(m-d)}
\eea
Therefore $d (E, \sigma(H_{\omega'}^l)) <\epsilon$ implies $d (E, \sigma(H_{\omega}^l)) <\epsilon+c r^{-(m-d)}$, 
which proves the lemma.
\end{proof}

Let us have a look on the implications of the preceding lemma for a concrete example. Assume that the single site potential is bounded below on the unit cube around zero by $\kappa >0$. Then we have by Theorem \ref{t-fv-DOS}
    \bea
    \PP \big \{\omega|  \sigma(H_\omega^l) \modcap [E -\epsilon,E+\epsilon ] \neq \emptyset \big \} \le C_W(I) \ \epsilon \ l^d
    \eea
for all $E, \epsilon$ such that $[E -\epsilon,E+\epsilon ]$ is contained in the open interval $ I $.
This Wegner estimate implies its uniform analog
    \be
    \label{e-unifWE}
    \PP \Big( \big \{\omega|  \sigma(H_\omega^l) \modcap [E -\epsilon,E ] \neq \emptyset \big \}_{l+r}^* \Big)
    \le
     C_W(I) \ (\epsilon+c r^{-(m-d)}) \ l^d
    \ee
for sufficiently large $r>0$. In the application in the multiscale analysis, both $\epsilon $ and $r$ are chosen as functions of $l$.  From the estimate in \eqref{e-unifWE} it might seem to be sufficient to choose $m>d$. This is also the minimal requirement to make the alloy type model with long range single site potentials well defined as a densely defined operator. However, for technical reasons, for the multiscale analysis to work one has to assume at least $m>2d$.
Under this assumption one can prove that the spectrum of $H_\omega$ is almost surely pure point near its bottom, and the corresponding eigenfunctions  decay faster than any polynomial, \cite{KirschSS-97,Zenk-02}. To obtain exponential decay of the eigenfunctions, one has to require $m>3d$ \cite{KirschSS-1998b}.

In the paper \cite{Zenk-02} by Zenk the above results have been extended to a model which incorporates random displacements of the single site potentials.

\section{Lipschitz  continuity of the IDS} \index{Lipschitz continuity}
\label{s-KSCH}
In \cite{KotaniS-87} Kotani and Simon extended to continuum alloy type models certain arguments previously used for the derivation of Wegner's estimate for the discrete Anderson model. They treated only the case where the single site potential is the characteristic function of the unit cube, but Combes and Hislop showed in \cite{CombesH-94b} that the same argument extends to non-negative single site potentials with uniform  lower bound on the unit cube. There also some steps of the proof have been streamlined.

One of the ideas in \cite{KotaniS-87} is that in the same way as rank one perturbations are used for discrete Laplacians, positive perturbations may be used in the continuum case. This is related to the Aronszajn-Donoghue Theory \cite{Aronszajn-57a,AronszajnD-1957,AronszajnD-1964,Donoghue-65}. See \cite{Carmona-83,Kotani-1986c,SimonW-86,Simon-95} for more background and references. This was essential, since a finite rank potential in the continuum  may be a Dirac distribution, but not a function.

\begin{thm}
\label{t-fv-DOS}
Let $H_\omega$  as in Definition \ref{d-AM} and assume additionally that there exists an $\kappa >0$ such that
\[
u \ge \kappa \chi_{[-1/2,1/2]^d}
\]
Then for all $E \in \RR$ there exists a constant $C_W=C_W(E)$ such that for all $l \in \NN$
and all intervals $I\subset ]-\infty,E]$
    \be
    \label{e-fv-DOS}
    \EE \big \{  \Tr \big [ P_\omega^l(I) \big ] \big \} \le C_W \ |I| \ l^d
    \ee
\end{thm}

\begin{rem}
(a)
It is sufficient to prove the theorem for the case $\kappa =1$. Since $\omega_0 u= \kappa\omega_0 \, \kappa^{-1} u $, the general case follows by rescaling the coupling constants and single site potentials.

(b)
The statement of the theorem remains true if one uses Neumann or periodic boundary conditions for $H_\omega^l$.

(c)
An explicit formula for the Wegner constant $C_W$ is given in \eqref{e-CWexpl}.
Since \eqref{e-fv-DOS} is linear in the volume it follows $ |N(E_2) -N(E_1)|\le C_W \,  |E_2-E_1|$.
Thus, as we discussed already in \S~\ref{ss-ContIDS}, the \index{density of states}density of states $n(E) := dN(E)/dE$ exists almost everywhere and is bounded by $n(E)\le C_W(E_2) $ for all $E\le E_2$.
\end{rem}

The next four paragraphs are devoted to the proof of Theorem \ref{t-fv-DOS}. Up to some modifications we follow the line of argument in Section 4 of \cite{CombesH-94b}.

\subsection{Partition of the trace into local contributions}
\label{ss-partTr}
In the present paragraph  we derive preparatory estimates on
    \be
    \EE \big \{ \Tr P_\omega^l(I) \big   \}
    \ee
where we do not yet use the specific alloy-type structure of the potential.
They have two aims. Firstly, to decompose the trace to contributions of unit cubes in $\Lambda_l$. This will later facilitate the averaging procedure with respect to random parameters, whose effect on the potential is felt only locally. Secondly, it allows us to reduce the averaging of the trace of the spectral projection to the averaging of the quadratic form of the resolvent. The latter is technically easier to perform.

Denote by $\Delta^l$ and $\Delta_N^l$ the Laplace operator on $\Lambda_l$ with Dirichlet, respectively Neumann boundary conditions.
In \S~\ref{ss-MN} we saw that the potential $V=V_\per+V_\omega$ is \index{infinitesimally bounded}infinitesimally bounded with respect to $-\Delta$ and that the constants in the bound can be chosen uniformly in $\omega\in\Omega$.
This implies that $V$ is infinitesimally form bounded with respect to any of the operators $-\Delta$, $-\Delta^l$ and $-\Delta_N^l$ with bounds uniform in $\omega\in\Omega, l\in \NN$ and the choice of Dirichlet or Neumann boundary conditions.
Consequently, there is a $C_0<\infty$ such that for all $\omega\in\Omega$ and $l\in \NN$
    \[
    |\langle \phi,V \phi \ra|\le \frac{1}{2} \langle \phi, -\Delta^l_N \phi \rangle + C_0 \|\phi\|^2
    \]
which implies
    \be
    \label{e-ulb}
    \langle \phi,H_\omega^l \phi \ra
    \ge \langle \phi,-\frac{1}{2}\Delta_N^l \phi \rangle - C_0 \|\phi\|^2
    \ge - C_0 \|\phi\|^2
    \ee
Thus $H_\omega^l + C_0 $ is a non-negative operator.

\begin{dfn}
\label{d-TrReg}
A monotone decreasing, convex function  $r\colon [0,\infty[ \to ]0,\infty[$ such that
\be
\label{fintrace}
C_\Tr:=C_\Tr(r):=\sum_{n \in \ZZ^d, n_j \ge 0} r\bigg (\frac{\pi^2}{2} \sum_{j=1}^d n_j^2 \bigg ) < \infty
\ee
will be called \index{trace!--- regularising}\emph{trace regularising}.
\end{dfn}

Throughout the rest of this section we denote by $\Lambda$ the unit cube centred at zero.
\begin{rem}
The bound \eqref{fintrace} means that the operator $r(-\frac{1}{2}\Delta_\Lambda^N)$ has finite trace. Namely, the eigenvalues of the Neumann Laplacian on the unit cube are given by
\[
\pi^2 \sum_{j=1}^d n_j^2 \ \text{ \ for all $n \in \ZZ^d$ such that  $ n_j \ge 0, j =1,\dots,d$ }
\]
cf.~for instance \cite{ReedS-78}, page 266. By the spectral mapping theorem the eigenvalues of $r(-\frac{1}{2}\Delta_\Lambda^N)$ are just $r\big ( \frac{1}{2} \pi^2 \sum_{j=1}^d n_j^2 \big )$.

Examples of functions $r$ which are trace-regularising are the exponential functions $r\colon x\mapsto e^{-tx}$ for $t >0$. They have been used in \cite{CombesH-94b} to implement the procedure outlined in this section.
Another choice for $r$ is a sufficiently high power of the resolvent $x\mapsto (x +1)^{-k}$ for $k> d/2$, which was used in \cite{KotaniS-87}. That the operator $x\mapsto (-\frac{1}{2}\Delta_\Lambda^N +1)^{-k}$ is actually 
\index{trace!--- class}trace class can be inferred from \cite{Simon-82c}.

The possibility to choose $r$ from a large class of functions is of interest if one wants to give explicit upper bounds on the density of states.
For instance, Section 3.2 of \cite{HupferLMW-2001a} is devoted to deriving such explicit upper estimates. However, there, following \cite{CombesH-94b}, the function $r(x)= e^{-tx}$ is used. Due to this  choice, the upper bound on the density of states is exponentially growing in the energy.
This can be improved to a merely polynomial growing bound.
Furthermore, if one studies \index{coupling constants!unbounded ---}coupling constants which may take on arbitrarily negative values, the choice of $r$
will determine which moment conditions one has to impose on the negative part of $\omega_0$, 
see also \S~\ref{ss-MagF}.
%
\end{rem}

\begin{prp} With $C_0$ as in \eqref{e-ulb}
\label{TrLoc}
    \[
    \EE \left \{ \Tr P_\omega^l(I) \right \}
    \le r(E_2+C_0)^{-1} \, C_{\Tr}(r) \, \sum_{j\in \tL_l} \left \|  \EE \{\chi_j P_\omega^l(I) \chi_j  \} \right \|
    \]
\end{prp}

\begin{proof}
Since $\frac{1}{r}$ is well-defined and bounded on the compact interval $I:=[E_1,E_2]$, we have
    \bea
    \label{regularize}
    \Tr\left [ P_\omega^l(I)\right ]  = \Tr\left [  r(H_\omega^l+C_0)^{-1}  \, P_\omega^l(I)  \, r(H_\omega^l+C_0) \right ]
    \eea
Furthermore, by spectral calculus and since for positive operators $A,B$ we have $\Tr(AB) \le \|A\| \ \Tr(B)$, the above line is bounded by
   \[
   r(E_2 +C_0)^{-1}  \Tr\left [ P_\omega^l(I)  \, r(H_\omega^l +C_0) \right ]
   \]
According to the direct sum decomposition
    \[
    L^2(\Lambda_l) = \bigoplus_{j \in \tL_l} L^2(\Lambda+j)
    \]
we consider the Laplace operators $-\Delta^{j,N}$ on $L^2(\Lambda+j)$ with Neumann boundary conditions.
Dirichlet-Neumann bracketing\index{Dirichlet-Neumann bracketing} implies
    \be
    \label{Neumann-senkt}
    H_\omega^l +C_0 \ge  -\frac{1}{2}\Delta^l_N \ge - \frac{1}{2} \bigoplus_{j \in \tL_l} \Delta^{j,N}  =: \dirH
    \ee
For a normalised eigenfunction $\phi$ of $H_\omega^l$ corresponding to the eigenvalue $\lambda$ we have by the spectral mapping theorem
    \beq
    \langle \phi, r(H_\omega^l+C_0) \phi\rangle
    =  r(\lambda+C_0)
    =  r(\langle \phi,(H_\omega^l+C_0) \phi\ra)
    \label{anti-Jensen}
    \le  r(\langle \phi,\dirH \phi\rangle )
    \eeq
Applying  Jensen's inequality to the spectral measure of $\dirH$ we estimate \eqref{anti-Jensen} from above by
$ \langle  \phi , r(\dirH) \phi \ra$.
Let $ \phi_n , n \in  \NN$ be an orthonormal basis of eigenvectors of  $H_\omega^l$ with corresponding eigenvalues $\lambda_n, n\in \NN$. We apply the above estimates to the trace \index{trace}
    \begin{multline*}
    \Tr \left [ P_\omega^l (I) r(H_\omega^l+C_0) \right ]
    \le
    \sum_{n\in \NN, \lambda_n \in I} \langle \phi_n, r(H_\omega^l+C_0) \phi_n\ra
    \\
    \le
    \sum_{n\in \NN, \lambda_n \in I} \langle \phi_n, r(\dirH) \phi_n\ra
    \le
    \Tr \left [ P_\omega^l (I) r(\dirH)  \right ]
    \end{multline*}
For the next step we write down the trace with respect to different basis. For each $j \in \tL$ let $\{\psi_n^j| \, n\in \NN\}$ be an orthonormal basis of $L^2(\Lambda+j)$, then  $\{\psi_n^j| \, n\in \NN, j \in \tL\}$ is an orthonormal basis of $L^2(\Lambda_l)$. Since $r(\dirH)\psi_n^j=\chi_j r(-\frac{1}{2}\Delta^{j,N})\chi_j\psi_n^j$ it follows for the trace
    \bea
    \Tr \left [ P_\omega^l (I) r(\dirH)  \right ]
    &=& \sum_{j \in \tL_l} \:  \sum_{n \in \NN}  \: \langle \psi_{j,n}, P_\omega^l (I) r(\dirH) \psi_{j,n} \ra
    \\
    &=& \sum_{j \in \tL_l} \:  \sum_{n \in \NN}  \: \langle \psi_{j,n}, \chi_j P_\omega^l (I)
    \chi_j r(- \half \Delta^{j,N}) \chi_j \psi_{j,n} \ra
    \\
    &=& \sum_{j \in \tL_l} \:  \Tr \left [ \chi_j P_\omega^l (I) \chi_j r(-\half\Delta^{j,N}) \chi_j \right ]  .
    \eea
Thus we have decomposed the trace to contributions from each unit cube. We summarize the estimates so far:
    \[
 \Tr P_\omega^l(I)  \le r(E_2+C_0)^{-1}   \sum_{j \in \tL_l} \:  \Tr \left [ \chi_j P_\omega^l (I) \chi_j r(-\half\Delta^{j,N} ) \chi_j \right ]
    \]
Since $I$ is a bounded interval and $V_\per + V_\omega$ is an infinitesimally small perturbation of $-\Delta^l$ independently of $\omega$, it follows that the dimension of $P_\omega^l(I) L^2(\Lambda_l)$ is bounded by a constant $C_3$ uniformly in $\omega$. Thus
    \bea
    \Tr \left [ \chi_j P_\omega^l (I) \chi_j r(-\half\Delta^{j,N} ) \chi_j \right ]
    \le
    C_3 \, r(0)    \text{ for all } \omega \in\Omega
    \eea
is an upper bound by an integrable majorant and we are able to interchange the trace and the expectation by Lebesgue's theorem on dominated convergence
    \begin{multline*}
    \EE \left \{ \Tr \left [ \chi_j P_\omega^l (I) \chi_j r(-\half\Delta^{j,N} ) \chi_j  \right ] \right \}
    =
    \Tr \left [ \EE \left \{ \chi_j P_\omega^l (I) \chi_j r(-\half\Delta^{j,N} ) \chi_j  \right \} \right ]
    \\
    =
    \Tr \left [ \EE \left \{ \chi_j P_\omega^l (I) \chi_j  \right \} \chi_j r(-\half\Delta^{j,N} ) \chi_j \right ]
    \le
    \left \| \EE \left \{ \chi_j P_\omega^l (I) \chi_j \right \} \right \| \ \Tr \left [ r(-\half\Delta^{0,N} ) \right ]
    \end{multline*}
By assumption, $r$ is \index{trace!--- regularising}trace regularising, so the trace in the last line is finite.
\end{proof}

\subsection{Spectral averaging of resolvents} \index{spectral!--- averaging}
\label{ss-SpecAvR}

Now we consider how resolvents are averaged when integrated over a random parameter. Together with the partition result in the previous paragraph \S~\ref{ss-partTr} this will enable us to complete in \S~\ref{ss-KSCHcompl} the proof of Theorem \ref{t-fv-DOS}.

Apart from this, the spectral averaging result bears in itself a meaning. Consider a nonnegative operator $H$ with discrete spectrum. Its resolvent $ R(E) =(H-E)^{-1}$ has singularities at the eigenvalues of $H$ which are of the form $(\lambda_n-E)^{-1}, \lambda_n \in \sigma(H)$ and thus are not integrable over the energy axis. In other words, for a general vector $\phi$ the function $E\mapsto\langle \phi, R(E)\phi\rangle $ will not have a convergent integral. Now, if $H=H_\lambda$ depends on a random parameter $\lambda$, we might hope that the averaged resolvent  $E\mapsto \int d \PP(\lambda) \langle \phi, R_\lambda(E)\phi\rangle $ will be integrable.
This would mean that the singularities of the resolvent have been smeared out sufficiently by the integral over $\lambda$. The lemma in this paragraph shows that this is actually the case for operators which depend in a specific way on the random parameter.

Consider the following operators on a Hilbert space $\mathcal H$. Let $H$ be a selfadjoint operator, $W$ symmetric and infinitesimally bounded with respect to $H$, and $J$ non-negative with $J^2\le W$. Choose two parameters
\begin{align*}
z & \in \CC_- := \{ z \in \CC | \, \Im z < 0\}
\\
\zeta & \in \overline{\CC_+} := \{ \zeta \in \CC | \, \Im \zeta \ge 0 \}
\end{align*}
and set
   \be
    \label{kzz}
    H(\zeta) := H + \zeta \, W,  \ K(\zeta, z) := J (H(\zeta) - z )^{-1} J
\ee
The following lemma is a slight generalisation of Lemma 4.1 in \cite{CombesH-94b}.
\begin{lem}
\label{AronD}
For all $z \in \CC_- $, all $t >0$ and any normalised $\phi \in \mathcal H$ we have
    \be
    \label{mit-t}
    \left | \int_\RR   \langle \phi, K(\zeta, z) \phi \rangle \, \frac {d\zeta }{ 1 +t \zeta^2}  \right | \le \pi
    \ee
\end{lem}
\begin{proof}
By Pythagoras we have $|\la\phi, (A+iB)\phi  \ra|^2=|\la \phi, A\phi   \ra|^2+ |\la \phi,B\phi \ra|^2$ for any two selfadjoint  operators $A,B$. Thus the norm of $K(\zeta, z)$ is bounded by $|\Im z|^{-1}\, \|J\|^2$.
On the other hand, the equation
\begin{eqnarray*}
- \Im K(\zeta,z)
= B[ ( H( {\bar \zeta} ) - {\bar z} )^{-1} [ (\Im \zeta) W - \Im z] (H(\zeta) -z )^{-1}] \,  B
\end{eqnarray*}
implies
     \be
    \label{e-minx2}
    \| K(\zeta, z) \| \le  |\Im \zeta |^{-1}
    \ee
Here we used that $W (H(\zeta) - z )^{-1}$ is a bounded operator.
Now observe that for all $z \in \CC_-$ the function $ \zeta \mapsto K(\zeta , z )$ is holomorphic and bounded on $\overline{\CC_+}$.
The residue theorem, integration over a closed curve in $\CC$ and the bounds on $K$ imply
    \be
    \label{residue}
     \left | \int_\RR   \langle \phi, K(\zeta, z) \phi \rangle \, \frac {d\zeta }{ 1 +t \zeta^2}  \right |
     = \frac{\pi}{\sqrt t} \,  \| K\big(i/\sqrt t, z\big)\|
    \ee
Together with \eqref{e-minx2}, this completes the proof.
\end{proof}

\begin{rem}
The lemma shows that for the particular family of operators
$H(\zeta)$ in \eqref{kzz}, where $\zeta$ is a random variable with measure $\frac{d\zeta}{1+t \zeta^2}$,
the $\zeta$-averaged resolvents are indeed integrable with respect to the energy. Thus the singularities of the resolvent have been smeared out.
\end{rem}

\subsection{Stone's formula and spectral averaging of projections} \index{Stone's formula}
\label{ss-Stone}
Stone's formula allows one to express the spectral projection in terms of the resolvent. This is handy because the resolvent has some nice analytic properties. In our case we use Stone's formula to derive the analog of \eqref{mit-t}
for spectral projections.

A sequence of bounded operators $A_n, n \in \NN$ on the Hilbert space $\cH$ converges \emph{strongly} (or in \mindex{strong topology}\emph{strong topology})  to $A$ if for every $\phi\in \cH$
\[
\lim_{n \to \infty} \|A\phi -A_n\phi\| =0
\]
\begin{lem}[Stone's formula]
\label{Stone-lem}
Let $H$ be a selfadjoint operator with spectral family denoted by $P(\cdot)$. Then the following limit holds in the strong topology
    \begin{multline*}
     \lim_{\delta \searrow 0} \frac{1}{2 \pi i}
    \int_{E_1}^{E_2}  \left [ (H  - E - i \delta)^{-1} - (H  - E + i \delta)^{-1} \right ]  dE
    \\
    = \frac{1}{2}  \Big [ P \big ([E_1, E_2] \big ) + P \big ( ]E_1, E_2[ \big)  \Big ]
    \end{multline*}
\end{lem}
\begin{proof}
The function
       \begin{align}
    \label{Stone-for}
    f_\delta (x)
    & :=      \frac{1}{ \pi}  \left ( \arctan \frac{x-E_1}{\delta}  - \arctan \frac{x-E_2}{\delta}  \right )
    \\
    \nn
    &=
    \frac{1}{2 \pi i}  \int_{E_1}^{E_2} \! \left [ (x  - E - i \delta )^{-1} - (x  - E + i \delta )^{-1} \right ] dE
    \\
    \nn
    &=
    -\frac{1}{ \pi }  \,  \Im  \int_{E_1}^{E_2} \! ( x - E + i \delta )^{-1} dE
    \end{align}
converges for $\delta \searrow 0$ to
    \[
    \frac{1}{2} \big ( \chi_{[E_1, E_2]} +  \chi_{]E_1, E_2[} \big) .
    \]
Now one applies the spectral theorem to $f_\delta (H) $.
\end{proof}
More details on Stone's formula can be found in \cite{ReedS-80}, or \cite{Weidmann-1980} where the spectral calculus is actually introduced in this way in Section 7.3.
\smallskip

Now let $H(\zeta)$ be as in the last paragraph  and $P(\zeta,I)$ the corresponding spectral projection onto an interval $I$. For a normalised vector $\psi$ in $\cH$ denote  ${\mathcal P}\,  (\zeta):= \langle \psi , J P(\zeta,I) J \psi \rangle $.
The next lemma contains a spectral averaging estimate for ${\mathcal P}$.
\begin{lem}
\label{lem-ProjSpecAv}
Let $\rho\in L^\infty(\RR) \cap L^1(\RR)$. Then
    \be
    \label{e-ProjSpecAv}
    \int_\RR \rho(\zeta) \, {\cP} \, (\zeta) \, d\zeta
    \le  \| \rho\|_\infty |I|
    \ee
\end{lem}
While Combes and Hislop \cite{CombesH-94b} considered only compactly supported $\rho$, it
was first observed in \cite{FischerHLM-1997} that densities with non-compact support can be treated.
There this extension was necessary to derive estimates for Gaussian random potentials.
\begin{proof}
We first consider the special density $\frac{d\zeta}{1+t \zeta^2}$ and an open interval $I$. By Stone's formula
    \begin{equation}
    \label{P2K}
    \int_\RR \frac {d\zeta }{ 1 +t \zeta^2} \,  {\cP}\,  (\zeta)
    \le  -
    \int_\RR \frac {d\zeta }{ 1 +t \zeta^2} \, \lim_{\delta \to 0}
    \frac{1}{\pi} \Im  \int_I dE \langle \psi ,  K(\zeta, E - i \delta)  \psi \ra
    \end{equation}
Note that $\frac{d\zeta}{1+t \zeta^2}$ is a finite Borel measure on $\RR$ and that \eqref{Stone-for} implies that $|f_\delta (\cdot)|$, and hence  $\|f_\delta (H(\zeta))\|$, is bounded by one. Thus we may apply the dominated convergence theorem to interchange the limit and the integration, and bound \eqref{P2K} by
    \be
    \label{nachLeb}
    \frac{1}{\pi} \lim_{\delta \to 0}
    \left | \int_I  dE  \int_\RR \frac{d\zeta}{1+t \zeta^2} \, \langle \psi, K(\zeta, E - i \delta) \psi \rangle  \right | \le |I|
    \ee
The last inequality follows from Lemma \ref{AronD}.
This implies for all $\rho\in L^\infty $ with compact support:
    \bea
     \int_\RR \rho (\zeta) \, {\cP} \, (\zeta) \, d\zeta
    &\le &
    \sup_{\supp \rho} \left [ \rho(\zeta) (1+ t\zeta^2) \right ] \, \
      \int_\RR \frac {{\cP} \, (\zeta)}{ 1 + t\zeta^2}  \, d\zeta
    \\
    & \le &
    \sup_{\supp \rho} \left [ \rho(\zeta) (1+ t\zeta^2) \right ] \, |I|
      \\
    & \to &
     \| \rho\|_\infty\, |I| \text{ for  } t \to 0
    \eea
Finally, assume only that $\rho\in L^\infty\cap L^1$. Set $\rho^y = \rho \, \chi_{\{ x | \, |x| < y \} }$ and decompose  $ \rho = \rho^y + \rho_y$. For $y \to \infty$, $\rho_y  $ tends to zero pointwise.
Since ${\cP}$ is bounded by one,   $\rho  \in L^1(\RR, d\zeta) $ is a $y$-uniform majorant for $\rho_y {\cP} $ and we may apply the dominated convergence theorem to conclude
    \be
    \label{pi}
    \int_\RR \rho(\zeta) \, {\cP} \, (\zeta) \, d\zeta
    =
    \lim_{y \to \infty}
    \int_\RR \rho^y (\zeta) \, {\cP} \, (\zeta) \, d\zeta   \le  \| \rho\|_\infty  \, |I|
    \ee
If $I$ is not open, we write it as an intersection of open, decreasing sets and use monotone convergence to conclude
\eqref{e-ProjSpecAv}.
\end{proof}

\subsection{Completion of the proof of Theorem \ref{t-fv-DOS}}
\label{ss-KSCHcompl}
The results on the localisation of the trace to unit cubes and spectral averaging of projections allow us to assemble the proof of Theorem \ref{t-fv-DOS}.

To estimate the operator norm appearing in Proposition \ref{TrLoc} we may as well bound the corresponding quadratic form since
    \be
    \label{quad-form}
     \left \| \EE \left \{ \chi_j P_\omega^l (I) \chi_j \right \} \right \|
     =
    \sup_{\|\phi\|=1}  \  \langle \phi,\EE \left \{ \chi_j P_\omega^l (I) \chi_j \right \} \phi \ra
    \ee
Now one can apply Fubini's Theorem and  Lemma \ref{lem-ProjSpecAv} with the choice $\rho=f$, ${\cH}=L^2(\Lambda_l)$, $J=\chi_j$, $H=H_0+ \sum_{k\in \tL\setminus j } \, \omega_k \, u(\cdot-k)$, $\zeta =\omega_j$ and $W=u(x-j)$:
    \[
    \langle \phi,\EE \left \{ \chi_j P_\omega^l (I) \chi_j \right \} \phi \ra
    \le \|f\|_\infty \,  |I|
    \]
This bound is $j$-independent and thus yields
    \be
    \label{e-CWexpl}
    \EE \left \{ \Tr P_\omega^l(I) \right \}
    \le
    r(E_2+C_0)^{-1}  \, C_{\Tr}(r) \, \|f\|_\infty \,  |I|  \, |\tL_l|
    \ee
\qed

Now it becomes clear why we introduced the operator $r(H_\omega^l+C_0)^{-1}r(H_\omega^l+C_0)=\Id$ in Proposition \ref{TrLoc}: without this regularisation of the trace we could have estimated
\[
\EE \{\Tr \chi_j P_\omega^l(I)\chi_j \} \le const. \, |I| \, |\tL_l|
\]
However, this would lead to a Wegner estimate with quadratic volume bound.
\smallskip

The role played by $r$ resembles the one of the function $g$ in paragraph \ref{ss-CHN} and Appendix \ref{a-SSF}.

\subsection{Single site potentials with changing sign}
\label{ss-indefWE}

In \S~\ref{ss-KHK} we saw an extension of the Wegner-Kirsch approach to single site potentials of changing sign.
The Kotani-Simon-Combes-Hislop proof of Wegner's estimate also allows such a generalisation \cite{Veselic-2001,Veselic-2002a}, which we present in this section. Its main shortcoming in comparison to the results in \S~\ref{ss-KHK} is that it is restricted to single site potentials 
which have a \index{generalised step function}(generalised) step function form. On the other hand, it is valid not only at spectral boundaries, but on the whole energy axis. Furthermore, it yields the existence of the density of states as a function and upper bounds on it.
\smallskip

\begin{thm}
\label{t-indf}
Let $ L_c^p(\RR^d) \ni w \ge \kappa \chi_{[0,1]^d}$ with  $\kappa>0 $ and $p(d)$ be as in \eqref{e-defp(d)}.
Let $ \Gamma \subset \ZZ^d$ be finite, the \index{convolution vector}\emph{convolution vector} $\alpha=(\alpha_k)_{k \in\Gamma}\in \RR^\Gamma$ satisfy
$\alpha^* := \sum_{k \not= 0} | \alpha_k | < |\alpha_0|$, and
the single site potential be of \emph{generalised step function} form:
\begin{equation}
\label{u}
u(x) = \sum_{k \in \Gamma} \alpha_k \ w(x -k).
\end{equation}
Assume that the density satisfies $f \in W_c^{1,1}(\RR)$.
Then for all $E \in \RR$ there exists a constant $C_W=C_W(E)$ such that
\begin{equation}
\label{e-indf-WE}
\EE \left \{ \Tr P_\omega^l( I) \right \}
\le C_W \  |I| \, l^d , \quad \text{ for all $ \, l \in \NN $ and $ I \subset ]-\infty, E]$}
\end{equation}
\end{thm}
The theorem implies that the \index{DOS|see{density of states}}DOS, the derivative of the IDS, exists for a.e.~$E$ and is locally uniformly bounded: $d N(E)/d E \le C(E_1)  $ for all $E \le E_1$.

\begin{proof}
For simplicity we assume $w= \chi_{[0,1]^d}$.
To estimate $ \EE \{ \langle \phi , \chi_j P_\omega^l(I) \chi_j \phi \rangle \}$ for any normalised  $\phi \in L^2 (\Lambda_l)$ we introduce a transformation of coordinates on the probability space $\Omega$.

For each cube $\Lambda= \Lambda_l$ denote $\Lambda^+ :=\{ \lambda - \gamma |
\ \lambda \in \tilde{\Lambda}, \gamma \in \Gamma \}$ and $L= \# \Lambda^+$. The operator $H_\omega^l$ depends only on the truncated random vector $(\omega_k)_{k \in\Lambda^+}\in \RR^{L}$.  On such vectors acts  a multi-level \index{Toeplitz matrix}Toeplitz matrix $A_\Lambda := \{ \alpha_{j-k} \}_{j, k \in \Lambda^+} $ induced by the convolution vector $\alpha$. The transformation has an inverse $B_\Lambda= \{b_{k,j}\}_{k,j \in \Lambda^+}= A_\Lambda^{-1}$ which is bounded  in the row-sum norm $ \| B_\Lambda \| \le \frac{1}{1-\alpha^*}$. Note that the bound is uniform in $\Lambda_l$. We drop now the subscript $\Lambda$ and denote with $\eta := A \omega$ the vector of the transformed random coordinates. They have the common density\index{common density}
\begin{equation}
\label{k}
 k( \eta)   = | \det B| \, F(A^{-1}\eta)
 \end{equation}
 where $
F(\omega)=\prod_{k \in \Lambda^+} f \left ( \omega_k \right )$ is the original density of the $\omega_k$.
We calculate the potential $V_\omega$ written as a function of $\eta$ (and $x \in \Lambda$):
$$
V_\omega (x)  = V_{B\eta}(x)= \sum_{j \in \tilde{\Lambda}} \eta_j \chi_j (x)
$$
In the new representation of the potential the single site potentials are non-negative, so we can make use of the spectral averaging formula in Lemma \ref{lem-ProjSpecAv}
 \begin{equation}
 \label{spav}
  \int_{\RR} d\eta_j \, k(\eta ) \, s(\eta )
 \le
 |I| \ \sup_{\eta_j} | k (\eta)|, \quad \mbox{ where }s(\eta ) := \langle \phi, \chi_j P_{B\eta}^l (I) \chi_j \phi \rangle
 \end{equation}
Fubini, (\ref{spav}), and the fundamental theorem of calculus  give
\begin{equation}
\label{sup}
\int_{\RR^L} d\eta \, k( \eta ) \, s( \eta )
\le |I| \, \int_{\RR^{L-1}} d\eta^{\bot j} \,
 \sup_{\eta_j} | k (\eta)|
   \le
|I| \, \int_{\RR^{L}} d\eta \       | (\partial_j k)(\eta) |
\end{equation}
Here $\eta^{\bot j}$ is an abbreviation for $\{\eta_k| \, k\in \Lambda^+ \setminus j\}$. The last integral equals
\[
|\det A| \, \int_{\RR^{L}} d\omega \,   | (\partial_j k ) (A \omega  ) |
\]
 which is bounded by
$\| f' \|_{L^1} \sum_{k \in \Lambda^+} |b_{k,j}|$. The proof of the theorem is finished by the  estimate
\begin{equation}
\label{indf-main-estimate}
\EE \left \{ \langle \phi , \chi_j P_\omega^l(I) \chi_j  \phi \rangle \right \}
\le
|I| \ \| f' \|_{L^1} \|B\|
\end{equation}
and Proposition \ref{TrLoc}.
\end{proof}

One drawback of Theorem \ref{t-indf} is the requirement of the weak differentiability of $f$. This excludes in particular the uniform distribution on an interval.
However, in a joint work \cite{KostrykinV-2002} with Kostrykin we have proven:
\begin{prp}
\label{prp-indf-gv}
Let the assumptions of Theorem \ref{t-indf} be satisfied with the only difference that $f$ is the uniform density on an interval.  Let $\Gamma \subset \{k \in \ZZ^d | \, k_i \ge 0 \ \forall \ i = 1, \dots,d  \}$.
Then \eqref{e-indf-WE} holds true.
\end{prp}

In fact, it turns out that the proof of Theorem  \ref{t-indf} can be extended
to density functions of finite total variation. This covers in particular linear combinations of functions in $W_c^{1,1}$ and (finite) step functions. More precisely:
 \begin{prp}
\label{prp-KosV2}
Let the assumptions of Theorem \ref{t-indf} be satisfied, but require $f$ merely to have finite total variation $\|f\|_{\Var}<\infty$. Then the Wegner estimate \eqref{e-indf-WE} holds.
 \end{prp}

The difference to Theorem \ref{t-indf} is that the constant $C_W$ now depends on $\|f\|_{\Var}$ instead of $\|f'\|_{L^1}$, cf.~\eqref{indf-main-estimate}.
The result in the Proposition \ref{prp-KosV2} is proven in \cite{KostrykinV-2003?}. Moreover, there we discuss how the condition $\sum_{k \not= 0} | \alpha_k | < |\alpha_0|$ can be relaxed using the theory of Toeplitz matrices.
Let us give an example in the one dimensional case.

The \index{Toeplitz matrix!symbol of ---}\emph{symbol} of the Toeplitz matrix $A$ is the function
\[
s_A \colon \TT\to \CC,
\quad
s_A(e^{i\theta})= \sum_{j\in\ZZ} \alpha_{j} \, e^{i \, j\theta} , \quad \theta \in ]\!-\!\pi,\pi]
\]
Since we assume that only finitely many components of $\alpha$ are different from zero, $s_A$ is actually a trigonometric polynomial, and thus uniformly continuous and bounded.
Invertibility criteria for $A_\Lambda$ as well as bounds on $B_\Lambda=A_\Lambda^{-1}$ and $B=A^{-1}$ may be established by studying the symbol $s_A$. Consider the case that the
symbol $s_A$ has no zeros  and the winding number  of $s_A$ with respect to $0\in\CC$ vanishes. A theorem of Baxter \cite{Baxter-1963}, see also \cite[Thm.~III.2.1]{GohbergF-1974}, states that
\begin{eqnarray}
\label{e-BH}
\sup_{l\in\NN} (\|B\|,\|B_{\Lambda_{\scriptstyle l}}\|  ) \le const. <\infty
\end{eqnarray}
In this case we have again:
 \begin{prp}
 Let $d=1$ and the assumptions of Theorem \ref{t-indf} be satisfied, but require for the convolution vector $\alpha=(\alpha_k)_{k \in\Gamma}$ merely that the symbol $s_A$ of the associated Toeplitz matrix has no zeros. Then the Wegner estimate \eqref{e-indf-WE} holds.
 \end{prp}

\begin{rem}[Anderson model] \index{Anderson model}
For the discrete Anderson model $ h_\omega = h_0 + V_\omega $ there is a result analogous to Theorem \ref{t-indf}.
Here $h_0$ is the finite difference Laplacian on $l^2(\ZZ^d)$ and $(V_\omega \psi)(n)= V_\omega(n) \psi(n), \, \forall \, n \in \ZZ^d $, a multiplication operator as in the continuum case.
 This is not surprising, since the arguments in \S~\ref{ss-SpecAvR} and \S\ref{ss-Stone} rely only on abstract functional analysis. If fact, as we mentioned earlier, Kotani and Simon were motivated in their treatment \cite{KotaniS-87} of the alloy type model by its discrete counterpart. Moreover, since on $l^2(\ZZ^d)$ the trace can be expressed using the canonical basis as
    \[
     \Tr [ P_\omega^l(I)  ] = \sum_{j \in \tL_l} \langle \delta_j, P_\omega^l(I) \delta_j \ra
    \]
the use of a trace regularising function is not necessary. Here $P_\omega^l$ denotes the spectral projection of the truncation $h_\omega^l$ of the Anderson model $h_\omega$. More precisely, $h_\omega^l$ is the finite matrix $\{\langle \delta_j, h_\omega \delta_k \ra\}_{j,k \in \tL_l}$.

Note that in the discrete case $\chi_j$ is just $\delta_j$. Under the assumptions of Theorem \ref{t-indf} on the coupling constants $\{\omega_j\}_j$ and the single site potential $u$ we have the following Wegner estimate for the Anderson model:
    \be
    \label{e-AM-Lip-Stet-fv}
    \EE \left \{ \Tr [ P_\omega^l(I)   \right \}  \le   \frac{\|f'\|_{L^1}}{1-\alpha^*} \ |I| \ |\tL_l|
    \ee
\end{rem}

\begin{rem}
Theorem \ref{t-indf} can also be understood as a Wegner estimate for the alloy type potential
    \[
    V_\eta (x) = \sum_{k \in \ZZ^d} \eta_k \, \chi_k
    \]
where the coupling constants $\{\eta_j\}_j$ are not any more independent, but 
\index{coupling constants!correlated ---}correlated satisfying certain conditions. See \S~4.2 in \cite{Veselic-03} for a precise formulation. Wegner estimates for correlated coupling constants can also be found in \cite{CombesHM-1998} (cf.~\cite{HupferLMW-2001a}, too).
\end{rem}

The use of the \index{common density}common density $F$, respectively $k$, in the proof of Theorem \ref{t-indf} is conceptually new. One could try to use \index{coupling constants!conditional densities of ---}conditional densities instead by considering the indefinite potential $V_\omega$ in its representation $V_{B\eta}$ as an alloy type potential with dependent coupling constants.
 However, this would require to have uniform upper bounds on the conditional densities, cf.~\cite{CombesHM-1998,HupferLMW-2001a}. They do not seem to be easy to establish for the model considered in this paragraph, and in fact sometimes fail to hold as can be seen in the following example.

\begin{exm}
It is sufficient to consider only one space dimension $d=1$. Let the density function be $f = \chi_{[0,1]}$ and the single site potential $u=\chi_{[0,1]}-\alpha\chi_{[1,2]}$ with $-\alpha\in]-1,0[$. To this model the results of Propositions \ref{prp-indf-gv} and \ref{prp-KosV2} apply.

The restriction of $H_\omega$ to the interval $]-1/2, l-1/2[$ of length $l$ depends only on the coupling constants $\omega_j$ with indices $j\in \{-1, \dots, l-1 \}=:\Lambda^+$. They are transformed by the Toeplitz matrix $A$ into new random variables $\eta_j, j\in \{-1, \dots, l-1 \}$, as in the proof of Theorem \ref{t-indf}. Here the convolution vector is given by $\alpha_0=1, \alpha_1 =-\alpha$.

The conditional density $\rho_j(\eta)=\rho_j^l(\eta)$ of the variable $\eta_j$ with respect to the remaining coupling constants  $\eta^{\bot j}=(\eta_k)_{k\in  \Lambda^+\setminus j}$ in $\Lambda^+$  is given by $\rho_j(\eta)=\frac{k(\eta)}{g_j(\eta)}$. Here $g_j(\eta)= \int k(\eta) d\eta_j$ denotes the marginal density.
The question is whether $\sup_j\rho(\eta)$ is finite.

One calculates the common density to be $ k(\eta) = \prod_{k=-1}^{l-1} \chi_{[0,1]} (\sum_{\nu=-1}^k \alpha^{k-\nu} \eta_\nu)$. For $   \eta_{j+1}\in [0,1], \eta_k =0, \forall k\not= j+1$ we have
    \[
    k(\eta)=
   \prod_{k=j+1}^{l-1} \chi_{[0,1]} (\alpha^{k-j-1} \eta_\nu)=1.
   \]
The marginal density
   \begin{eqnarray*}
   g_j(\eta)
   &=& \prod_{k=-1}^{j-1} \chi_{[0,1]} \left (\sum_{\nu=-1}^k \alpha^{k-\nu} \eta_\nu \right)\, \int
   \prod_{k=j}^{l-1} \chi_{[0,1]} \left(\sum_{\nu=-1}^k \alpha^{k-\nu} \eta_\nu\right) \, d\eta_j
   \\
   &\le&
   \int  d\eta_j \, \prod_{k=j}^{j+1} \chi_{[0,1]} \left (\sum_{\nu=-1}^k \alpha^{k-\nu} \eta_\nu \right )
   \end{eqnarray*}
has for $\eta_{j+1}\in [0,1], \eta_k =0, \forall k\not\in \{j, j+1\}$ the upper bound
   \[
    \int_0^1 \chi_{[0,1]} (\alpha \eta_j +\eta_{j+1})d\eta_j \le \alpha^{-1} (1-\eta_{j+1})
   \]
Particularly, $g_j(\eta) \searrow 0$  for $\eta_{j+1}\nearrow 1$ and thus
   \begin{eqnarray*}
   \sup_\eta \rho_j(\eta) =\infty
   \end{eqnarray*}
Therefore, proofs of a Wegner estimate which require the conditional density to be bounded cannot be applied to this alloy type potential. See also \S~4.3 of \cite{Veselic-03} for another example.
\end{exm}

\subsection{Unbounded coupling constants and magnetic fields} \index{magnetic potential} 
\index{coupling constants!unbounded ---}
\label{ss-MagF}
Motivated by certain physical applications, e.g.~the study of the quantum hall effect (see for instance \cite{BellissardvESB-93,Nakano-98,FroehlichGW-00,Schulz-BaldesKR-00,KellendonkRSB-02,ElbauG-02}), it is desirable to extend the results on the continuity of the IDS to include Hamiltonians with magnetic fields. This is, for instance, done in the papers  \cite{CombesHN-2001,HislopK-02,HupferLMW-2001a}.

We discuss here the results on alloy type potentials obtained in \cite{HupferLMW-2001a} by Hupfer, Leschke, M\"uller, and Warzel,
since they are build on the method presented in the preceding \S\S~\ref{ss-partTr}--\ref{ss-KSCHcompl}. Moreover, their result  allows  the coupling constants to be unbounded, as long as very negative fluctuations are exponentially rare. Actually, the primary interest of their research are Hamiltonians with Gaussian random potentials, so they need to cope with unbounded fluctuations of the potential. The proof is based on earlier techniques from \cite{FischerHLM-1997} --- which in turn use \cite{CombesH-94b} --- and \index{Dirichlet-Neumann bracketing}Dirichlet-Neumann bracketing for magnetic Schr\"o\-dinger operators, as discussed in Appendix A of \cite{HupferLMW-2001a}. The results concerning alloy type potentials are summarised in \S~4.1 of their paper, which we review shortly.
\medskip

Let   $A\colon \RR^d \to \RR^d$ be a measurable vector potential with the property $|A|^2 \in L_{\loc}^1(\RR^d)$. Denote with $ H_0$ the selfadjoint closure of $ \sum_{i=j}^d (i\partial_j + A_j)^2$ defined on smooth functions with compact support. The alloy type Schr\"o\-dinger operator $H_\omega = H_0 + V_\omega$ now incorporates a magnetic field. In \cite{HupferLMW-2001a} it is proven that Theorem \ref{t-fv-DOS} essentially remains true if the magnetic field is included. (Their conditions on the single site potential are slightly different.)

Moreover, two cases are discussed, where the coupling constants are unbounded random variables, and the theorem still remains true. In the first one, it is assumed that $\omega_0$ is non-negative and a certain moment condition is satisfied, roughly  $\EE \{\omega_0^{2d+2}\}<\infty$.

The second one concerns the case where $\omega_0$ is distributed according to the Laplace distribution: $\PP\{\omega_0 \in I \} = \frac{1}{a} \int_I dx\,  e^{|x|/a}$.
Note that the probability that $\omega_0$ assumes very negative values is exponentially small. In fact, in \cite{HupferLMW-2001a} it is noted, that this is a necessary requirement for their techniques to work. The reason for this is that they use $r(x)= e^{-tx}$ as the \index{trace!--- regularising}trace regularising function, cf.~Definition \ref{d-TrReg}. A different choice of $r$ would allow for more general distributions unbounded from below.
\medskip

We conclude this section by listing further literature on random Hamiltonians with magnetic fields.
Works treating the regularity of the IDS of random Schr\"o\-dinger operators with magnetic field include
\cite{Wang-93,BarbarouxCH-97a,Wang-00,HupferLW-2001,HislopK-02}, while the question of the (in)dependence of the IDS on boundary conditions for these models has been treated in \cite{Nakamura-2001,DoiIM-2001,HupferLMW-2001a,HundertmarkS-02?}. A related problem is the analysis of the \index{semigroup!kernel of ---}semigroup kernels of magnetic operators \cite{BroderixHL-2000,BroderixLM-02}.
In \cite{KirschR-00,Raikov-01} the behaviour of the IDS in a strong magnetic field is identified.

The asymptotic behaviour of the IDS near the boundaries of the spectrum in the presence of random magnetic fields was the object of study of the articles
\cite{Nakamura-00a,Nakamura-00b,Nakamura-02a}
which prove high energy and Lifshitz asymptotics for certain models. The high energy asymptotics has been analysed already in \cite{Matsumoto-93,Ueki-94}.

For Schr\"o\-dinger operators with constant magnetic field and random potential generated by
a Poissonian process the different possible behaviours  of the IDS at the bottom of the spectrum are analysed
 in \cite{BroderixHLK-95,Erdos-98,HupferLW-99a,HupferLW-00a,Erdos-01,Warzel-01,HundertmarkKW}.

The analysis of Landau Hamiltonians in the single band approximation is done in \cite{DorlasMP-95,DorlasMP-96,PuleS-97,PuleS-00,HupferLW-2001}. 
Examples of localisation proofs which allow for magnetic fields can be found in \cite{DorlasMP-95,CombesHM-95,DorlasMP-96,DorlasMP-97,Wang-97,BarbarouxCH-97a,
GerminetDB-98b,DorlasMP-1999,FischerLM-00,PuleS-00,AizenmanENSS}.

\appendix

\section{Properties of the spectral shift function} \index{spectral!--- shift function}
\label{a-SSF}
For a exposition of the theory of the spectral shift function (SSF) see \cite{BirmanY-93,Simon-95} or the last chapter of \cite{Yafaev-92}. We review here the relevant facts in our context.

For two \index{trace!--- class}trace class operators $A, B$ the SSF $\xi(\cdot, A, B)$ may be defined by the formula
\be
\label{e-KTF}
\Tr (f(A)-f(B)) =\int f'(\lambda) \, \xi(\lambda, A,B) \, d\lambda
\ee
for functions $f\in C_c^\infty$. Actually, it holds for more general functions, too. By Theorem 8.3.3 in \cite{Yafaev-92}, it is sufficient to assume $f \in C^1(\RR)$ and that $f'$ is the Fourier transform of a finite complex measure. One can define the SSF also via the perturbation determinant from scattering theory
\[
\xi(\lambda, A, B):= \frac{1}{\pi} \lim_{\epsilon\searrow 0} \arg \det [1+(A-B)(B-\lambda -i\epsilon)^{-1})]
\]
In this case the equality \eqref{e-KTF} is called \index{Krein trace formula}\emph{Krein trace formula}.
The SSF can be bounded in terms of the properties of $A-B$, namely
\be
\label{e-TRbound}
\|\xi(\lambda, A, B) \|_{1}\le \|A-B\|_{ J_1}
\ee
Here $J_1$ denotes the ideal of trace class operators and $\|\cdot\|_{ J_1}$ the \emph{trace norm}.
On the other hand, if $A-B$ is \mindex{finite rank}finite rank
\be
\label{e-RANKbound}
\|\xi(\lambda, A, B) \|_{\infty}\le \rank{A-B}
\ee
Since we have an estimate on $\xi$ in the $L^1$ and $L^\infty$-norms, it is natural to ask whether an estimate for the $L^p$-norm, $p\in] 1,\infty[$, may be derived. This indeed turns out to be true and can be understood as an interpolation result, cf.~the proof of Theorem 2.1 in \cite{CombesHN-2001}.

To formulate this bound we have to introduce ideals of 'better than trace class' operators. For a compact operator $C$ denote by $\mu_n(C), n \in \NN$ its \index{singular values}singular values, in non-increasing order. If $C$ is trace class, the sum of the singular values is finite  and equals $\|C\|_{ J_1}$.
We denote by $J_\beta$ the class of compact operators such that
\be
\label{e-Jpnorm}
\|C\|_{J_\beta}:=\Big ( \sum_{n\in \NN} \mu_n(C)^\beta \Big )^{1/\beta} < \infty
\ee
The theory of such operators is classical for $\beta \ge 1$. However, since we want to interpolate between \eqref{e-TRbound} and \eqref{e-RANKbound}, we need to consider operators whose singular values converge not slower,
but \emph{faster} than a
$l^1$-sequence to zero. This leads us to consider operators such that $ \|C\|_{J_\beta}$ is finite, for $\beta$ \emph{smaller than one}. In particular, all such operators are trace class, which explains why they are sometimes called  \index{super-trace class}\emph{super-trace class}. It follows that the SSF may be defined for such operators. They have been studied in \cite{GohbergK-69,BirmanS-77,BirmanS-87}, while their relevance in the present context was recognised in \cite{CombesHN-2001}.

Form these sources we infer the following properties of $J_\beta$. Since for any compact operator $A$ and bounded  $B$ the singular values of the products obey
\be
\label{e-ideal}
\mu_n(AB) \le \|B\| \, \mu_n(A) \quad \text{ and }  \quad \mu_n(BA) \le \|B\| \, \mu_n(A)
\ee
the set $J_\beta$ is an two-sided ideal in the algebra of bounded operators for all $\beta>0$. For $\beta\ge 1$ the functional $A\mapsto \|A\|_{J_\beta}$  is a norm, which is not true for $\beta<1$. More precisely, in this case we have only
\[
\|A+B\|_{J_\beta}^\beta \le \|A\|_{J_\beta}^\beta + \|B\|_{J_\beta}^\beta
\]
This property implies that $\|\cdot \|_{J_\beta}$ is a \index{quasi-norm}\emph{quasi-norm} and that
\[
\dist_\beta(A,B)=\|A-B\|_{J_\beta}^\beta
\]
is a well defined metric on $J_\beta$. The pair $(J_\beta, \dist_\beta)$ forms a complete, separable linear metric space, in which the \mindex{finite rank}finite rank operators form a dense subset.

In \cite{CombesHN-2001} the following $L^p$-bound on the SSF was proven.

\begin{athm}
\label{t-LpSSF}
Let $p\ge 1$ and $A,B$ be selfadjoint operators whose difference is in $J_{\beta}$
where $\beta=1/p$.
Then the spectral shift function $\xi(\cdot,A, B)$ is in $L^p(\RR)$ and
\be
\label{e-LpSSF}
\|\xi(\cdot,A, B)\|_{L^p} \le \big\|A-B\big\|_{ J_{\beta}}^{\beta}
\ee
\end{athm}
This estimate is sufficient for our purposes. There exists a sharp version proven by Hundertmark and Simon in \cite{HundertmarkS-02}. It is used in the result described in Remark \ref{r-HKV}.

\begin{athm}
\label{t-optSSF}
Let $F\colon [0,\infty[\to[0,\infty[$ be a convex function such that $F(0)=0$.
Let $A,B$ be bounded and $C$ a non-negative compact operator such that  for all $N\in \NN$
\be
\label{e-100DMScond}
\sum_{n=N}^\infty \mu_n(|A-B|) \le \sum_{n=N}^\infty \mu_n(C)
\ee
Then
\begin{multline}\nn
\int F\big(|\xi(\lambda,A,B)|\big) d\lambda
\le
\int F\big(|\xi(\lambda,C,0)|\big) d\lambda
=
\sum_{n\in\NN} \big[ F(n)-F(n-1)\big] \, \mu_n(C)
\end{multline}
\end{athm}
Condition \eqref{e-100DMScond} is in particular satisfied if $|A-B|\le C$.
Of course, to apply Theorem \ref{t-LpSSF}, we need a criterion for the operators which arise in our situation
to be in $J_\beta$ for $\beta=1/p \le 1$. So, let's have a closer look at the application of the theory of the SSF to Schr\"o\-dinger operators.

\medskip

Since we are studying Schr\"o\-dinger operators, we cannot expect to deal with trace class perturbations.
However, the theory extends to operator pairs such that the difference of a sufficiently high power of their resolvents is trace class.
More precisely, assume that $H+u,H$ is a pair of lower bounded operators such that
\be
\label{e-trclass}
H+u \ge C_0,H\ge C_0 \text{ and } g(H+u)- g(H) \in J_1
\ee
where $g(x)=(x-C_0+1)^{-k}$ for some sufficiently large $k>0$.
Following \cite{CombesHN-2001}, we denote $g(H+u)- g(H)$ by $V_{\eff}$. This is the 'effective' perturbation, although it is obviously not a multiplication operator.
One defines the SSF of the pair $H+u,H$ as
\be
\label{e-IP}
\xi(\lambda, H+u,H):= - \xi\big(g(\lambda), g(H+u),g(H)\big) \quad
\text{ for $\lambda \ge 0$}
\ee
and $\xi=0$ otherwise. This definition of $\xi$ is independent of the choice of the exponent $k>0$ in $g$. By Theorem 8.9.1 in \cite{Yafaev-92} the trace formula \eqref{e-KTF} holds if $f\in C^2(\RR)$ and $f'$ has compact support.
This conditions are clearly satisfied by the switch function $\rho$ we use in \S~\ref{ss-SAtraceSP} 
and \S~\ref{ss-CHN}.

\smallskip

The purpose of the theorem we are heading to now is twofold: firstly, to establish that $V_{\eff}$ is in $J_\beta$ for  suitable $\beta<1$. Thus, we will be able to apply Theorem \ref{t-LpSSF}. Secondly, to control the upper bound $\|V_{\eff}\|_{J_{\beta}}^{\beta}$ appearing in \eqref{e-LpSSF}.
\medskip

It is well known that operators which may be formally written as $f(x)g(-i\nabla)$ are in the 
\index{Hilbert-Schmidt class}Hilbert-Schmidt class if $f,g \in L^2(\RR^d)$. The product of such two operators is \index{trace!--- class}trace class.
Extending this idea, we want to show for certain operators that they are in some (super-trace)  ideal $J_\beta$, $\beta<1$  by writing them as a product of sufficiently many operators of the type $f(x)g(-i\nabla)$.
For this purpose it is useful to note that the H\"older inequality extends also to the case of exponents smaller than one: let
$a_i\colon \NN \to \CC$, $i =1, \dots, N $ be such that $|a_i(n)|^{p_i}$ is summable, where $p_i >0$ for all $i =1, \dots, N $, and set $ \frac{1}{r}:= \sum_{i=1}^N\frac{1}{p_i}$. Then the pointwise product $\prod_{i=1}^N a_i$ is in $l^r(\NN)$ and
\[
\Big \|\prod_{i=1}^N a_i \Big\|_r     \le \prod_{i=1}^N  \|a_i\|_{p_i}
\]
By applying this to the sequence of \index{singular values}singular values of compact operators, we obtain the following
\begin{alem}
Let $A_i\in J_{p_i}$ for $i = 1, \dots, N $, then $\prod_{i=1}^N A_i$ is in $J_r$ where  $ \frac{1}{r}:= \sum_{i=1}^N\frac{1}{p_i}$ and
\be
\label{e-Hoelder}
\Big \|\prod_{i=1}^N A_i \Big\|_{J_r}     \le \prod_{i=1}^N  \|A_i\|_{ {J_{p_i}}}
\ee
\end{alem}
See also \cite{BirmanS-87} Corollary 11.11.
The following result is taken from \cite{Nakamura-2001}, cf.~Lemma 11 and its proof. For $l \in \NN$ we abbreviate
$\Lambda=\Lambda_l$.
\begin{alem}
\label{l-Baustein}
Let $q > d/2, q \in 2 \NN$, $f\in L_c^\infty(\RR^d)$ and $C_0\in \RR$ be such that $V_{\per}+V_\omega\ge C_0$ for all $\omega$. Then the operator product $f\,(H_\omega^l-C_0+1)^{-1}$ is in the ideal $J_{q}$ and
\[
\| f\chi_{\Lambda}\,(H_\omega^l-C_0+1)^{-1} \|_{J_{q}} \le \|f\,(-\Delta+1)^{-1} \|_{J_{q}}  \le  C(q) \ \|f\|_q
\]
 \end{alem}
\begin{proof}
There exist a bounded \index{extension!operator@--- operator}extension operator $\cE \colon W^{2,2}(\Lambda_l) \to W^{2,2}(\RR^d) $ and its norm is independent of $l \in \NN$, cf.~Section IV.3.2 in \cite{Stein-70}.
Thus we have
\[
f\chi_{\Lambda}(H_\omega^l-C_0+1)^{-1} =f\chi_{\Lambda}(H_\omega-C_0+1)^{-1}(H_\omega-C_0+1)\cE (H_\omega^l-C_0+1)^{-1}
\]
Since $\cE$ is an extension operator, $\chi_\Lambda \, \cE$ is the identity on $W^{2,2}(\Lambda_l)$.
By the ideal property of $J_{q}$ and the boundedness of $\cE$ we have
\begin{multline*}
\|f\chi_\Lambda\,(H_\omega^l-C_0+1)^{-1} \|_{J_{q}}
\\
\le
\|f(H_\omega-C_0+1)^{-1}\|_{J_{q}}\, \|(H_\omega-C_0+1)\cE(H_\omega^l-C_0+1)^{-1}\|
\\
\le
const. \, \|f(H_\omega-C_0+1)^{-1}\|_{J_{q}}
\end{multline*}
By the \index{Kato-Simon inequality}\emph{Kato-Simon} \cite{Kato-73,Simon-79a} or  
\mindex{diamagnetic inequality}\emph{diamagnetic inequality} we have
\be
\label{e-dm}
|f(H_\omega-C_0+1)^{-1} \psi|
\le
f(-\Delta+1)^{-1} |\psi|
\ee
for $\psi \in L^2(\RR^d)$, cf.~%
proof of Theorem 3.3 in \cite{HundertmarkS-02?}. For the 'free' case we know by  Theorem 4.1 in \cite{Simon-79}
$f\, (-\Delta+1)^{-1}  \in J_q$ and
\[
\|f\, (-\Delta+1)^{-1} \|_{J_q} \le \|f\|_q \| h\|_q
\]
where $h(x) = (x^2+1)^{-1}$.
The pointwise inequality \eqref{e-dm} implies by Theorem 2.13 in \cite{Simon-79}
$f\, (H_\omega-C_0+1)^{-1} \in J_q$ and
\[
\|f\, (H_\omega-C_0+1)^{-1} \|_{J_q}
\le
\|f\, (-\Delta+1)^{-1} \|_{J_q}
\]
\end{proof}
\begin{arem}
The result remains true if we consider Neumann boundary conditions instead of Dirichlet ones, and if we include a bounded \index{magnetic potential}magnetic vector potential in the background operator $H_0$, see Lemma 10 in \cite{Nakamura-2001}.
This fact relies on the existence of an appropriate \index{extension!operator@--- operator}extension operator which takes (magnetic) Sobolev functions on $\Lambda$ to (magnetic) Sobolev functions on $\RR^d$.
\end{arem}

The following result establishes that $V_{\eff}=g(H_\omega^l+\omega_+u)- g(H_\omega^l+\omega_-u)$ is indeed super-trace class and that its quasi-norm  $\|V_{\eff}\|_{J_p}$ can be bounded independently of the cube $\Lambda$.

\begin{athm}
\label{t-NakPain}
Let $H_1=-\Delta+V$ and $H_2=H_1+u$, with $\frac{1}{2}C_0\le V, u \in L_{\loc, \unif}^p(\RR^d)$, where $p$ is as in \eqref{e-defp(d)}. Denote by $H_1^l,H_2^l$ the corresponding Dirichlet restrictions to the cube $\Lambda_l$. Assume
\[
\|V\|_{p, \, \unif, \loc} \le C_1, \quad \text{ and } \quad  \|u\|_{p, \, \unif, \loc}\le C_2
\]
and that the support of $u$ is contained in the ball $B_R(x)$. For any
$\beta >0$ choose $k \in \NN$ with $k>\frac{d+4}{2\beta}$. For $g(x)= (x -C_0 +1 )^{-k}$ set
$ V_{\eff}:= g(H_2)-g(H_1)$.

Then $V_{\eff}\in J_\beta$ and $\|V_{\eff}\|_{J_\beta} $ is bounded by a constant which  is independent of $\Lambda_l$ and $x$ and depends on $V$ and $u$ only trough $C_0,C_1,C_2$ and $R$.
\end{athm}
\begin{proof}
Choose a function $f\in C_c^\infty(B_{2R}(x))$ such that $f\equiv 1$ on $B_{R}(x)$.
An iteration of the resolvent formula yields
\begin{align}
\nn
V_{\eff} &= -\sum_{m=0}^{k-1} (H_2-C_0+1)^{-(k-m)} \, u \, (H_1-C_0+1)^{-(m+1)}
\\
\label{e-VeffRG}
 &= -\sum_{m=0}^{k-1} \big[ f^{k-m} \, (H_2-C_0+1)^{-(k-m)}\big ]^* \, u \, \big [f^{m+1} \, (H_1-C_0+1)^{-(m+1)}\big]
\end{align}
By Lemma 9 and Appendix A in \cite{Nakamura-2001}  we have the following representation
\[
f^{\nu} \, (H_1-C_0+1)^{-\nu} = \sum_{i=1}^{N}\prod_{j=1}^{\nu} f_{ij}  \, (H_1-C_0+1)^{-1} B_{ij}
\]
and analogously  for $H_2$. Here $N=N(\nu)$ is an integer which depends only on $\nu$, the functions $f_{ij} \in C_c^\infty(B_{2R}(x)) $ are linear combinations of derivatives of $f$, and $B_{ij}$ are bounded operators with norms independent of $l$. Set
\[
C_{q,f,B} = C(q) \, \max\{\|B_{ij} \| \, \|f_{ij}  \|_{L^q}  \, | \,i=1, \dots,N, j=1, \dots,\nu\}
\]
where $ q \in ]\frac{d}{2},\frac{d}{2}+2]$ and $C(q)$ are as in Lemma \ref{l-Baustein}.
By the same lemma, the ideal property of $J_\beta$ and the H\"older inequality \eqref{e-Hoelder}
we have
\begin{align}
\label{e-NakHoe}
\|f^{\nu} \, (H_1-C_0+1)^{-\nu} \|_{J_t}^t
 \le
\sum_{i=1}^{N}\prod_{j=1}^{\nu} \|f_{ij}  \, (H_1-C_0+1)^{-1} \|_{J_q}^t  \, \|B_{ij} \|^t
 \le
N\, C_{q,f,B}^q
\end{align}
if $t=q/\nu<1$. For $t\ge 1$, $\|\cdot\|_{ J_t }$ is even a norm and a similar estimate holds. Since $u$ is infinitesimally bounded with respect to the Laplacian
\[
C_u := \max_{i,j} \big \{\|u f_{ij}  \, (H_1-C_0+1)^{-1}B_{ij}\| \big \}
\]
 is finite.  Thus, in  analogy to \eqref{e-NakHoe}
\[
\|u f^{\nu+1} \, (H_1-C_0+1)^{-\nu+1} \|_{J_t}^t
 \le
N \, C_u^t \, C_{q,f,B}^{q}
\]
\smallskip

From the preceding we see that \eqref{e-VeffRG} factorises $V_{\eff}$ as a product of bounded operators and $k$ operators which are in $J_{q}$. All the involved operator and super-trace class norms can be bounded independently of $\Lambda_l$ and $x$.
Using the ideal property and H\"older's inequality we see that $V_{\eff}$ is in $J_{r}$ for all $r\ge q/k$ and
$ \| V_{\eff} \|_{J_r} $ is bounded by a constant which is independent of $\Lambda_l$ and $x$.
The way we choose $k$ and $q$ makes it possible to take $r=\beta$.
\end{proof}
\smallskip

\begin{arem}[Properties and relevance of the SSF] \index{spectral!--- shift function}
An exposition of the role played by the SSF in scattering theory can be found in \cite{Yafaev-92}.
The SSF has proven useful in the study of random operators, 
particularly in problems related to \mindex{surface model}surface models,
e.g.~the definition of the density of surface states \cite{Chahrour-99a,Chahrour-00,KostrykinS-00a,KostrykinS-01b}.

Various of its properties are discussed in the literature: monotonicity and concavity \cite{GeislerKS-95,GesztesyMM-99,Kostrykin-00}, the asymptotic behaviour in the large coupling constant \cite{Pushnitski-00,Safronov-01,PushnitskiR-02} and semiclassical limit \cite{Nakamura-99}, and some other bounds \cite{Pushnitski-97,Pushnitski-98,Pushnitski-99}.
\end{arem}

\def\cprime{$'$}

\begin{theindex}

  \item absolutely continuous
    \subitem --- IDS, 132
    \subitem --- coupling constants, 101
    \subitem --- spectrum, 103--105
  \item Aizenman-Molchanov technique, 
        \see{fractional moment method}{134}
  \item alloy type
    \subitem --- model, 99, 101, 104, 130
    \subitem --- potential, 101, 102
  \item amenable group, 113, 116, 125
  \item Anderson model, 129, 131, 160

  \indexspace

  \item b.c., \see{boundary conditions}{100}
  \item Bernoulli
    \subitem --- -Anderson potential, 99
    \subitem --- disorder, 136
    \subitem --- random variables, 99
  \item Bethe-Sommerfeld conjecture, 108
  \item Betti number, 132
  \item binary alloy, 99
  \item Bishop's theorem, \see{volume comparison theorem}{121}
  \item bound state, 103
  \item boundary
    \subitem --- tube, 114
    \subitem --- conditions, 100, 110, 129
      \subsubitem --- independence, 129
  \item bounded random operator, 112
  \item Brownian motion, 120

  \indexspace

  \item common density, 159, 161
  \item convolution vector, 158
  \item coupling constants, 101
    \subitem Bernoulli distributed ---, 131
    \subitem conditional densities of ---, 161
    \subitem correlated ---, 161
    \subitem H\"older continuous ---, 150
    \subitem locally continuous ---, 145
    \subitem unbounded ---, 102, 154, 162
  \item covering manifold
    \subitem abelian ---, 110

  \indexspace

  \item decay
    \subitem --- of Green function, 135
    \subitem --- of eigenfunctions, 103--104
    \subitem --- of heat kernel, 119
  \item delocalised states, 104
  \item density of states, 129, 131, 153
    \subitem --- measure, 106
  \item Dirichlet form, 120
  \item Dirichlet-Neumann bracketing, 106, 127, 148, 154, 162
  \item discontinuous IDS, 132
  \item distribution
    \subitem Bernoulli ---, 131
    \subitem finite dimensional ---s, 102
    \subitem --- function, 106--107, 113, 125
    \subitem --- of coupling constant, 101
    \subitem --- of eigenvalues, 133
    \subitem --- of potential values, 131
  \item domain monotonicity, 121
  \item DOS, \see{density of states}{159}
  \item doubling property, \see{Tempelman property}{114}

  \indexspace

  \item effective Hamiltonian on a manifold, 107
  \item equivariance, 109
  \item ergodic, 99
    \subitem --- action, 101
    \subitem --- operator, 102
    \subitem --- potential, 102
    \subitem --- theorem, 106, 125, 127
      \subsubitem pointwise ---, 114
  \item exhaustion, 114, 115
    \subitem admissible ---, 114--115
  \item extended state, \see{scattering state}{103}
  \item extension
    \subitem Friedrichs ---, 110
    \subitem --- operator, 165, 166
    \subitem --- property, 129
  \item extra-potential, 107

  \indexspace

  \item F\oe lner sequence, 113
  \item Feynman-Kac formula, 120
  \item finite volume integrated density of states, 
        \see{normalised eigenvalue counting function}{106}
  \item fluctuation boundaries of the spectrum, 105
  \item fractional moment method, 134
  \item fundamental domain, 110

  \indexspace

  \item $\Gamma$-trace, 113, 132
  \item generalised step function, 158
  \item Gronwall's Lemma, 148

  \indexspace

  \item H\"older continuity
    \subitem --- of IDS, 131, 142, 149
    \subitem --- of coupling constants, 150
  \item heat equation, 122
    \subitem fundamental solution of ---, 120
    \subitem --- kernel, 119, 121
      \subsubitem off-diagonal decay of ---, 119
  \item Heisenberg group, 111
  \item Hellmann-Feynman formula, 140, 145, 147
  \item Hilbert-Schmidt class, 104, 165
  \item holomorphic family of operators, 140
  \item hyperbolic space, 115

  \indexspace

  \item IDS, \see{integrated density of states}{100}
  \item infinitesimally bounded, 101, 153
  \item initial scale estimate, 134, 135
  \item integrated density of states, 106, 115
  \item interlacing theorem, 149

  \indexspace

  \item Kato-Rellich Theorem, 101
  \item Kato-Simon inequality, 165
  \item Krein trace formula, 163

  \indexspace

  \item Laplace transform, 106, 125
  \item Laplace-Beltrami operator, 108, 109, 132
  \item level statistics, 133
  \item Lifshitz tails, 105, 129
  \item Lipschitz continuity, 131, 147, 152
  \item localisation
    \subitem Anderson ---, 103, 133
    \subitem dynamical ---, 104
    \subitem exponential ---, 103
    \subitem --- interval, 103
    \subitem spectral ---, 103
  \item log-H\"older continuity, 132
  \item long range
    \subitem --- correlations, 103
    \subitem --- single site potentials, 151

  \indexspace

  \item magnetic potential, 149, 162, 166
  \item maximum principle, 122
  \item measurability, 116
  \item measurable family of operators, 116
  \item measure preserving transformations, 102
  \item mobility edge, 104, 129
  \item multiscale analysis, 131, 151

  \indexspace

  \item non-randomness of spectra, 112
  \item normalised eigenvalue counting function, 106

  \indexspace

  \item one-electron Hamiltonian, 98
  \item operator domain, 100

  \indexspace

  \item Pastur-\v Subin convergence criterion, 125
  \item Pastur-\v Subin trace formula, 106, 115
  \item percolation, 132
  \item periodic
    \subitem --- operator on manifold, 108, 115
    \subitem --- potential, 105
  \item points of increase, 113
  \item polyhedral domain, 110
  \item principle of not feeling the boundary, 119
  \item pure point spectrum, 103, 133

  \indexspace

  \item quadratic form, 110, 118
    \subitem --- domain, 117
  \item quantum
    \subitem --- percolation model, 132
    \subitem --- waveguides, 107
  \item quasi-norm, 164

  \indexspace

  \item random metric, 108
  \item relative form-boundedness, 101
  \item relatively bounded, 100
  \item resonance, 136
  \item Ricci curvature, 109

  \indexspace

  \item scattering state, 103
  \item Schr\"odinger equation, 103
  \item Schr\"odinger operator
    \subitem --- on manifold, 107
    \subitem random ---, 98, 109
    \subitem semigroup generated by ---, 143
  \item self-averaging, 106
  \item semigroup, 119, 122, 126
    \subitem contraction ---, 119
    \subitem heat ---, 124
    \subitem --- kernel, 120
    \subitem kernel of ---, 162
    \subitem Markov ---, 120
    \subitem positivity preserving ---, 120
    \subitem Schr\"odinger ---, 143
    \subitem ultracontractive ---, 120
  \item Shulman property, 114
  \item single site
    \subitem --- distribution, 101, 134
    \subitem --- potential, 101, 134
      \subsubitem --- of changing sign, 136, 150
      \subsubitem --- of long range, 151
  \item singular values, 143, 163, 165
  \item spectral
    \subitem --- averaging, 139, 155
    \subitem --- gaps, 108, 110
    \subitem --- measure, 113
    \subitem --- projection, 104, 139
    \subitem --- shift function, 142, 148, 163, 167
    \subitem  --- localisation, 103
  \item SSF, \see{spectral shift function}{142}
  \item stationary, 102
  \item Stone's formula, 156
  \item subadditive process, 106, 127
  \item super-trace class, 143, 163
  \item superadditive process, 106, 127

  \indexspace

  \item Tempelman property, 114
  \item tempered sequence, 114
  \item Toeplitz matrix, 159
    \subitem symbol of ---, 160
  \item trace, 154
    \subitem --- class, 139, 142, 154, 163, 165
    \subitem --- regularising, 153, 155, 162

  \indexspace

  \item unique continuation property, 131, 133, 149

  \indexspace

  \item van Hove property, 115, 124
  \item volume
    \subitem --- comparison theorem, 121
    \subitem --- density, 109
  \item von Neumann algebra, 112

  \indexspace

  \item Wegner constant, 130
  \item Wegner estimate, 129, 134, 135
  \item Weyl asymptotics, 127, 134, 136, 141

\end{theindex}

\end{document}